\documentclass[twocolumn,showpacs,floats,floatfix,superscriptaddress,aps,pra]{revtex4-2}
\usepackage{amsfonts,amssymb,amsmath}
\usepackage{color,calc}
\usepackage[dvips]{graphicx}
\usepackage{bm}

\def\e{\,\text{e}}

\def\e{e}

\def\to{\rightarrow}

\newcommand{\ket}[1]{\vert #1\rangle}
\def\d{\text{d}}

\def\ket#1{| #1 \rangle}

\def\to{\rightarrow}

\begin{document}

\author{Daniel Louzon}
\thanks{These authors contributed equally to this work. \\Corresponding author: \href{mailto:genko.genov@uni-ulm.de}{genko.genov@uni-ulm.de}}
\affiliation{Institute for Quantum Optics, Ulm University, Albert-Einstein-Allee 11, 89081 Ulm, Germany}
\affiliation{Racah Institute of Physics, The Hebrew University of Jerusalem, Jerusalem, 91904, Givat Ram, Israel}

\author{Genko T. Genov}
\thanks{These authors contributed equally to this work. \\Corresponding author: \href{mailto:genko.genov@uni-ulm.de}{genko.genov@uni-ulm.de}}
\affiliation{Institute for Quantum Optics, Ulm University, Albert-Einstein-Allee 11, 89081 Ulm, Germany}

\author{Nicolas Staudenmaier}
\thanks{These authors contributed equally to this work. \\Corresponding author: \href{mailto:genko.genov@uni-ulm.de}{genko.genov@uni-ulm.de}}
\affiliation{Institute for Quantum Optics, Ulm University, Albert-Einstein-Allee 11, 89081 Ulm, Germany}

\author{Florian Frank}
\affiliation{Institute for Quantum Optics, Ulm University, Albert-Einstein-Allee 11, 89081 Ulm, Germany}

\author{Johannes Lang}
\affiliation{Institute for Quantum Optics, Ulm University, Albert-Einstein-Allee 11, 89081 Ulm, Germany}
\affiliation{Diatope GmbH, Buchenweg 23, 88444 Ummendorf, Germany}

\author{Matthew L. Markham}
\affiliation{Element Six Global Innovation Centre, Fermi Avenue, Harwell Oxford, Didcot, Oxfordshire, OX11 0QR, United Kingdom}

\author{Alex Retzker}
\affiliation{Racah Institute of Physics, The Hebrew University of Jerusalem, Jerusalem, 91904, Givat Ram, Israel}
\affiliation{AWS Center for Quantum Computing, Pasadena, CA 91125, USA}

\author{Fedor Jelezko}
\affiliation{Institute for Quantum Optics, Ulm University, Albert-Einstein-Allee 11, 89081 Ulm, Germany}

\title{Robust Noise Suppression and Quantum Sensing by Continuous Phased Dynamical Decoupling}

\date{\today}

\begin{abstract}
We propose and demonstrate experimentally continuous phased dynamical decoupling (CPDD), where we apply a continuous field with discrete phase changes for quantum sensing and robust compensation of environmental and amplitude noise. CPDD does not use short pulses, making it particularly suitable for experiments with limited driving power or nuclear magnetic resonance at high magnetic fields. It requires control of the timing of the phase changes, offering much greater precision than the Rabi frequency control needed in standard continuous sensing schemes. We successfully apply our method to nanoscale nuclear magnetic resonance and combine it with quantum heterodyne detection, achieving $\mu$Hz uncertainty in the estimated signal frequency for a 120\,s measurement.
Our work expands significantly the applicability of dynamical decoupling and opens the door for a wide range of experiments, e.g., in NV centers, trapped ions or trapped atoms.

\end{abstract}

\maketitle

%%%%%%%%%%%%%%%%%%%%%%%%%%%%%%%%%%%%%%%%%%%%%%%%%%%%%%%%%%%%%%%%%%%%%%%%%%%%%%%%%%%%%%%%%%%%%%%%%%%%%%%%%%%%%%%%%%%%%%%%%%%%%%%
%%%%%%%%%%%%%%%%%%%%%%%%%%%%%%%%%%%%%%%%%%%%%%%%%%%%%%%%%%%%%%%%%%%%%%%%%%%%%%%%%%%%%%%%%%%%%%%%%%%%%%%%%%%%%%%%%%%%%%%%%%%%%%%
%\section{Introduction}\label{Section:Introduction}
%%%%%%%%%%%%%%%%%%%%%%%%%%%%%%%%%%%%%%%%%%%%%%%%%%%%%%%%%%%%%%%%%%%%%%%%%%%%%%%%%%%%%%%%%%%%%%%%%%%%%%%%%%%%%%%%%%%%%%%%%%%%%%%
%%%%%%%%%%%%%%%%%%%%%%%%%%%%%%%%%%%%%%%%%%%%%%%%%%%%%%%%%%%%%%%%%%%%%%%%%%%%%%%%%%%%%%%%%%%%%%%%%%%%%%%%%%%%%%%%%%%%%%%%%%%%%%%
%
Developments in quantum technologies are increasingly important for multiple applications in quantum computing, quantum sensing, and quantum communication. However, decoherence remains a major challenge. Pulsed dynamical decoupling (DD) addresses this problem by using sequences of short pulses to nullify the average effect of the unwanted qubit-environment interactions \cite{Viola1999PRL,Suter2016RevModPhys}. However, the pulses should ideally be instantaneous to avoid spurious effects \cite{LoretzPRX2015}, which is challenging, e.g., due to heating at low temperatures or with biological samples.

Continuous dynamical decoupling (CDD) is an alternative that does not require strong pulses but uses a continuous protecting field and has been successfully demonstrated in multiple applications  \cite{Suter2016RevModPhys,WrachrupNano2013,DegenARPC2014,DegenRMP2017,BalasubramanianNatMat2009,deLangeScience2010,NaydenovPRB2011,SriarunothaiQST2018,GenovPRR2020,StarkSciRep2018,StarkNatComm2017,KnowlesNatMat2014,McguinnessNatNano2011,SolomonPRL1959,WalsworthNature2013,KucskoNature2013,BalasubramanianOpinBio2014,HirosePRA2012,AielloNatComm2013,Baumgart2016PRL}.
The continuous field opens an energy gap in the dressed state basis for first order noise protection, perpendicular
to the field. However, the energy gap
suffers from driving field fluctuations, introducing additional noise.
Concatenated CDD overcomes this issue by adding more dressing fields to iteratively compensate for amplitude noise \cite{CaiNJP2012}.
Major drawbacks of this approach are increased complexity
and limited gate speed due to the small final energy gap.
Other approaches include phase modulated double drive CDD, where the continuous first field is phase modulated to mimic the effect of a noiseless second field \cite{CohenFP2017}, combining continuous and pulsed DD \cite{Genov2019MDD}, or using the noise correlations between driving fields to suppress noise \cite{Salhov2024prl}. Robust CDD schemes are also possible in multi-level systems \cite{StarkSciRep2018,CohenFP2017,TimoneyNature2011,AharonPRL2013,AharonNJP2016,BarfussNPhys2018,TimoneyNature2011,Baumgart2016PRL} but these typically involve additional overhead in terms of control parameters. This is often challenging due to, e.g., drift of the magnetic field or spatial inhomogeneity, limiting the coherence time especially at high magnetic fields, where small relative changes can lead to fast decoherence.

%%%%%%%%%%%%%%%%%%%%%%%%%% FIGURE %%%%%%%%%%%%%%%%%%%%%%%%%%%%%
\begin{figure*}[t!]
\includegraphics[width=\textwidth]{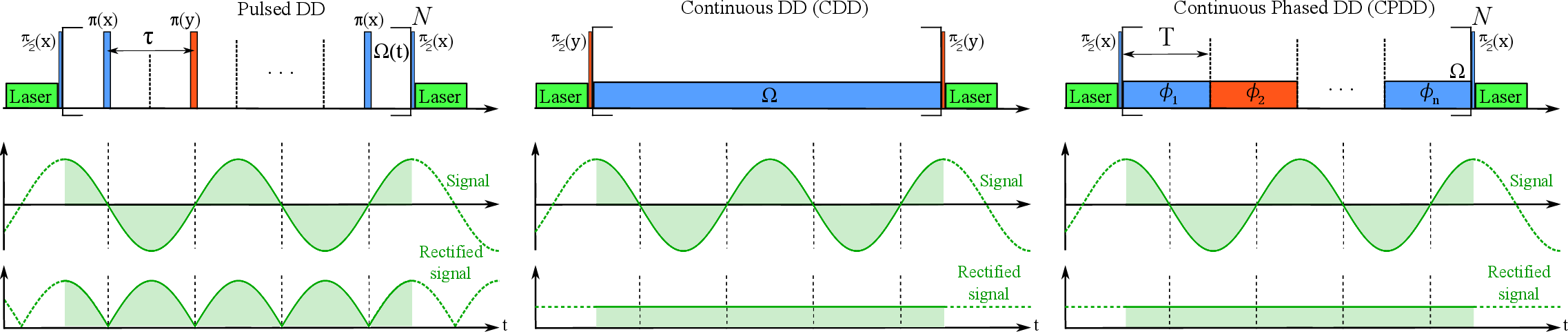}
\caption{Experimental schemes for quantum sensing with pulsed, standard continuous and continuous phased dynamical decoupling with signal $g\cos{(\omega_{L}t+\xi)}$. Pulsed DD uses the phases of the pulses for improved robustness but each pulse should be much shorter than the half period of the signal. CDD is applicable with $\Omega=\omega_{L}$ but is sensitive to amplitude noise. CDD is applicable with $\Omega=\omega_{L}$ but is sensitive to amplitude noise.
The phase changes of CPDD improve robustness. In contrast to pulsed DD, a uniform rectified signal is obtained for CDD and CPDD on resonance when the initial signal phase $\xi=0$ and the phase changes are applied at intervals $T=\pi/\omega_{L}\approx\pi/\Omega$ for CPDD.
}
\label{Fig:Exp_scheme}
\end{figure*}
%%%%%%%%%%%%%%%%%%%%%%%%%% FIGURE %%%%%%%%%%%%%%%%%%%%%%%%%%%%%

In this work, we propose and implement experimentally continuous phased DD (CPDD), using a strong continuous field to suppress the effect of environmental noise to first order and apply phase changes at well defined time intervals that compensate both amplitude and  environmental frequency noise errors to higher order.
CPDD does not require short pulses and strong driving fields, which makes it particularly suitable for experiments that require low driving power due to temperature or setup limitations.
The rotary echo protocol, where the periodic phases are $0$ and $\pi$ \cite{SolomonPRL1959,HirosePRA2012,AielloNatComm2013}, is one known example of CPDD. However, its performance is not optimal due to frequency noise, e.g., at low driving fields. Windowless sequences of $\pi/2$ pulses and composite pulses have also been applied for decoupling and refocusing of pulse errors \cite{BurumJMR1981,NanzJMR1995,ShakaJMR1983,Levitt1986,GenovPRL2014}.
In this letter, we apply phase changes for continuous decoupling that follow robust sequences of $\pi$ pulses, previously used for pulsed DD only \cite{RDD_review12Suter,TorosovPRL2011,GenovPRL2014,GenovPRL2017}, improving robustness to both frequency and amplitude noise.

We demonstrate quantum sensing of a signal from an external bath of hydrogen nuclear spins at low and high magnetic fields with a single NV center in diamond. We use CPDD with the phases of the popular XY8 sequence $\left(0,1,0,1,1,0,1,0\right)\pi/2$ \cite{Gullion1990jmr,Maudsley1986jmr,RDD_review12Suter}, labeling the new sequence continuous XY8 (CXY8).
All DD protocols perform expectedly well at low field.
However, only CPDD successfully detects a signal at high magnetic field as the technique is robust against amplitude errors, while imprecise control of the Rabi frequency proves detrimental for conventional CDD.
Finally, we successfully combine our method with quantum heterodyne detection \cite{SchmittScience2017}, achieving
$\mu$Hz uncertainty in the  estimated frequency.
Our work expands significantly the applicability of continuous dynamical decoupling, e.g., in NV centers, trapped ions or trapped atoms.
%

%%%%%%%%%%%%%%%%%%%%%%%%%%%%%%%%%%%%%%%%%%%%%%%%%%%%%%%%%%%%%%%%%%%%%%%%%%%%%%%%%%%%%%%%%%%%%%%%%%%%%%%%%%%%%%%%%%%%%%%%%%%%%%%
%%%%%%%%%%%%%%%%%%%%%%%%%%%%%%%%%%%%%%%%%%%%%%%%%%%%%%%%%%%%%%%%%%%%%%%%%%%%%%%%%%%%%%%%%%%%%%%%%%%%%%%%%%%%%%%%%%%%%%%%%%%%%%%
\label{Section:The system}
%%%%%%%%%%%%%%%%%%%%%%%%%%%%%%%%%%%%%%%%%%%%%%%%%%%%%%%%%%%%%%%%%%%%%%%%%%%%%%%%%%%%%%%%%%%%%%%%%%%%%%%%%%%%%%%%%%%%%%%%%%%%%%%
%%%%%%%%%%%%%%%%%%%%%%%%%%%%%%%%%%%%%%%%%%%%%%%%%%%%%%%%%%%%%%%%%%%%%%%%%%%%%%%%%%%%%%%%%%%%%%%%%%%%%%%%%%%%%%%%%%%%%%%%%%%%%%%
\emph{The system.}---%
Magnetometry experiments measure the frequency and amplitude of a signal \cite{DegenRMP2017,HirosePRA2012,AielloNatComm2013,StarkNatComm2017,StarkSciRep2018}.
We demonstrate CPDD for sensing of an oscillating field (see Fig. \ref{Fig:Exp_scheme}).
%
%%%%%%%%%%%%%%%%%%%%%%%%%%%%%%%%%%%%%%%%%%%%%%%%%%%%%%%%%%%%%%%%%%%%%%%%%%%%%%%%%%%%%%%%%%%%%%%%%%%%%%%%%%%%%%%%%%%%%%%%%%%%%%%
%\subsection{Error compensation by CPDD}\label{Subsection:Error_compensation}
%%%%%%%%%%%%%%%%%%%%%%%%%%%%%%%%%%%%%%%%%%%%%%%%%%%%%%%%%%%%%%%%%%%%%%%%%%%%%%%%%%%%%%%%%%%%%%%%%%%%%%%%%%%%%%%%%%%%%%%%%%%%%%%
%
We consider the Hamiltonian
\begin{align}\label{H_rot_basis_w_0}
H=&\frac{\omega_0}{2}\sigma_{z}+\Omega\cos{(\omega_0 t+\phi)}\sigma_{x}+g\cos{(\omega_{L}t+\xi)}\sigma_{z},
\end{align}
where $\omega_0$ is the Bohr transition frequency, $\Omega$ is the Rabi frequency of the driving field with phase $\phi$, $\omega_{L}$ is the (angular) frequency of the signal,
$g$ is its amplitude, and $\xi$ its unknown phase.
We move to the interaction basis with respect to $H_0^{(1)}=\omega_0\sigma_{z}/2$ and apply the rotating-wave approximation, using $\Omega\ll\omega_0$, and obtain
\begin{align}\label{H_sensing_1s}
H_{\text{1}}&=\frac{\Omega}{2}\left(\cos{(\phi)}\sigma_{x}+\sin{(\phi)}\sigma_{y}\right)+
g\cos{(\omega_{L}t+\xi)}\sigma_{z}.
\end{align}
We move to the interaction basis with respect to the driving field $\frac{\Omega}{2}\left(\cos{(\phi)}\sigma_{x}+\sin{(\phi)}\sigma_{y}\right)$
and obtain
\begin{align}\label{H_sensing_2s}
H_{\text{2}}&=-\frac{\Delta}{2}\sigma_{x}+\frac{g}{2}\left(\cos{(\xi)}\sigma_{z}-\sin{(\xi)}\sigma_{y}\right),
\end{align}
where the detuning $\Delta\equiv\omega_{L}-\Omega$ and we neglect the fast oscillating terms at frequency $2\omega_{L}$, assuming $g\ll\omega_{L}$ and $\Delta\ll\omega_{L}$.
Due to this rotating-wave approximation, the detection signal effective amplitude is $g/2\equiv \kappa g$, i.e., it is attenuated by a factor $\kappa = 1/2$ in comparison to Eq. \eqref{H_sensing_1s}. We note that attenuation is slightly higher than that of pulsed DD where $\kappa_{\text{pulsed}}=2/\pi$~\cite{DegenRMP2017}.
When $\omega_{L}=\Omega$, known as Hartmann-Hahn resonance \cite{Hartmann1962pr},
we observe oscillations in the population of the $x$ state in this basis with a frequency $g$. However, amplitude noise in $\Omega$ leads to unwanted detuning $\Delta$ that quickly causes decoherence for standard continuous dynamical decoupling (CDD)
\cite{Suter2016RevModPhys,WrachrupNano2013,DegenARPC2014,DegenRMP2017,BalasubramanianNatMat2009,deLangeScience2010,NaydenovPRB2011,
KnowlesNatMat2014,McguinnessNatNano2011,WalsworthNature2013,LeSageNature2013,BalasubramanianOpinBio2014,HirosePRA2012,AielloNatComm2013,StarkNatComm2017,StarkSciRep2018}.
Since amplitude noise is relative, its effects worsen at high magnetic fields, where stronger driving is needed to match the Larmor frequency.
Achieving $\Omega=\omega_{L}$ is then practically impossible, leading to detuned oscillations, loss of contrast, and estimation bias.

\begin{figure*}[t!]
\includegraphics[width=\textwidth]{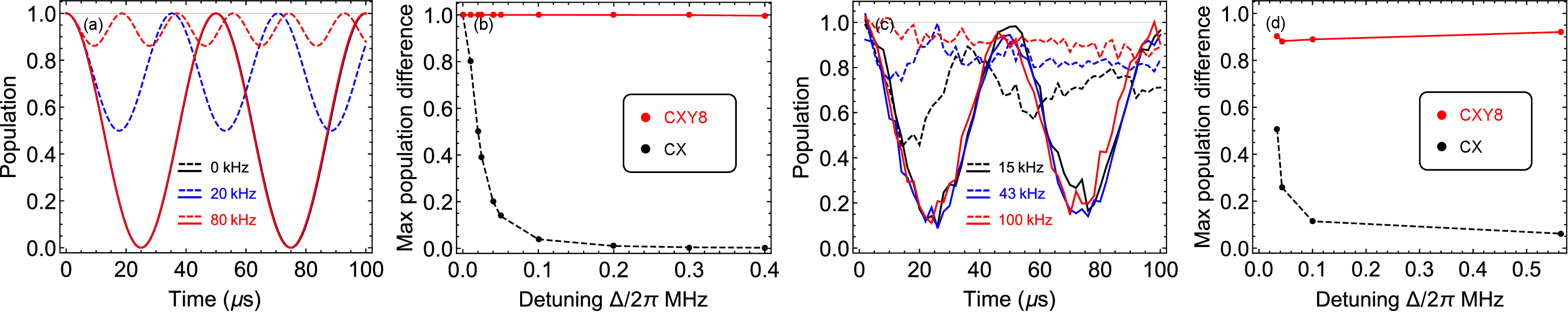}
\caption{Simulation and experimental performance of CDD (CX sequence, dashed lines) and CPDD (CXY8 sequence, solid lines) with phase changes at $T=62.5$ ns for a signal with $\omega_L = (2\pi)\,8$ MHz. (a) Population of state $|0\rangle$ of the NV electron spin vs. interaction time for different detunings $\Delta$ (shown in the legend). (b) Maximum contrast of the oscillations in (a) vs. detuning $\Delta$ for CX and CXY8. The data for all simulations are taken stroboscopically at intervals of $0.5$ $\mu$s. (c) and (d) show experimental data, which correspond to the simulations in (a) and (b).
}
\label{Fig3:Sensing_simulation}
\end{figure*}

CPDD applies phase changes in $\phi$ at time intervals $T=\pi/\Omega$ to compensate errors (see Fig. \ref{Fig:Exp_scheme}), using the phases of robust pulsed DD sequences, e.g., XY, Knill DD, or the universally robust (UR) DD sequences \cite{RDD_review12Suter,TorosovPRL2011,GenovPRL2014,GenovPRL2017}.  Thus, CPDD can be regarded as a sequence of phase shifted $\pi$ pulses with zero pulse separation. These allow for the fastest inversion rate of the qubit, given a peak Rabi frequency, leading to a narrowband filter for environmental noise. We sense the signal field by its effect during the pulses, as opposed to standard pulsed DD.
CPDD is robust to amplitude errors as its resonance condition is $T=\pi/\omega_{L}$, akin to pulsed DD where it is $\tau=\pi/\omega_{L}$ (see Fig. \ref{Fig:Exp_scheme}). Frequency selectivity is obtained by scanning the time interval $T$ of the phase changes, as opposed to CDD where it is necessary to scan $\Omega$.
The Hartmann-Hahn resonance condition $\Delta=0$ needs to be satisfied only approximately for CPDD, so errors can be compensated, e.g., $\Delta$ can be around 5\% of $\Omega$ for a total duration 1600$\,T$  \cite{Supplemental}.
We analyze the effect of errors without the sensed field.
CDD applies a continuous drive  $\sim \sigma_{x}$ (see Eq. \eqref{H_sensing_1s}), which we label CX. Its fidelity is $F_{\text{CX}}=1-O(\epsilon)$,
where $\epsilon$ is the typically small error in the transition probability of one $\pi$ pulse, e.g., $\epsilon= 1-\cos^2{(\Delta t)}\approx 0.006$ when $\Delta=0.05 \,\omega_{L}$\cite{Supplemental}.
The fidelity of CPDD with the CXY8 sequence is $F_{\text{CXY8}}=1-O(\epsilon^3)$, which is much closer to $1$ than with the standard continuous decoupling.

The effect of the phase changes on the rectified signal is not trivial as only the component $\frac{g}{2}\cos{(\xi)}\sigma_{z}$ in Eq. \eqref{H_sensing_2s} is typically unaffected by them \cite{Supplemental}.
Thus, CPDD has selectivity with respect to the signal initial phase, similarly to pulsed DD \cite{DegenRMP2017},
making it directly applicable to quantum heterodyne detection (Qdyne) \cite{SchmittScience2017}.
Similarly to pulsed DD, the signal can produce sequence-dependent spurious harmonics \cite{LoretzPRX2015},
which can be comparable to the main harmonic because the signal is sensed by its effect during the driving. Such unwanted signals can be suppressed by randomized phase changes \cite{Supplemental}, as in pulsed DD \cite{WangPRL2019}.

%%%%%%%%%%%%%%%%%%%%%%%%%%%%%%%%%%%%%%%%%%%%%%%%%%%%%%%%%%%%%%%%%%%%%%%%%%%%%%%%%%%%%%%%%%%%%%%%%%%%%%%%%%%%%%%%%%%%%%%%%%%%%%%
%%%%%%%%%%%%%%%%%%%%%%%%%%%%%%%%%%%%%%%%%%%%%%%%%%%%%%%%%%%%%%%%%%%%%%%%%%%%%%%%%%%%%%%%%%%%%%%%%%%%%%%%%%%%%%%%%%%%%%%%%%%%%%%
%\section{Experimental implementation.}\label{Section:Experiment}
\emph{Experimental implementation.}\label{Section:Experiment}---%
%%%%%%%%%%%%%%%%%%%%%%%%%%%%%%%%%%%%%%%%%%%%%%%%%%%%%%%%%%%%%%%%%%%%%%%%%%%%%%%%%%%%%%%%%%%%%%%%%%%%%%%%%%%%%%%%%%%%%%%%%%%%%%%
We performed experiments on two different single Nitrogen-Vacancy centers (NV) in diamond with an isotopically enriched ${}^{12}\!$C (99.999\%) layer \cite{Supplemental}.
The negative charge state of the NV center has $S=1$ triplet ground state \cite{DohertyPRB2012}. We used a permanent neodymium magnet to create an external magnetic field bias aligned with the NV axis to remove the degeneracy between the $\ket{\pm 1}$ states of the NV electron spin and to control the Larmor frequency of the hydrogen bath. We placed a copper wire on the diamond surface and used it to drive the $\ket{0}\rightarrow\ket{-1}$ transition resonantly, creating an effective two-level system.
An arbitrary waveform generator is used for direct generation of the sequences and precise control of the phase changes.
Illumination with a green laser ($518\,$nm), used for initialization and readout of the NV spin state,  reveals a spin dependent fluorescence and favorable relaxation into state $\ket{0}$ \cite{RobledoNJP2011,HarrisonDRM2006}.
All experimental details can be found in \cite{Supplemental}.

First, we demonstrate the robustness of CPDD for sensing the amplitude $g\approx 2\pi~11.7$ kHz of a weak oscillating field. The results from a numerical simulation and experiments are shown in Fig. \ref{Fig3:Sensing_simulation}.
When the Rabi frequency of the driving field is equal to the angular frequency of the sensed field, i.e., $\Omega=\omega_{L}=2\pi~8$ MHz, we observe oscillations of the population of state $|0\rangle$ for both CX and CXY8 with $P=\cos(\Theta(t)/2)$, where $\Theta=gt$. The oscillations are quickly lost with CX even with small $\Delta\approx 2\pi~20$ kHz, i.e., of the order of the coupling strength $g$. On the contrary, with CXY8 the robustness is improved by more than twenty times, as we observe oscillations even for $\Delta = 2\pi~400$ kHz \cite{Supplemental}. The experimental results match the simulation very well with the worse performance of CDD in the experiment mainly due to amplitude fluctuations, which are not included in the simulation.

%%%%%%%%%%%%%%%%%%%%%%%%%% FIGURE %%%%%%%%%%%%%%%%%%%%%%%%%%%%%
\begin{figure*}[t!]
\includegraphics[width=\textwidth]{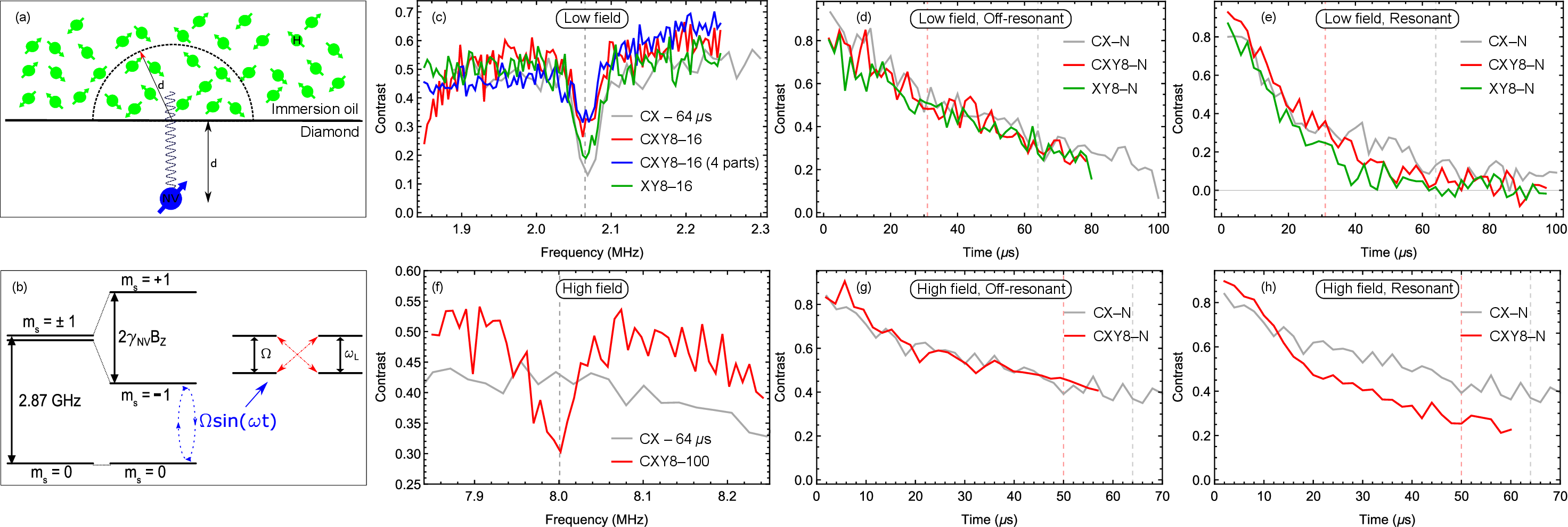}
\caption{(a) Sketch of a shallow NV center at depth $d$. It interacts with the hydrogen bath on the diamond surface within a volume of roughly $d^3$ \cite{MuellerNatComm2014}.
(b) A simplified level scheme of the NV center ground state.
When $\Omega=\omega_L$ the interaction between the NV and the hydrogen bath leads to a dip in fluorescence for CX. The resonance condition for XY8 (CXY8) is $\tau=\pi/\omega_L$ ($T=\pi/\omega_L$), see Fig. \ref{Fig:Exp_scheme}.
Experimental results for (c) DD spectra in the low field regime. The larger dip for CX is due to its longer interaction time.
The label (C)XY8-$N$ shows that the sequence is repeated $N$ times. CXY8-16-4 denotes a CXY8-16 sequence where we did not vary $\Omega$ continuously as we varied $T$ but it took only four values $2\pi\{1.9, 2, 2.1, 2.2\}$ MHz, demonstrating CXY8 robustness.
DD order scan (d) off-resonant and (e) resonant with the Larmor frequency at low field; (f) DD spectrum in the high field regime. No contrast is observed for CX even though a longer interaction time is used than for CXY8. DD order scan (g) off-resonant and (h) resonant with the Larmor frequency at high field. In (h) we included the off-resonant CX plot due to negligible difference from CX on resonance, as shown in (f). The respective interaction times used in (c) and (f) are marked in plots (d,e) and (g,h) with gray (CX) and red (CXY8), vertical, dashed lines.
}
\label{Fig:Exp_results}
\end{figure*}
%%%%%%%%%%%%%%%%%%%%%%%%%% FIGURE %%%%%%%%%%%%%%%%%%%%%%%%%%%%%

In a second experiment, we detected a nano-NMR signal from the external hydrogen bath of immersion oil with a shallow (depth $\approx 10$nm) NV center (Fig.\ref{Fig:Exp_results}(a)).
We performed the experiment at two bias fields, parallel to the NV axis: $B_z=(485\pm 1)$G (low field) to polarize the inherent nitrogen nuclear spin, with the Larmor frequency of the hydrogen nuclear spins  $\omega_L\approx2\pi~2.065$ MHz. Second, we applied
$B_z=(1882\pm 3)$G (high field) where the Larmor frequency is $\omega_L\approx2\pi~8.005$ MHz. The inherent nitrogen nuclear spin is not polarized then but the effect is negligible due to the high Rabi frequency and the decoupling sequences \cite{Vetter2022prapplied}. At low field, the Rabi frequency was roughly $2\pi~14$ MHz for pulsed DD. Hence,  $\frac{\omega_L}{\Omega}\approx 0.14$, giving a good approximation for instantaneous pulses. Figure \ref{Fig:Exp_results}(b) shows the level scheme and the Hartmann-Hahn resonance condition for CDD. Figure \ref{Fig:Exp_results}(c-e) shows the results of three different sequences in the low field regime. In all cases the hydrogen dip is clearly visible with a difference in dip size as expected from theory \cite{Supplemental}.

%%%%%%%%%%%%%%%%%%%%%%%%%% FIGURE %%%%%%%%%%%%%%%%%%%%%%%%%%%%%
\begin{figure}[b!]
\includegraphics[width=\columnwidth]{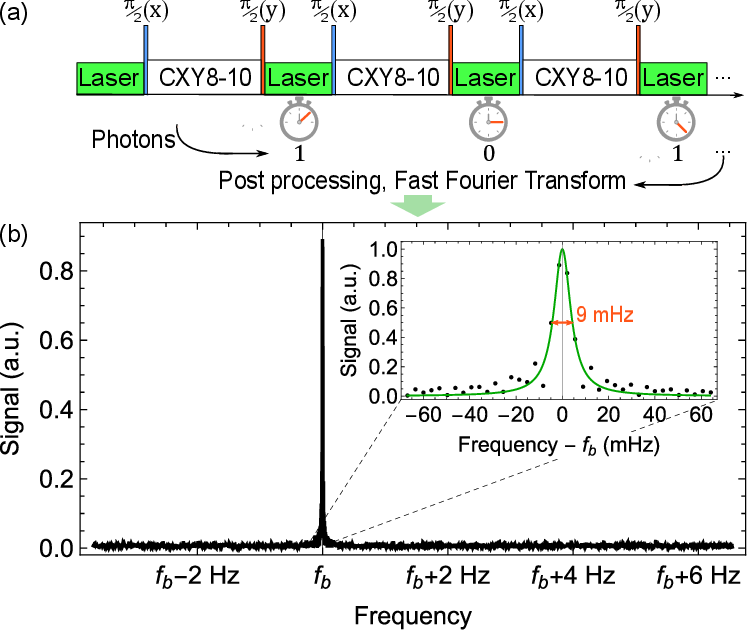}
\caption{(a) Scheme of nanoscale high-frequency sensing with Qdyne, using CXY8. (b) Fast Fourier transform of the photon trace to detect a signal at the beat frequency $f_b$, which corresponds to the signal frequency of $8\,$MHz.
}
\label{Fig:Qdyne}
\end{figure}
%%%%%%%%%%%%%%%%%%%%%%%%%% FIGURE %%%%%%%%%%%%%%%%%%%%%%%%%%%%%

At high field, the necessary shorter time spacing between pulses limited the microwave Rabi frequency to around $2\pi~12$ MHz due to heating. Therefore, we could not assume instantaneous pulses for pulsed DD.
Thus, we only performed continuous drive sequences at high field.
Figure \ref{Fig:Exp_results}(f-h) shows a dip in fluorescence due to hydrogen, which was visible only with CXY8, even though we used a longer interaction time for standard CX. This is
due to the better frequency selectivity of CXY8 as its control parameter is the time interval of the phase changes $T$. In contrast, the CX requires $\omega_{L}=\Omega$,
depending on the microwave Rabi frequency, which suffers from amplitude noise, mainly due to power fluctuations~\cite{FedderAPB2011}.
In our experiments, the vertical resolution of the arbitrary waveform generator is $8-bit$, resulting in a $30-40$~kHz error in the Rabi frequency, making it practically impossible to lock the driving frequency on a signal with a frequency width of about $40$~kHz. On the contrary, the CXY8 resonance condition allows a Rabi frequency error of more than $\pm 5\%$ around $\omega_{L}$ (see Fig. \ref{Fig3:Sensing_simulation} and \cite{Supplemental}). Thus, the frequency selectivity is determined by the \emph{time} digitization of the driving signal, i.e.,  $65$~GHz, resulting in a much smaller error than the experimental requirements.
We note that the broader width of the proton dip in the high field experiment is due to the larger magnetic noise and diffusion broadening \cite{Staudenmaier2022}.

%%%%%%%%%%%%%%%%%%%%%%%%%%%%%%%%%%%%%%%%%%%%%%%%%%%%%%%%%%%%%%%%%%%%%%%%%%%%%%%%%%%%%%%%%%%%%%%%%%%%%%%%%%%%%%%%%%%%%%%%%%%%%%%
%%%%%%%%%%%%%%%%%%%%%%%%%%%%%%%%%%%%%%%%%%%%%%%%%%%%%%%%%%%%%%%%%%%%%%%%%%%%%%%%%%%%%%%%%%%%%%%%%%%%%%%%%%%%%%%%%%%%%%%%%%%%%%%

Finally, we combined CPDD with quantum heterodyne detection (Qdyne), achieving $\mu$Hz frequency uncertainty with a 120\,s measurement for nano-scale high-frequency sensing. Figure \ref{Fig:Qdyne}(a) shows the scheme of Qdyne.
We apply CXY8 on resonance with $\Omega\approx 2\pi\,8$ MHz and $T=\pi/\Omega=62.5$ ns. During the sequence the qubit acquires a phase, which depends on the signal phase $\xi$ at the beginning of each CXY8 sequence \cite{Supplemental}.
The spin population is sampled at fixed
time intervals $T_L$ such that the signal phase $\xi$ changes by a constant increment. We record individually the number of detected photons and the time of detection during the laser pulses. The experiment takes 120 s and consists of 13.9$\times 10^6$ measurements. We perform a fast Fourier transform of the recorded photon trace and observe a signal with a beat frequency of $f_b=19.4328391$ kHz $\pm 4.4\,$mHz (see Fig. \ref{Fig:Qdyne}(b)), which corresponds to an estimated frequency of $8000001.8883\,$Hz $\pm 4.4\,$mHz. We achieve $\delta\nu_0 \approx 28~\mu$Hz frequency uncertainty for $T_\mathrm{tot} = 120$\,s measurement time, which can be further reduced by prolonging $T_\mathrm{tot}$ and is mainly limited by the coherence time of the clock. The frequency sensitivity of CXY8 with Qdyne is $\eta_{\nu_0} = \delta\nu_0\, T_\mathrm{tot}^{3/2}=$ 32 mHz/Hz$^{3/2}$ and the amplitude sensitivity is $\eta_B = \frac{B_0}{\mathrm{SNR}} T_\mathrm{tot}^{1/2}=118 \pm 8$ nT/$\sqrt{\text{Hz}}$ \cite{Supplemental}. These values are comparable to previous results with Qdyne \cite{Staudenmaier2021pra} and can be further improved by optimizing the measurements for reaching maximum sensitivity, which goes beyond the scope of our proof-of-principle demonstration.

\emph{Discussion.}---%\label{Section:Discussion}
The sensitivity of CPDD and pulsed DD differ by their signal attenuation factor $\kappa=1/2$ and $\kappa_\text{pulsed}=2/\pi$, respectively (see Eq. \eqref{H_sensing_2s} and \cite{DegenRMP2017,Supplemental}). The dynamic range of CPDD is similar to the one of dressed states methods e.g., continuous dynamical decoupling \cite{DegenRMP2017}, and is 0.1-100 MHz for electronic spins or superconducting qubits \cite{DegenRMP2017} with the additional advantage of improved robustness. It can be expanded by incorporating a second driving field to sense GHz frequencies, similar to double drive continuous decoupling \cite{CaiNJP2012,Salhov2024prl}. The sensitivities can be improved even further by engineering of the host diamond properties to obtain NV centers with longer coherence times \cite{orwa2011annealing,osterkamp2015stabilizing}, enhancing photon collection efficiency  \cite{hadden2010strongly,siyushev2010monolithic,Supplemental} or with ensembles of NV centers \cite{Barry2020sensitivity}.
Furthermore, CPDD can be modified to perform robust quantum gates. As their speed is limited by the frequency difference of the eigenstates of the last dressed qubit, CPDD will be faster than alternative continuous robust schemes like concatenated CDD \cite{CaiNJP2012}.

%%%%%%%%%%%%%%%%%%%%%%%%%%%%%%%%%%%%%%%%%%%%%%%%%%%%%%%%%%%%%%%%%%%%%%%%%%%%%%%%%%%%%%%%%%%%%%%%%%%%%%%%%%%%%%%%%%%%%%%%%%%%%%%
%%%%%%%%%%%%%%%%%%%%%%%%%%%%%%%%%%%%%%%%%%%%%%%%%%%%%%%%%%%%%%%%%%%%%%%%%%%%%%%%%%%%%%%%%%%%%%%%%%%%%%%%%%%%%%%%%%%%%%%%%%%%%%%
%\section{Conclusion}\label{Section:Conclusion}
%%%%%%%%%%%%%%%%%%%%%%%%%%%%%%%%%%%%%%%%%%%%%%%%%%%%%%%%%%%%%%%%%%%%%%%%%%%%%%%%%%%%%%%%%%%%%%%%%%%%%%%%%%%%%%%%%%%%%%%%%%%%%%%
%%%%%%%%%%%%%%%%%%%%%%%%%%%%%%%%%%%%%%%%%%%%%%%%%%%%%%%%%%%%%%%%%%%%%%%%%%%%%%%%%%%%%%%%%%%%%%%%%%%%%%%%%%%%%%%%%%%%%%%%%%%%%%%
\emph{Conclusion.}---%\label{Section:Conclusion}
We proposed and demonstrated experimentally robust noise suppression and quantum sensing by continuous phase shifted dynamical decoupling. We applied a continuous driving field to reduce the effect of environmental noise to first order. Discrete phase changes at well defined time intervals compensate both amplitude and environmental noise errors to higher order and increase the robustness of the technique.
CPDD does not require short pulses and high driving fields unlike pulsed DD, making it suitable for applications with limited driving power due to thermal constraints, or when high magnetic fields are required, and the Larmor frequency of the sensed spins approaches the maximum achievable Rabi frequency, rendering pulsed DD inefficient.
The presented technique can also compensate for detuning and microwave field inhomogeneity in ensemble NV sensing applications. A major advantage of CPDD is that it requires control of the time intervals of phase changes, which is much easier than control of the Rabi frequency, as demonstrated in our experiments. Finally, we have shown that CPDD is compatible with high-resolution, synchronized measurement schemes like Qdyne, demonstrating quantum sensing with $\mu$Hz frequency uncertainty.

%\acknowledgments
\emph{Acknowledgment.}---
This work was supported by the German Federal Ministry of Research (BMBF) by future cluster QSENS and projects DE-Brill (No. 13N16207), SPINNING, DIAQNOS (No. 13N16463), quNV2.0 (No. 13N16707), QR.X and Quamapolis (No. 13N15375), DLR via project QUASIMODO (No.50WM2170), Deutsche Forschungsgemeinschaft (DFG) via Projects No. 386028944, No. 445243414, and No. 387073854 and Excellence Cluster POLiS, BMBF via project EXTRASENS (No. 13N16935), European Union’s HORIZON Europe program via projects QuantERA II (No. 101017733), QuMicro (No. 101046911), SPINUS (No. 101135699), CQuENS (No. 101135359), QCIRCLE (No. 101059999), and FLORIN (No. 101086142), European Research
Council (ERC) via Synergy grant HyperQ (No. 856432), Carl-Zeiss-Stiftung via the Center of Integrated Quantum Science and technology (IQST) and project Utrasens-Vir, as well as the Baden-W{\"u}rttemberg Foundation.
A.R. acknowledges the support of ERC grant QRES, project number 770929, Quantera grant MfQDS, ISF, the Schwartzmann university chair and the Israeli Innovation Authority under the project ``Quantum Computing Infrastructures''.


%\bibliography{references}

%\end{document}

\clearpage
\onecolumngrid
\section*{Supplemental Material to ``Robust Noise Suppression and Quantum Sensing by Continuous Phased Dynamical Decoupling''}
\vspace{1cm}
\twocolumngrid
\renewcommand{\thefigure}{S.\arabic{figure}}
\renewcommand{\theequation}{S.\arabic{equation}}
\renewcommand{\thesection}{S.\arabic{section}}
\renewcommand{\thepage}{S.\arabic{page}}
\setcounter{figure}{0}
\setcounter{equation}{0}
\setcounter{section}{0}
\setcounter{page}{1}

%%%%%%%%%%%%%%%%%%%%%%%%%%%%%%%%%%%%%%%%%%%%%%%%%%%%%%%%%%%%%%%%%%%%%%%%%%%%%%%%%%%%%%%%%%%%%%%%%%%%%%%%%%%%%%%%%%%%%%%%%%%%%%%
\section{Detailed Theory of CPDD}\label{Appendix:CDD_CPDD}
%%%%%%%%%%%%%%%%%%%%%%%%%%%%%%%%%%%%%%%%%%%%%%%%%%%%%%%%%%%%%%%%%%%%%%%%%%%%%%%%%%%%%%%%%%%%%%%%%%%%%%%%%%%%%%%%%%%%%%%%%%%%%%%

%%%%%%%%%%%%%%%%%%%%%%%%%%%%%%%%%%%%%%%%%%%%%%%%%%%%%%%%%%%%%%%%%%%%%%%%%%%%%%%%%%%%%%%%%%%%%%%%%%%%%%%%%%%%%%%%%%%%%%%%%%%%%%%
\subsection{Continuous Dynamical Decoupling}\label{Subsection:CDD}
%%%%%%%%%%%%%%%%%%%%%%%%%%%%%%%%%%%%%%%%%%%%%%%%%%%%%%%%%%%%%%%%%%%%%%%%%%%%%%%%%%%%%%%%%%%%%%%%%%%%%%%%%%%%%%%%%%%%%%%%%%%%%%%

In order to introduce the mechanism of CPDD, we first describe the theory of sensing by continuous dynamical decoupling (CDD)
\cite{Suter2016RevModPhys,DegenARPC2014,DegenRMP2017,BalasubramanianNatMat2009,deLangeScience2010,NaydenovPRB2011,StarkSciRep2018,StarkNatComm2017,Viola1999PRL,WrachrupNano2013,
KnowlesNatMat2014,McguinnessNatNano2011,WalsworthNature2013,LeSageNature2013,BalasubramanianOpinBio2014,HirosePRA2012,AielloNatComm2013,StarkSciRep2018}.

We consider the Hamiltonian
\begin{align}\label{Supp_H_rot_basis_w_0}
H=&\frac{\omega_0}{2}\sigma_{z}+\Omega\cos{(\omega t+\phi)}\sigma_{x}+g\cos{(\omega_{L}t+\xi)}\sigma_{z},
\end{align}
where $\omega_0$ is the Bohr transition frequency, $\Omega$ and $\omega$ are the Rabi frequency and carrier (angular) frequency of the driving field with a phase $\phi$, and $\omega_{L}=\Omega+\Delta$ is the (angular) frequency of the signal (e.g., equal to the Larmor frequency of the proton spins in the hydrogen bath in our nano-NMR experiment) and $\Delta$ is its detuning from the Rabi frequency $\Omega$, $g$ is the signal amplitude, and $\xi$ is its unknown (random) phase.
In the case of an artificial signal, the latter could be generated by an oscillating magnetic field $B\cos{(\omega_{L}t+\xi)}$. The value of $g=\gamma_\text{NV} B/2$, where $\gamma_\text{NV}$ is the gyromagnetic ratio of the NV center electron spin, when the detector qubit consists of the $|0\rangle$ and $|-1\rangle$ ground states of the electron spin of an NV center, as in our case (or is a spin $1/2$ system, in general). In comparison, $g=\gamma_\text{NV} B$ when the detector qubit consists of the $|\pm 1\rangle$ ground states of the NV center on the so called double-quantum transition \cite{Myers2016prl}.

We then move to the interaction basis with respect to $H_0^{(1)}=\omega\sigma_{z}/2$. Typically, the carrier frequency is equal to the qubit transition frequency $\omega=\omega_0$, as assumed in the main text, but we do not assume this below to account for possible carrier frequency detuning $\delta=\omega-\omega_0$ in the following analysis. We obtain after applying the rotating wave approximation ($\Omega\ll\omega_0$)
\begin{align}\label{Supp_H_sensing_1s}
H_{\text{1}}&=U_0^{(1)}(t)^{\dagger}H_{\text{s}}U_0^{(1)}(t)-i U_0^{(1)}(t)^{\dagger}\partial_{t}U_0^{(1)}(t)\\
&=-\frac{\delta}{2}\sigma_{z}+\frac{\Omega}{2}\left(\cos{(\phi)}\sigma_{x}+\sin{(\phi)}\sigma_{y}\right)\notag\\
&~~~~+
g\cos{(\omega_{L}t+\xi)}\sigma_{z},\notag
\end{align}
where $U_0^{(1)}(t)=\exp\left(-i H_0^{(1)} t\right)$. We can then rotate our basis to take into account the phase $\phi$ and obtain
\begin{align}\label{H_sensing_1f}
H_{\text{1,$\phi$}}&=R_{z}(\phi)^{\dagger}H_{\text{1}}R_{z}(\phi)\notag\\
&=-\frac{\delta}{2}\sigma_{z}+\frac{\Omega}{2}\sigma_{x}+g\cos{(\omega_{L}t+\xi)}\sigma_{z},
\end{align}
where the rotation operator is defined as $R_{k}(\phi)\equiv\exp\left(-i \phi\sigma_{k}/2\right),k=x,y,z$. The next step involves moving to the rotating frame with respect to $H_0^{(2)}=\omega_{L}\sigma_{x}/2=(\Omega+\Delta)\sigma_{x}/2$ and we obtain
\begin{align}\label{H_sensing_2s_supplemental}
H_{\text{2}}&=U_0^{(2)}(t)^{\dagger}H_{\text{1,$\phi$}}U_0^{(2)}(t)-i U_0^{(2)}(t)^{\dagger}\partial_{t}U_0^{(2)}(t)\\
&=-\frac{\Delta}{2}\sigma_{x}+\frac{g}{2}\left(\cos{(\xi)}\sigma_{z}-\sin{(\xi)}\sigma_{y}\right),\notag
\end{align}
where $U_0^{(2)}(t)=R_{x}\left(\omega_{L}t\right)$ and we neglected the fast oscillating terms at frequency $2\omega_{L}$, assuming $g\ll\omega_{L}$, $\delta\ll\omega_{L}$ and $\Delta\ll\omega_{L}$.
Since the Hamiltonian is constant in this basis, we can obtain an exact analytic solution.
The time evolution of the system resembles detuned Rabi oscillations with the difference that the basis is rotated around the $y$ axis from the standard case, which one can obtain by the transformation $\sigma_{x}\to -\sigma_{z}$, $\sigma_{z}\to \sigma_{x}$.

In the case of Hartmann-Hahn resonance $\omega_{L}=\Omega$, i.e., the detuning $\Delta=0$, we observe oscillations in the population of the $x$ state of this basis with a frequency of $g$.
Specifically, the time evolution of the Hamiltonian in Eq. \eqref{H_sensing_2s_supplemental} is determined by the propagator
\begin{align}\label{Eq:propagator_CDD}
U_{2}(t)&=\exp\left(-i H_{2} t\right) \notag\\
&= R_{-\xi(z,y)}\left(\Theta\right)=R_{x}\left(\xi\right)R_{z}\left(\Theta\right)R_{x}\left(-\xi\right)\notag\\
&= R_{z}\left(\Theta\right)=\cos{\left(\Theta/2\right)}\sigma_0- i \sin{\left(\Theta/2\right)}\sigma_{z},
\end{align}
where $R_{-\xi(z,y)}\left(\Theta\right)=\exp\left[-i \Theta (\cos{(\xi)}\sigma_{z}-\sin{(\xi)}\sigma_{y}\right]$ is an operator for rotation around an axis at angle $-\xi$ in the $z-y$ plane of the Bloch sphere in this basis, $\Theta=g t$, and $\sigma_0$ is the identity operator. In the last two equalities we assumed $\xi=0$ for simplicity of presentation and without loss of generality.
We note that the rectified signal for CDD and CPDD differs significantly from the case of standard pulsed DD where the accumulated phase is given by $\Theta\equiv\int_{0}^{t}2g|\cos{(\Delta t^{\prime})}|\d t^{\prime}\approx\frac{4}{\pi}g t$ (see Fig. 1 in the main text). The main reason is that the effect of the signal is sensed during the application of the driving field for CDD and CPDD, in contrast to pulsed DD, where the signal is sensed mainly by its effect during the free evolution time between the pulses.

The oscillations due to the sensed field can be observed stroboscopically in the bare basis at times that correspond to $\omega_{L} t=n\pi,~n$ is even. Specifically, the propagator in the rotating basis at $\omega_0$, defined in Eq. \eqref{Supp_H_sensing_1s}, takes the form
\begin{align}\label{Eq:H_CDD}
U_{1}(t)&=R_{z}(\phi)U_0^{(2)}(t)U_{2}(t)U_0^{(2)\dagger}(0)R_{z}^\dagger(\phi)\notag\\
&=R_{z}(\phi)R_{x}\left(\omega_{L}t\right)U_{2}(t)R_{x}\left(0\right)R_{z}(-\phi)\notag\\
&=(-1)^{k}R_{z}(\phi)R_{z}(\Theta)R_{z}(-\phi)=(-1)^{n/2}R_{z}(\Theta),
\end{align}
where we assumed that the evolution started at time $t=0$ for all equations, and $\xi=0$ and $\omega_{L}t=n\pi,~n$ is even, in the last two equalities.
Thus, on resonance the time evolution in the rotating basis at $\omega_0$, defined by the Hamiltonian in Eq. \eqref{Supp_H_rot_basis_w_0} corresponds to the time evolution in the basis of Eq. \eqref{H_sensing_2s_supplemental} up to an overall phase shift.
If the phase $\phi$ is constant, it can be accounted for and the amplitude $g$ can be inferred stroboscopically from the time evolution in the rotating basis at frequency $\omega_0$.

%\vspace{0.5cm}
%%%%%%%%%%%%%%%%%%%%%%%%%%%%%%%%%%%%%%%%%%%%%%%%%%%%%%%%%%%%%%%%%%%%%%%%%%%%%%%%%%%%%%%%%%%%%%%%%%%%%%%%%%%%%%%%%%%%%%%%%%%%%%%
\subsection{Continuous Phased Dynamical Decoupling}\label{Subsection:CPDD}
%%%%%%%%%%%%%%%%%%%%%%%%%%%%%%%%%%%%%%%%%%%%%%%%%%%%%%%%%%%%%%%%%%%%%%%%%%%%%%%%%%%%%%%%%%%%%%%%%%%%%%%%%%%%%%%%%%%%%%%%%%%%%%%

CPDD uses the phase $\phi$ of the driving field as a control parameter to compensate for amplitude and frequency noise, e.g., due to environment or magnetic field changes. The effect of the phase changes is not trivial for sensing and we first analyze it for the simplest case of a single phase change from $\phi_1$ to $\phi_2$ at time $t_1$. Then, the propagator from $t=0$ to time $t>t_1$ in the rotating basis at $\omega_0$ takes the form
\begin{align}
U_{1}(t)
&=R_{z}(\phi_2)U_0^{(2)}(t)U_{2}(t-t_1)U_0^{(2)\dagger}(t_1)R_{z}^\dagger(\phi_2)\\
&~~~~R_{z}(\phi_1)U_0^{(2)}(t_1)U_{2}(t_1)U_0^{(2)\dagger}(0)R_{z}^\dagger(\phi_1)\notag\\
&=R_{z}(\phi_2)R_{x}\left(\omega_{L}t\right)U_{2}(t-t_1)R_{x}\left(-\omega_{L}t_1\right)\notag\\
&~~~~R_{z}(\phi_1-\phi_2)R_{x}\left(\omega_{L}t_1\right)U_{2}(t_1)R_{x}\left(0\right)R_{z}(-\phi_1).\notag
\end{align}
In order to understand the effect of a phase change we consider the commutator
\begin{align}
&\left[R_{x}\left(-\omega_{L} t_1\right)R_{z}(-\Delta\phi_1)R_{x}
\left(\omega_{L} t_1\right),U_{2}(t_1)\right]=\notag\\
&=2 i \sin{\left(\omega_{L} t_1+\xi\right)}\sin{\left(\frac{\Delta\phi_1}{2}\right)}\sin{\left(\frac{\Theta}{2}\right)}
\end{align}
where $\Delta\phi_1=\phi_2-\phi_1$. It is evident that the commutator is zero when the phase is constant, i.e., $\Delta\phi=0$, which corresponds to the standard case of CDD or when there is no sensed field, i.e., $\Theta = 0$.

In the more general case of arbitrary phase changes $\Delta\phi_1\ne 0$ and in the presence of a sensed field, the commutator will be zero when $\omega_{L} t_1+\xi=0~(\text{mod}~\pi)$, i.e., the phase changes $\Delta\phi$ should take place at the times when the waveform of the signal $g(t)$ has a phase $0~(\text{mod}~\pi)$.
On the contrary, the commutator will be non-zero for the $\sigma_{y}$ signal component in Eq. \eqref{H_sensing_2s_supplemental}, e.g., for a signal with an initial phase $\xi=\pi/2$. Then, this component of the signal will experience more complex time-evolution and
will usually be suppressed by the CPDD sequence.
Thus, the application of arbitrary phase changes in CPDD allows for selective detection of signals with initial phase $\xi=0$ and suppression of signals with initial phase $\xi=\pi/2$, in contrast to standard CDD.
For example, the condition that the commutator is zero is satisfied for an initial phase $\xi=0$ and $t_1=T=\pi/\omega_{L}=\pi/\Omega$. Then, the propagator becomes
\begin{align}
U_{1}(t)
&=R_{z}(\phi_2)R_{x}\left(\omega_{L}t\right)U_{2}(t-t_1)R_{x}\left(-\pi\right)\\
&~~~~R_{z}(-\Delta\phi_1)R_{x}\left(\pi\right)U_{2}(t_1)R_{x}\left(0\right)R_{z}(-\phi_1)\notag\\
&=R_{z}(\phi_2)R_{x}\left(\omega_{L}t\right)R_{z}(\Theta-\Theta_1)\notag\\
&~~~~R_{z}(\Delta\phi_1)R_{z}(\Theta_1)R_{x}\left(0\right)R_{z}(-\phi_1).\notag\\
&=R_{z}(\phi_2)R_{x}\left(\omega_{L}t\right)R_{z}(\Delta\phi_1)R_{z}(\Theta)R_{x}\left(0\right)R_{z}(-\phi_1)\notag\\
&=-R_{z}(2\Delta\phi_1)R_{z}(\Theta),\notag
\end{align}
where
we assumed that the evolution started at time $t=0$ for all equations, and $\xi=0$ and $\omega_{L}t=2\pi$ in the last equality.

In case of multiple phase changes, the subsequent ones should then be done at equal time intervals $\pi k/\omega_{L},k\in \mathbb{N}$ after the first phase change. Hence, the propagator becomes
\begin{align}\label{Eq:propagator_CPDD}
U_{1}(t)=(-1)^{n}R_{z}\left(2\sum_{k=1}^{n/2}\Delta\phi_{2k}\right)R_{z}(\Theta),~n~\text{is even,}
\end{align}
where $\Delta\phi_{2k}=\phi_{2k}-\phi_{2k-1}$ and we assumed that we have an even number $n$ of the constant phase intervals. It is evident that the propagator is the same as the one of CDD in Eq. \eqref{Eq:H_CDD}, apart from a cumulative phase shift, which can easily be accounted for or made zero by the choice of the relative phase shifts.

In summary, the possibility to apply arbitrary phase changes in CPDD leads to selectivity with respect to the initial phase of the signal and the selectivity condition is similar as for standard pulsed DD schemes \cite{DegenRMP2017}. Finally, we note that we can choose the “good” initial phase $\xi$ of the signal by
shifting the whole CPDD sequence in time. We only need to change the duration of the first pulse period to satisfy
$\omega_{L} t_1+\xi = 0~(\text{mod}~\pi)$ and apply the subsequent phase changes at intervals of $T=\pi k/\omega_{L},~k\in\mathbb{N}$. The phase selectivity of the protocol implies that, unlike standard CDD, it is directly applicable in synchronized sensing schemes, e.g., the Qdyne protocol, which are typically limited only by the coherence time of the signal or a classical clock \cite{SchmittScience2017}.

%%%%%%%%%%%%%%%%%%%%%%%%%% FIGURE %%%%%%%%%%%%%%%%%%%%%%%%%%%%%
\begin{figure}[t!]
\includegraphics[width=\columnwidth]{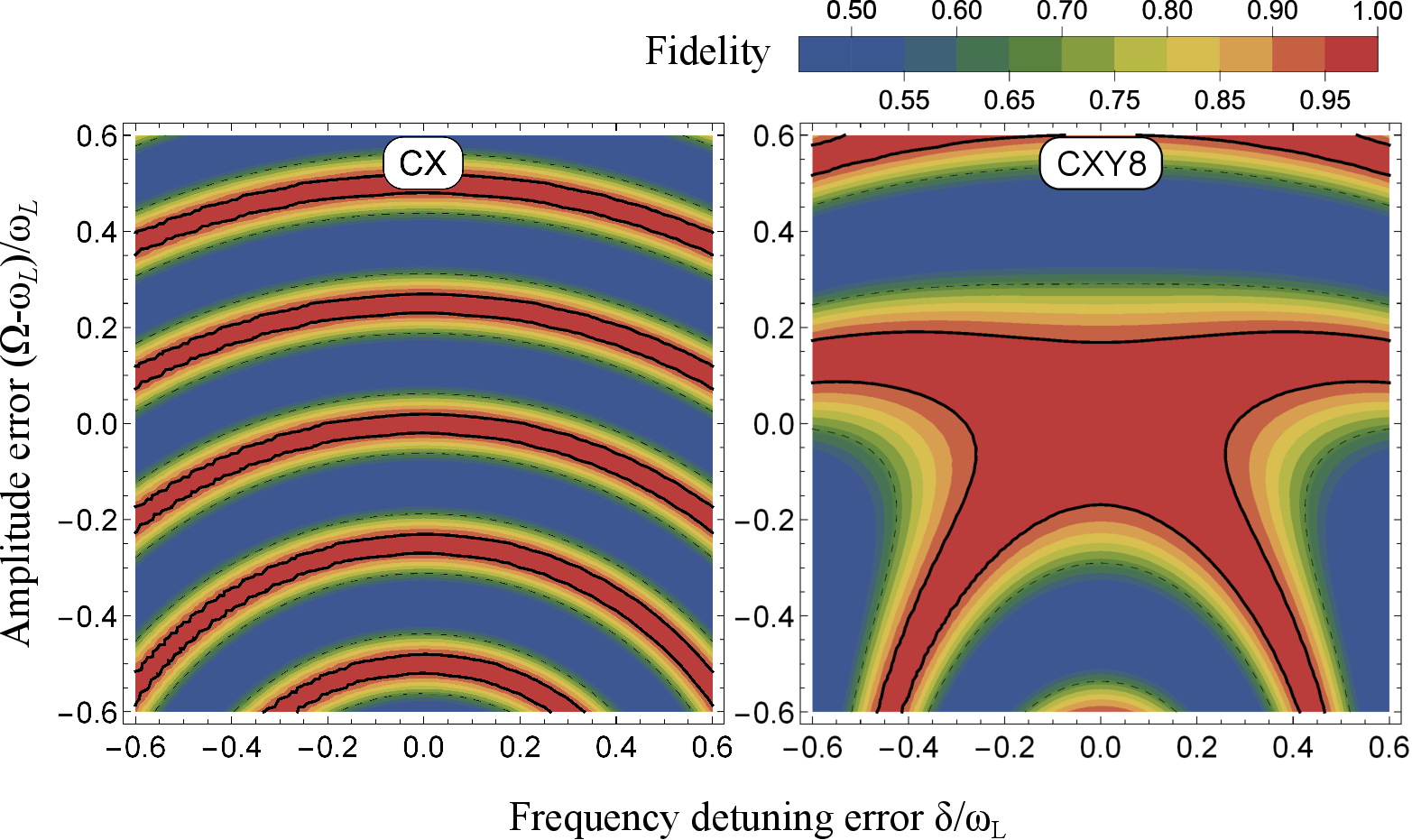}
\caption{Simulation of fidelity vs. variation in the detuning error of the carrier frequency $\omega$ of the applied driving field from the transition frequency $\omega_0$ of the sensor qubit, i.e.,  $\delta/\omega_{L}=(\omega-\omega_0)/\omega_{L}$ and the amplitude error of the driving field from the target Rabi frequency $\omega_{L}$, i.e., $(\Omega-\omega_{L})/\omega_{L}=-\Delta/\omega_{L}$ for: (left) continuous drive (labeled CX) with a target pulse area $A=\omega_{L}t=8\pi$ and (right) the respective CXY8 robust sequence with the same target pulse area. The fidelity calculation assumes the approximation of static frequency and amplitude noise and no sensed field.
}
\label{Fig2:2D_plots}
\end{figure}
%%%%%%%%%%%%%%%%%%%%%%%%%% FIGURE %%%%%%%%%%%%%%%%%%%%%%%%%%%%%

\section{Analysis of the effect of pulse errors on CDD and CPDD}\label{Appendix:Effect_pulse_errors}

We analyze the impact  of pulse errors by considering their effect in the absence of the sensed field. Then, the imperfect propagator determines the interaction frame of the sensed field in the Hamiltonian in Eq. \eqref{H_sensing_2s_supplemental} and pulse errors lead to loss of contrast. This is also the case when the detuning $\Delta$ is much larger than $g$, so the effect of the sensed field on the evolution due to the Hamiltonian in Eq. \eqref{H_sensing_2s_supplemental} can be neglected.
We consider the Hamiltonian of the qubit with eigenstates $|0\rangle$ and $|-1\rangle$ in Eq. \eqref{Supp_H_sensing_1s} without a sensed field.
In general, $\delta$, $\Omega$ and $\phi$ can also be time dependent.
This Hamiltonian corresponds to the Hamiltonian in Eq. (2) in the main text but includes the additional noise term $\delta$ and no sensed field.

The propagator that characterizes the evolution of the qubit  can in general be parameterized by \cite{GenovPRL2014,GenovPRL2017,Bruns2018PRA}
\begin{equation} \label{Eq:U_bare}
\mathbf{U} = \left[\begin{array}{cc} \sqrt{\epsilon} \e^{i\alpha}  & \sqrt{1-\epsilon} \e^{-i\beta} \\  -\sqrt{1-\epsilon} \e^{i\beta} & \sqrt{\epsilon} \e^{-i\alpha}  \end{array} \right],
\end{equation}
where $p\equiv 1-\epsilon$ is the transition probability, i.e., the probability that the qubit will be transferred to state $|-1\rangle$ if it was initially in state $|0\rangle$, $\epsilon\in[0,1]$ is the unknown error in the transition probability, $\alpha$ and $\beta$ are unknown phases.
In case of a perfect pulse, the transition probability becomes $p=1$ and $\epsilon=0$. However, this is often not the case, e.g., due to frequency or amplitude drifts or inhomogeneity,
which make $\epsilon\ne 0$. For example, an amplitude error of $\Delta/\omega_{L}=(\omega_{L}-\Omega)/\omega_{L}=5\,\%$ from the target Rabi frequency of $\omega_{L}$ (Hartmann-Hahn resonance) leads to $\epsilon=1-\cos^2 \left(\Delta t/2\right)\approx 0.06$ for a target $\omega_{L} t=\pi$ pulse, assuming $\delta=0$.
Such errors can be compensated by applying phased sequences of pulses, where the phases of the subsequent pulses are chosen to cancel the errors of the individual pulses up to a certain order \cite{GenovPRL2014,GenovPRL2017,Bruns2018PRA}.

\begin{figure*}[t!]
\includegraphics[width=0.8\textwidth]{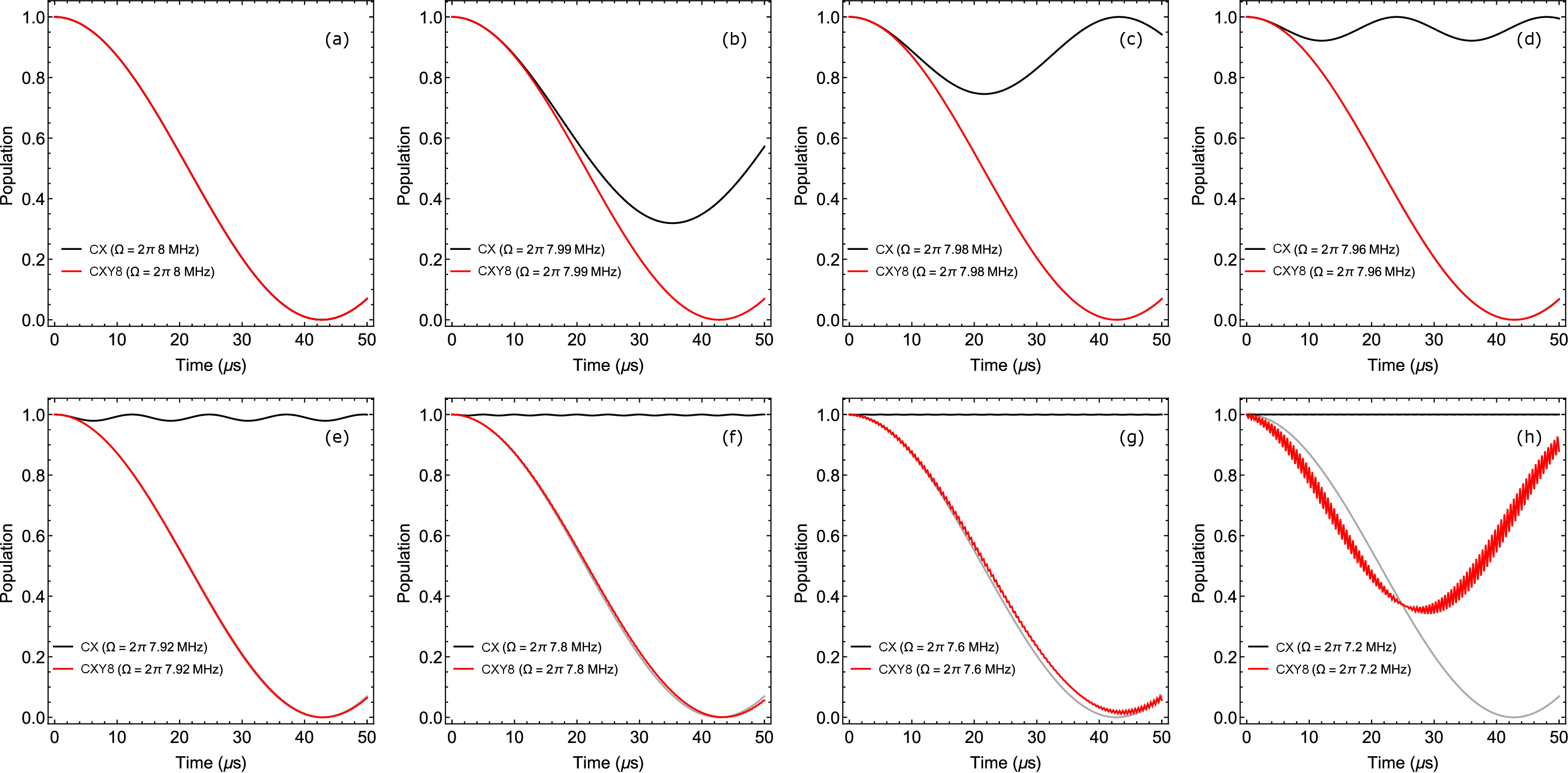}
\caption{
Simulation of CDD and CPDD for sensing,
showing the population of state $|0\rangle)$ for standard CDD (labeled CX) and CPDD (with the CXY8 sequence) with phase changes applied at time intervals of $T=62.5$ ns, corresponding to the parameters for high field sensing in the experimental section. The gray curve shows the expected oscillation in the ideal case when $\Delta=0$ for both protocols. The sensed field has an amplitude of $g=2\pi~11.7$ kHz, angular frequency $\omega_{L} = 2\pi~8$ MHz, and initial phase $\xi=0$. The sensor qubit is initially prepared in the $|0\rangle$ state, e.g., by optical pumping. (a) When the Rabi frequency of the driving field $\Omega=\omega_{L}=2\pi~8$ MHz, one can observe oscillations of the population $P=\cos(\Theta(t)/2)$, where $\Theta=gt$, for both the CX and CXY8 sequences. (b-h) When the Rabi frequency differs from the sensed frequency, i.e., $\Delta = \Omega-\omega_{L}\ne 0$, the oscillations due to the sensed field are quickly lost with CDD, i.e. for the CX sequence, even with small $\Delta\approx 2\pi~10$ kHz, which is of the order of the coupling strength $g$. On the contrary CPDD with the CXY8 sequence allows for a much more robust performance and efficient sensing, e.g., even for $\Delta = 2\pi~400$ kHz, by using the time intervals of the phase changes as a control parameter. The data for all simulations are taken stroboscopically at intervals of $0.5$ $\mu$s, which correspond to the duration of one CXY8 sequence on resonance.
}
\label{FigS2:Sensing_simulation}
\end{figure*}

If the pulses are time separated, the propagator of the whole cycle $\left[\right.$free evolution for time $\tau/2-\text{pulse}-$ free evolution for time $\tau/2\left.\right]$ changes by taking $\alpha\to\widetilde{\alpha}=\alpha+\delta\tau$, where we assumed that $\delta$ is constant during one $\left[\tau/2-\text{pulse}-\tau/2\right]$ period. Although the pulse separation is ideally zero with CPDD, we will use $\widetilde{\alpha}$ in the following analysis to emphasize that the error compensation is present even if the pulses are time-separated. Additionally, a shift in the phase $\phi_{k}$ at the beginning of a pulse leads to $\beta\to\beta+\phi_{k}$ \cite{GenovPRL2014,GenovPRL2017}.
We note that we made no assumption of the pulse shape and detuning time dependence during this analysis, except for the RWA, coherent evolution, and the assumption that effect of the pulse and free evolution before and after the pulse on the qubit is the same during each pulse (except for the effect of the phase $\phi_{k}$).
Thus, the propagator of the $k$-th pulse in the bare basis takes the form
\begin{align}\label{Eq:U_bare_phased}
U(\phi_k)=\left[\begin{array}{cc} \sqrt{\epsilon}\e^{i\widetilde{\alpha}}  & \sqrt{1-\epsilon}\e^{-i(\beta+\phi_{k})}\\
-\sqrt{1-\epsilon}\e^{i(\beta+\phi_{k})} &\sqrt{\epsilon}\e^{-i\widetilde{\alpha}} \end{array} \right].
\end{align}
Assuming coherent evolution during a sequence of $n$ pulses with different initial phases $\phi_{k}$, the propagator of the composite sequence then becomes
\begin{equation}\label{Eq:U_phased_sequence}
U^{(n)}=U(\phi_{n})\dots U(\phi_{1}),
\end{equation}
and the phases $\phi_{k}$ of the individual pulses can be used as control parameters to achieve a robust performance.
We can evaluate the latter by considering the fidelity \cite{RDD_review12Suter,GenovPRL2017}
\begin{equation}\label{Eq:Fid_phased_sequence}
F=\frac{1}{2}\text{Tr}\left[\left(U_0^{(n)}\right)^{\dagger}U^{(n)}\right],
\end{equation}
where $U_0^{(n)}$ is the propagator of the respective pulse sequence when $\epsilon=0$, i.e., when the pulse performs a perfect population inversion. For example, the fidelity of a single pulse is given by $F=\sqrt{1-\epsilon}$. We note that this measure of fidelity does not take into account variation in the phase $\beta$, which is important when we apply an odd number of pulses. However, the latter is fully compensated when we apply an even number of pulses with perfect transition probability. Thus, we use the fidelity measure in Eq. \eqref{Eq:Fid_phased_sequence} as it usually provides a simple and sufficient measure of performance when we apply an even number of pulses.
We can obtain the fidelity of a
sequence of eight pulses with zero phases, i.e., $\phi_{k}=0$, which is given by
\begin{equation}\label{Eq:Fid_zero8}
F_{(\phi_{k}=0)}=1-32 \cos{(\widetilde{\alpha})}^4\epsilon-O(\epsilon^2).
\end{equation}
Additionally, the fidelity of the widely used XY8 sequence \cite{RDD_review12Suter} with phases $\left(0,1,0,1,1,0,1,0\right)\pi/2$ is
\begin{equation}\label{Eq:Fid_XY8}
F_{\text{XY8}}=1-4\left[\cos{(\widetilde{\alpha})}+\cos{(3\widetilde{\alpha})}\right]^2\epsilon^3-O(\epsilon^4).
\end{equation}
Usually the transition probability error is quite small, i.e., $\epsilon \ll 1$, e.g., $\epsilon=1-\cos^2 \left(\Delta t/2\right)\approx 0.006$ for a target $\omega_{L} t=\pi$ pulse and $\Delta/\omega_{L}=0.05$, assuming $\delta=0$, so the error in the fidelity ($1-F$) of the XY8 sequence ($\sim \epsilon^3$) will be much smaller than the one of the sequence with constant zero phases ($\sim \epsilon$). A numerical simulation of the fidelity of the continuous (labeled CX) drive and the XY8 sequence with zero pulse separation (labeled CXY8) is shown in Fig. \ref{Fig2:2D_plots}. The simulation shows that CXY8 is significantly more robust than the standard continous decoupling. Similarly, one can show that we can obtain a robust performance and even better fidelity with other sequences of phased pulses, e.g., by using the KDD or UR sequences \cite{RDD_review12Suter,GenovPRL2014,GenovPRL2017,Bruns2018PRA}.

Finally, we demonstrate the robustness of CPDD for sensing the amplitude $g\approx 2\pi~11.7$ kHz of a weak oscillating field. The results from a numerical simulation and experiments are shown in Fig. 2 in the main text and the details for the simulation results are demonstrated in Fig. \ref{FigS2:Sensing_simulation}.
Figure \ref{FigS2:Sensing_simulation} shows that when the Rabi frequency of the driving field is equal to the frequency of the sensed field, i.e., $\Omega=\omega_{L}=2\pi~8$ MHz, we observe oscillations of the population of state $|0\rangle$ for both CX and CXY8 with $P=\cos(\Theta(t)/2)$, where $\Theta=gt$. The oscillations are quickly lost with CX even with small $\Delta\approx 2\pi~20$ kHz, i.e., of the order of the coupling strength $g$. On the contrary, with CXY8 the robustness is improved by more than twenty times, as we observe oscillations even for $\Delta = 2\pi~400$ kHz.

%%%%%%%%%%%%%%%%%%%%%%%%%%%%%%%%%%%%%%%%%%%%%%%%%%%%%%%%%%%%%%%%
\section{Slope and variance detection with CPDD}
\label{Appendix:Slope_Variance_Detection}
%%%%%%%%%%%%%%%%%%%%%%%%%%%%%%%%%%%%%%%%%%%%%%%%%%%%%%%%%%%%%%%%
In this section, we analyze how CPDD can be used for the detection of a weak oscillating signal.
We consider a measurement, which consists of the microwave pulse sequence $\pi/2(x)$ pulse - dynamical decoupling (DD) - $\pi/2(\phi)$ pulse, where DD can be pulsed, continuous, or CPDD and the phase $\phi$ of the last $\pi/2$ pulse can be varied.
If we apply the dynamical decoupling sequence on resonance, we observe Rabi oscillations, characterized with a pulse area $\Theta$, which can be observed directly in the bare basis, as evident from Eq. \eqref{Eq:propagator_CDD} and Eq. \eqref{Eq:propagator_CPDD}. On resonance, the pulse area $\Theta=2\kappa gt\cos{\left(\xi\right)}$, where the attenuation factor $\kappa=2/\pi$ for pulsed dynamical decoupling and $\kappa=1/2$ for CPDD~\cite{DegenRMP2017}.
In the following, we consider the application of CPDD for slope/linear and variance detection \cite{DegenRMP2017} and compare its performance to standard techniques.

%\subsection{Detection of a signal with a constant phase}
\subsection{Slope/linear detection}

Slope/linear detection is usually applicable when the initial phase $\xi$ of the sensed field, e.g.,  $B\cos{\left(\omega_s t+\xi\right)}$, is fixed at the beginning of each measurement \cite{DegenRMP2017}. The measurement protocol for determining the amplitude $B$ is as described above with the phase of the last $\pi/2$ pulse alternating between $90^{\circ}$ and $-90^{\circ}$, i.e., we apply $\pi/2(\pm y)$ pulses.
The population of state $\ket{0}$ of the NV electron spin with the last $\pi/2(\pm y)$ pulses is then $p_0 =  \frac{1}{2}\mp\frac{1}{2}\sin\left(\Theta\right)$.
The normalized signal contrast from the fluorescence difference from the two measurements is then \cite{Staudenmaier2022,Staudenmaier2023prl}
\begin{align}
c&=c_\text{max}e^{-\Gamma(t)}\sin\left(\Theta\right)\notag\\
&=c_\text{max}e^{-\Gamma(t)}\sin\left(\kappa\gamma_{\text{NV}}Bt\cos{\left(\xi\right)}\right)\notag\\
&\approx c_\text{max}e^{-\Gamma(t)}\kappa\gamma_{\text{NV}}Bt\cos{\left(\xi\right)},
\end{align}
where $c_\text{max}$ is the maximum contrast, $\Gamma(t)$ accounts for decoherence during the duration $t$ of the decoupling sequence, and $\Theta=2\kappa gt\cos{\left(\xi\right)}=\kappa\gamma_{\text{NV}}Bt\cos{\left(\xi\right)}$, as $g=\gamma_{\text{NV}}B/2$ in our experiment. The last approximation is valid only for small $\Theta\ll 1$.
One can estimate the amplitude of the oscillating magnetic field by fitting the observed Rabi oscillations, similar to the experiment in Fig. 2 in the main text with $\Theta=\kappa\gamma_{\text{NV}}Bt\cos{\left(\xi\right)}$.

Another approach is to detect a change in the signal from a reference point of the measurement, which corresponds to a well-known value of $B$. Slope or linear detection is typically used to detect a perturbation in a signal with a constant initial phase  \cite{DegenRMP2017}. The optimal reference point is when $p_0=1/2$ \cite{DegenRMP2017}, i.e., $\Theta=\pi k,k\in \mathbb{Z}$. Then, the expected change in the contrast $d c$ due to a small perturbation $d B$ is given by
\begin{align}
d c&=-c_\text{max}e^{-\Gamma(t)}\cos\left(\Theta=\pi k\right)d \Theta\notag\\
&=\mp c_\text{max}e^{-\Gamma(t)}\kappa \gamma_{\text{NV}}  t\cos{\left(\xi\right)}\, dB,
%\approx c_\text{max}\Theta,
%=c_\text{max}\sin\left(2\kappa gt\cos{\left(\xi\right)}\right)\approx c_\text{max}\,2\kappa gt\cos{\left(\xi\right)},
\end{align}
where the chosen measurement time duration is typically close to $t=T_2/2$ for the best sensitivity, given a typical decay function $\Gamma(t)$ with coherence time $T_2$ \cite{DegenRMP2017}. It is evident that the detection mechanism of CPDD is similar to the one of the standard pulsed techniques with the main difference being the attenuation factor $\kappa$.

\subsection{Variance detection}

When the initial phase and/or amplitude of the detected oscillating signal fluctuate from measurement to measurement it is advantageous to detect the signal variance.
We then consider a measurement, which consists of the microwave pulse sequence $\pi/2(x)$ pulse - dynamical decoupling (DD) - $\pi/2(\pm x)$ pulse, where DD can be pulsed, continuous, or CPDD.
If we apply the respective sequence on resonance, we observe Rabi oscillations with a pulse area $\Theta$, which can be observed directly in the bare basis, as evident from Eq. \eqref{Eq:propagator_CDD} and Eq. \eqref{Eq:propagator_CPDD}.
The population of state $\ket{0}$ of the NV electron spin with the last $\pi/2(\pm x)$ pulses is then $p_0 =  \frac{1}{2}\mp\frac{1}{2}\cos\left(\Theta\right)$.
The normalized signal contrast from the fluorescence difference from the two measurements is \cite{Staudenmaier2022}
\begin{align}
c=c_\text{max}e^{-\Gamma(t)}\cos\left(\Theta\right).
\end{align}
Multiple readouts are performed and we average the result, which leads to
\begin{align}
\overline{c}=c_\text{max}e^{-\Gamma(t)}\left\langle\cos\left(\Theta\right)\right\rangle.
\end{align}
First, we consider the case of detecting the amplitude of a signal $B\cos{\left(\omega_s t+\xi\right)}$ with a constant amplitude $B$ and random initial phase $\xi$. The variance of the magnetic field is $\left\langle B(t)^2\right\rangle\equiv B_{\text{rms}}^2=\langle \left(B\cos{\left(\omega_s t+\xi\right)}\right)^2\rangle=B^2/2$ with $B_{\text{rms}}^2$ being the root-mean-square magnetic field. The contrast takes the form
\begin{align}\label{Eq:Decay_const_phase}
\overline{c}&=c_\text{max}e^{-\Gamma(t)}\left\langle\cos\left(\kappa\gamma_{\text{NV}}Bt\cos{\left(\xi\right)}\right)\right\rangle\\
&=c_\text{max}e^{-\Gamma(t)}J_0\left(\kappa\gamma_{\text{NV}}Bt\right)\notag\\
&\approx c_\text{max}e^{-\Gamma(t)}(1-\kappa^2\gamma_{\text{NV}}^2 B^2 t^2/4)\notag,\\
&= c_\text{max}e^{-\Gamma(t)}(1-\kappa^2\gamma_{\text{NV}}^2 B_{\text{rms}}^2 t^2/2)\notag,
\end{align}
where $J_0$ is the Bessel function of the first kind of zeroth order and the approximation is valid only for small $\gamma_{\text{NV}}Bt\ll 1$. We can detect the amplitude of the oscillating signal by fitting the experimental values to the expected decay in Eq. \eqref{Eq:Decay_const_phase} when we vary the duration $t$ of the decoupling sequence, e.g., as demonstrated in Fig. \ref{Fig:CXY8order}.

Next, we consider the case when both the amplitude and the phase of the signal vary from experiment to experiment. For example, this is the case when the proton signal we observe is due to statistical polarization in the detection volume of the NV center. We assume that the magnetic field, generated by the sample due to statistical polarization is given by \cite{Gefen2019,Staudenmaier2022}
\begin{align}
B(t)=A\cos{(\omega_s t)}+D\sin{(\omega_s t)},
\end{align}
where $A,D\sim N(0,B_{\text{rms}})$ the normal distribution with zero mean and standard deviation $B_{\text{rms}}$, so the variance of the magnetic field is $\left\langle B(t)^2\right\rangle=B_{\text{rms}}^2$. When we apply a DD sequence on resonance we can typically sense only $A\cos{(\omega_s t)}$.
The accumulated phase $\Theta$ follows a Gaussian distribution, centred at zero with a variance $\langle \Theta^2\rangle$ , so we obtain
\begin{align}\label{Eq:Decay_stat_polarization}
\overline{c}&=c_\text{max}e^{-\Gamma(t)}\langle\cos\left(\Theta\right)\rangle=c_\text{max}e^{-\Gamma(t)}\exp\left(-\langle\Theta^2\rangle/2\right)\notag\\
&=c_\text{max}e^{-\Gamma(t)}\exp\left(-\kappa^2\gamma_{\text{NV}}^2 B_{\text{rms}}^2 t^2/2\right)\notag\\
&\approx c_\text{max}e^{-\Gamma(t)}\left(1-\kappa^2\gamma_{\text{NV}}^2 B_{\text{rms}}^2 t^2/2\right),
\end{align}
where we used that
$\langle \Theta^2\rangle\equiv\Theta_{\text{rms}}^2=\kappa^2\gamma_{\text{NV}}^2 \langle B^2\rangle t^2$ with $\langle B^2\rangle=B_{\text{rms}}^2$. The last approximation is valid only for small $\langle\Theta^2\rangle$. %Similar to the previous cases,
We detect $B_{\text{rms}}$ by fitting the experimental values to the decay function in Eq. \eqref{Eq:Decay_stat_polarization} when we vary the duration $t$ of the decoupling sequence.
Similar to linear detection, the performance of CPDD for variance detection is similar to the one of the standard pulsed techniques with the main difference being the attenuation factor $\kappa$.

%%%%%%%%%%%%%%%%%%%%%%%%%% FIGURE %%%%%%%%%%%%%%%%%%%%%%%%%%%%%
\begin{figure}[t!]
\includegraphics[width=0.95\columnwidth]{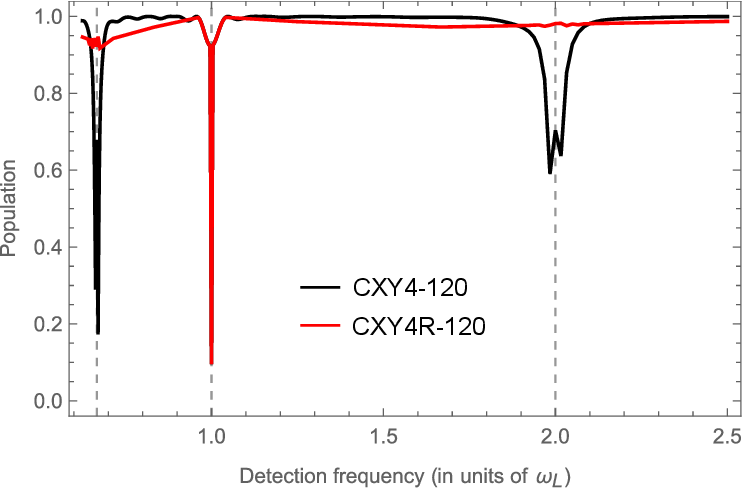}
\caption{
Simulated spectra showing the population of state $|0\rangle$ as a function
of the detection (angular) frequency of CPDD  $\pi/T$ in units of $\omega_{L}$ with phase changes at time intervals $T$, which varied between 100 and 400 ns with the Rabi frequency $\Omega=\pi/T$. The sensed field has an angular frequency $\omega_{L} = 2\pi~2$ MHz, an amplitude of $g=2\pi~20$ kHz, and initial phase $\xi=0$. The ratio $g/\Omega=0.01$ on resonance and the total number of $480$ pulses correspond to the respective parameters in simulations of spurious harmonics with pulsed DD in \cite{LoretzPRX2015}. The sensor qubit is initially prepared in state $|0\rangle$, e.g., by optical pumping. CPDD with the CXY4 sequence shows the expected spurious harmonics at $2\omega_{L}/3$ and $2\omega_{L}$, similar to pulsed DD with XY4 \cite{LoretzPRX2015}. CPDD with the CXY4R, which uses phase randomization, allows one to suppress these harmonics almost completely, similar to pulsed DD \cite{WangPRL2019}. The simulation with randomized phases is performed by adding a random phase to each of the $120$ sequences in a single experiment and running the simulation $640$ times. The incomplete suppression around $2\omega_{L}/3$ is most likely due to the limited number of runs and the higher ratio $g/\Omega$.
}
\label{Fig:XY4andXY4R}
\end{figure}
%%%%%%%%%%%%%%%%%%%%%%%%%% FIGURE %%%%%%%%%%%%%%%%%%%%%%%%%%%%%

\section{Spurious harmonics}\label{Appendix:Spurious_harmonics}

Similar to pulsed dynamical decoupling \cite{LoretzPRX2015}, the effect of the signal during sensing can produce spurious harmonics, which depend on the applied sequence of phase changes (see Fig. \ref{Fig:XY4andXY4R}).
Since the signal is sensed by its effect during the sequence, the spurious harmonics can be comparable to the main one. This is evident in the simulation in Fig. \ref{Fig:XY4andXY4R}, where the black curve shows the expected spurious harmonics with CPDD with the CXY4 sequence at $2\omega_{L}/3$ and $2\omega_{L}$, in a similar way to pulsed dynamical decoupling with XY4 \cite{LoretzPRX2015}. CPDD with the CXY4R, which uses phase randomization, allows one to suppress these harmonics almost completely, in a similar way to pulsed DD \cite{WangPRL2019}.  The simulation parameters correspond to the respective spurious harmonics with pulsed DD in Ref. \cite{LoretzPRX2015}.

\section{Sample Preparation}\label{Appendix:Sample_Preparation}

Two diamond samples where used for the experiments. For the measurements shown Figure~3 of the main text the diamond sample is an electronic grade, type IIa synthetic diamond featuring an isotopically enriched $^{12}$C layer of $\approx 250~\mathrm{\mu m}$ thickness. The layer with an isotopic purity of 99.999 \% was grown  by Element Six using chemical vapor deposition. To create shallow NV centers suitable for the detection of hydrogen spins on the diamond surface, ion implantation was employed using nitrogen ions with 98 \% $^{15}$N isotopic abundance. An ion energy of $2.5~\mathrm{keV}$ was chosen to yield NV centers at typical depths of $8\pm2~\mathrm{nm}$ \cite{Findler2020SciRep}. A nitrogen ion dose of $5 \times 10^8~\frac{^{15}N^+}{cm^2}$ results in single, isolated NV centers. After ion implantation the sample was annealed in ultra high vacuum with pressures below $1\times10^{-7}~\mathrm{mbar}$ at $1000~\mathrm{^{\circ}C}$ for 3 hours to mobilize the created vacancies and allow for NV center creation with a yield of 0.7 \%. Before each processing step, the sample was boiled several hours in a 1:1:1 mixture of sulfuric, perchloric and nitric acid to remove any graphitic or organic residues from the surface. The NV is about $d=(10.1\pm 0.5)$ nm deep in the diamond, which we estimated by measuring the variance of the fluctuating NMR magnetic field at its location with XY8 dynamical decoupling \cite{MuellerNatComm2014}.

For the other measurements a diamond sample that is fabricated into a solid immersion lens by Element Six is used. It was overgrown with a 100-nm-thick layer of isotopically purified $^{12}$C (99.999 \%). The NV centers here were incorporated during the growth of the $^{12}$C layer.

%%%%%%%%%%%%%%%%%%%%%%%%%% FIGURE %%%%%%%%%%%%%%%%%%%%%%%%%%%%%
\begin{figure*}[t!]
\includegraphics[width=0.9\textwidth]{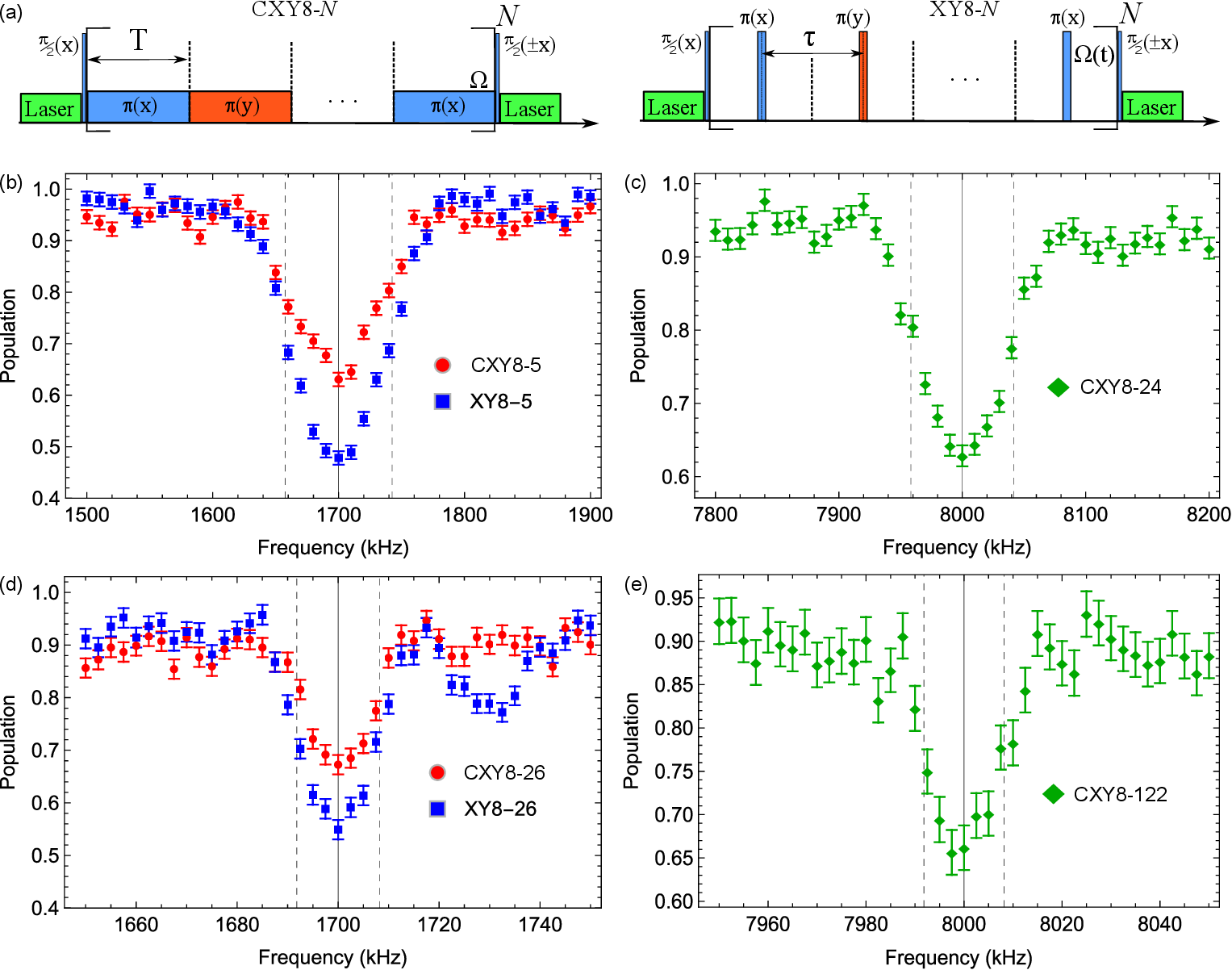}
\caption{
(a) Pulse scheme for the CXY8-$N$ and XY8-$N$ measurements for frequency sensing. We use a laser to prepare the electron spin of an NV center in state $|0\rangle$ by optical pumping and read out its final state. We prepare a coherent superposition state by a microwave $\pi/2$ pulse, apply the control scheme (CXY8 or XY8), and then another $\pi/2$ pulse with a phase that alternates, to map coherences back onto populations. The final population of state $|0\rangle$ is estimated from the difference of the signals from the two alternating measurements, reducing errors due to charge-state and count-rate fluctuations \cite{haberle2017nuclear}.We vary $T$ ($\tau$) for CXY8-$N$ (XY8-$N$) with fixed repetition number $N$ to detect the frequency of the AC signal. (b) Measured signal, corresponding to the population of state $|0\rangle$ of the electron spin of a single NV center, vs. the detected signal frequency $f=\frac{1}{2T}$ ($f=\frac{1}{2\tau}$) %\nicefrac
for CXY8 (XY8). We observe a dip at the frequency $1700\,$kHz of the artificial signal for both CXY8-5 and XY8-5 sequences, as expected from theory. The slightly lower population with XY8 than CXY8 on resonance is expected and due to the difference in the rectified signal amplitudes of the two sequences (see text). The vertical dashed lines for all subfigures show the expected full width at half maximum (FWHM) of the dip, i.e., the inverse of the sequence duration, which is FWHM $\approx 85\,$kHz in this case. (c) We observe the expected a dip in the population at $8000\,$kHz of a high frequency artificial signal, where only CXY8 is applicable. FWHM $\approx 83.3\,$kHz. (d) The measurement is the same as in (b) but with a lower artificial signal amplitude, which requires a higher sequence order and total sequence duration to observe a similar dip; FWHM $\approx 16.3\,$kHz. The additional sideband dip with XY8 does not correspond to the frequency of a real signal and is likely due to pulse errors of the non-instantaneous rectangular pulses. (e) The measurement is same as in (d) but with a lower artificial signal amplitude; FWHM $\approx 16.4\,$kHz due to the slightly shorter sequence duration than in (d). All experimental results show excellent correspondence to theory.
}
\label{Fig:CXY8_XY8_tau_scan}
\end{figure*}
%%%%%%%%%%%%%%%%%%%%%%%%%% FIGURE %%%%%%%%%%%%%%%%%%%%%%%%%%%%%

\section{Experimental Setup}\label{Appendix:Experimental_setup}

We perform the experiments at two different magnetic fields. The low magnetic field was chosen to be $B_z=(485\pm 1)\,\mathrm{G}$,  close to the excited state level anti-crossing, as the inherent nitrogen is polarized, and the fluorescence is more sensitive to field fluctuations. This makes it a convenient field to work with experimentally. Additionally, as we compare results for different detuning from the signal frequency, it is expedient to reduce any possible effect of another hyperfine transition which could result in a somewhat different Rabi frequency. We note that this effect is likely to be small even for non-polarized nitrogen at our higher magnetic field where we apply a Rabi frequency of $(2\pi)\,8$\,MHz as the hyperfine splitting is only $\approx(2\pi)\,3$\,MHz for \textsuperscript{15}N.

All MW pulses for the experimental sequences were generated in direct synthesis by a Keysight (M8195A) and Tektronix (AWG70001A) arbitrary waveform generators, using a Python based in house written software (Qudi) \cite{Binder2017qudi}. The measurement sequences are programmed to keep the phase relation between all pulses in the rotating frame and the high sampling rate of the AWGs (50 and 60\,GS/s) allows to have precise direct control of the signal phases.
The instantaneous rise time of the devices allows for a bandwidth as the inverse of twice the sampling rate.

For the CDD sequence, we first prepare  $\phi_1=\pi/2$ with respect to the phase ($\phi=0$) of the initial $\pi/2$ pulse used to prepare the electron spin in a superposition state. For both the XY8-$N$ and CXY8-$N$ sequences, $\phi_{1...8}=[0,\frac{\pi}{2},0,\frac{\pi}{2},\frac{\pi}{2},0,\frac{\pi}{2},0]$ and then the phase repeats itself $N$ times.

An external test signal is applied via the same antenna as the microwave control for the NV center manipulation. To apply the test signal two different sources were used depending on whether phase control over the test signal was necessary. For the measurements with phase control, as for the ones of Figure~2(c),(d) in the main text, the signal was generated with the same AWG as the microwave control for NV center. We use an external signal from another microwave source (Rohde\&Schwarz SMIQ04B) for the other measurements with an artificial oscillating signal. In this case there is no phase relationship between the measured signal and the control sequence.

We use a FAST ComTec P7887 for the Qdyne measurement. The recorded photons are saved with an individual time tag of their arrival during the sweep when they occur. This information is stored in a file that is accessible for further analysis. In post processing the photons are filtered for the readout window of the excitation laser pulse. The (undersampled) oscillation of the test signal is imprinted onto this time trace. Because of the readout noise of the individual readout the oscillation is not directly visible but becomes evident after performing, e.g., a Fast Fourier Transform of the data,  as shown in Figure~4b of the main text.

\section{Additional Experimental Results} \label{Appendix:Additional_Experimental_Results}

\subsection{Quantum sensing of the frequency of an artificial AC field}\label{Appendix:Experimental_AC_sensing}

In a first series of experiments, we use a single NV center to detect the frequency of an artificially generated AC signal at frequency $1700\,$ kHz ($8000\,$ kHz), which we label low (high) frequency for two different signal amplitudes.
Figure \ref{Fig:CXY8_XY8_tau_scan}(a) shows the pulse scheme for the CXY8-$N$ and XY8-$N$ measurements. We use a laser to prepare the electron spin of the NV center in state $|0\rangle$ by optical pumping and read out its final state. We prepare a coherent superposition state by a microwave $\pi/2$ pulse, apply the control scheme (CXY8 or XY8), and then use another $\pi/2$ pulse with a phase that alternates between $0$ and $180^{\circ}$, to map coherences back onto populations. We estimate the final population of state $|0\rangle$ from the difference between the signals, reducing errors due to charge-state and count-rate fluctuations \cite{haberle2017nuclear}. All control fields originate from the same arbitrary waveform generator and amplifier. We generate the artificial external signal to be sensed from a second microwave source, so the phase of the signal at the beginning of the interaction is not fixed with respect to the control fields. We vary the interval of phase changes $T$ (the interpulse duration $\tau$) for CXY8 (XY8) to detect the frequency of the AC signal, using the relation $f=\frac{1}{2T}$ ($f=\frac{1}{2\tau}$). %\nicefrac

Figure \ref{Fig:CXY8_XY8_tau_scan}(b) demonstrates frequency sensing of a relatively low frequency (1700 kHz) AC signal, which is typical for pulsed DD applications for NV centers. As expected from theory, the measured signal shows a dip in the population of state $|0\rangle$ at the frequency $1700\,$kHz of the artificial signal, for both CXY8-5 and XY8-5 sequences. The slightly lower population with XY8 than CXY8 on resonance is expected from theory and is due to the corresponding difference in the rectified signal amplitudes of the two sequences, which are $g/2$ ($2g/\pi$) for CXY8 (XY8) with $g$ -- the amplitude of the artificial signal in angular frequency units (see the main text). The expected full width at half maximum (FWHM) of the dip also matches theory and is approximately equal to the inverse of the sequence duration $8N\times\frac{1}{2f}$ %$8N\times\nicefrac{1}{2f}$
ms on resonance. Thus, FWHM $\approx \frac{2f}{8N}=\frac{2\times1700}{8\times 5}=85\,$kHz %$\approx \nicefrac{2f}{8N}=\nicefrac{2\times1700}{8\times 5}=85\,$kHz
for both CXY8 and XY8.

Figure \ref{Fig:CXY8_XY8_tau_scan}(c) demonstrates the application of CXY8-24 for detecting a high frequency AC field, where standard pulsed DD, e.g., with the XY8 sequence, cannot be applied. We observe the expected dip in the population at the frequency $8000\,$kHz of the artificial signal with a FWHM $\approx \frac{2f}{8N}=\frac{2\times 8000}{8\times 24}=83.3\,$kHz.
%$\approx \nicefrac{2f}{8N}=\nicefrac{2\times 8000}{8\times 24}=83.3\,$kHz.
%

\begin{figure}[t!]
\includegraphics[width=\columnwidth]{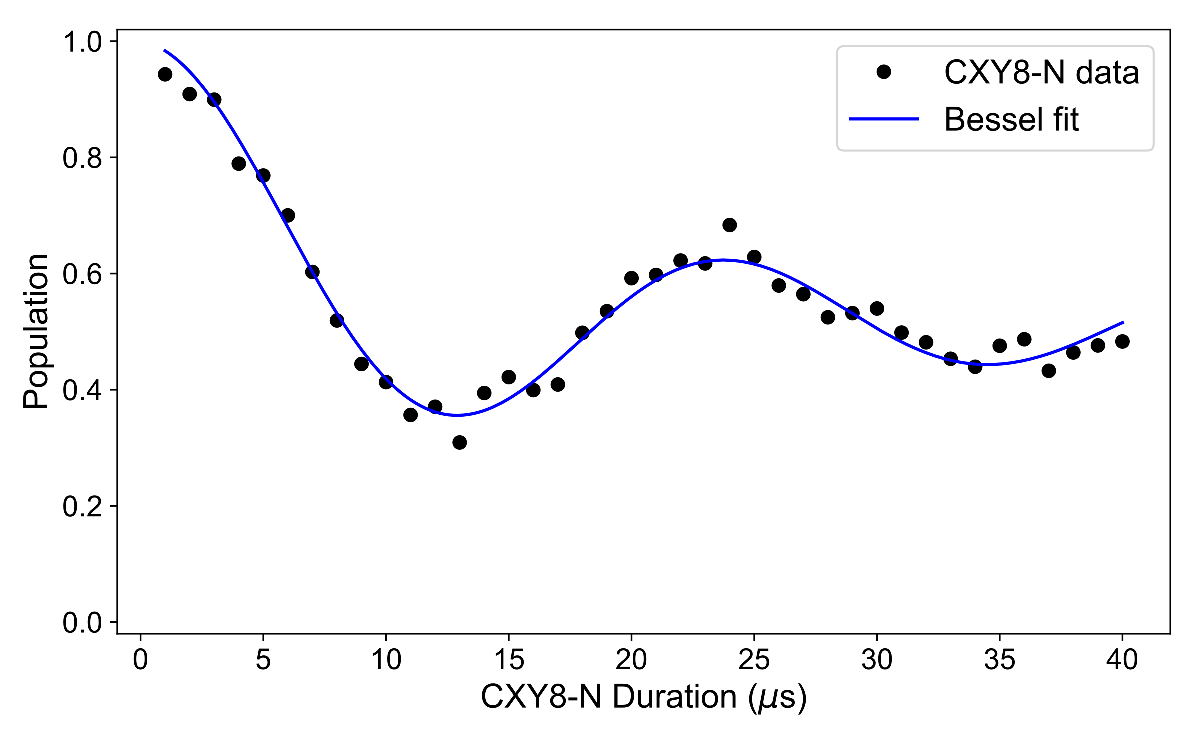
%referee reply/Fig_SM_HHXY8-order.eps
}
\caption{
CXY8 order measurement with an 8 MHz test signal. The order is swept from 2 to 80 in steps of 2 with $T=0.5\,\mu$s, resulting in decoupling times from 1 $\mu$s to 40 $\mu$s. With the frequency of the oscillation the amplitude of the signal at the NV center is estimated from a fit to the function $\bar{c}(t) = c_\mathrm{max} \, J_0(gt) e^{-(t/T_2)^b} + \mathrm{offset}$, as per Eq.~\eqref{Eq:Decay_const_phase}. We obtain an estimate of $g=(2\pi)46.7\pm\,0.5$ kHz, where the error interval is given $\pm$ the standard error. Thus, the estimate of the AC magnetic field amplitude is $B=2g/\gamma_{\mathrm{NV}}=(3.33 \pm 0.04) \mu$T.
}
\label{Fig:CXY8order}
\end{figure}

In the next two measurements, we detected the frequency of an AC signal with a lower amplitude. The measurement details for Figs. \ref{Fig:CXY8_XY8_tau_scan}(d),(e) are the same as for Figs. \ref{Fig:CXY8_XY8_tau_scan}(b),(c), respectively, with the only difference being the order $N$ of the respective sequence, i.e., the number of times $N$ the whole sequence is repeated. The higher order is necessary to observe a similar dip as in (b) and (c), given a lower signal amplitude. We observe the expected dip in the population at the frequency $1700\,$kHz in Fig. \ref{Fig:CXY8_XY8_tau_scan}(d) for both CXY8 and XY8 with a lower FWHM $\approx 16.4\,$kHz due to the longer total duration of both sequences. Similarly, Fig. \ref{Fig:CXY8_XY8_tau_scan}(e) shows a population dip at $8000\,$kHz with FWHM $\approx 16.4\,$kHz. All experimental results match theory very well.

\subsection{Qdyne}\label{Appendix:Qdyne}
We also perform quantum heterodyne (Qdyne) measurements with the single NV center and an 8\,MHz test signal. For details on the measurement technique see the main text and references \cite{SchmittScience2017, Staudenmaier2021pra, Staudenmaier2023prl}.

First, we estimate the signal strength at the location of the NV center. For this we perform an order ($N$) scan of the CXY8-$N$ measurement with fixed $T=1/(2f_{L})$ on resonance, see Figure~\ref{Fig:CXY8order}.
From a fit of the Bessel function as in equation~\eqref{Eq:Decay_const_phase} and the relation $\Theta=gt=1/2 \gamma_{\mathrm{NV}} B t$ we determine that $g=(2\pi)46.7\pm\,0.5$ kHz, where the error interval is given $\pm$ the standard error. Thus, the estimate of the AC magnetic field amplitude is $B=2g/\gamma_{\mathrm{NV}}=(3.33 \pm 0.04) \mu$T.

\begin{figure*}[t!]
\includegraphics[width=0.95\textwidth]{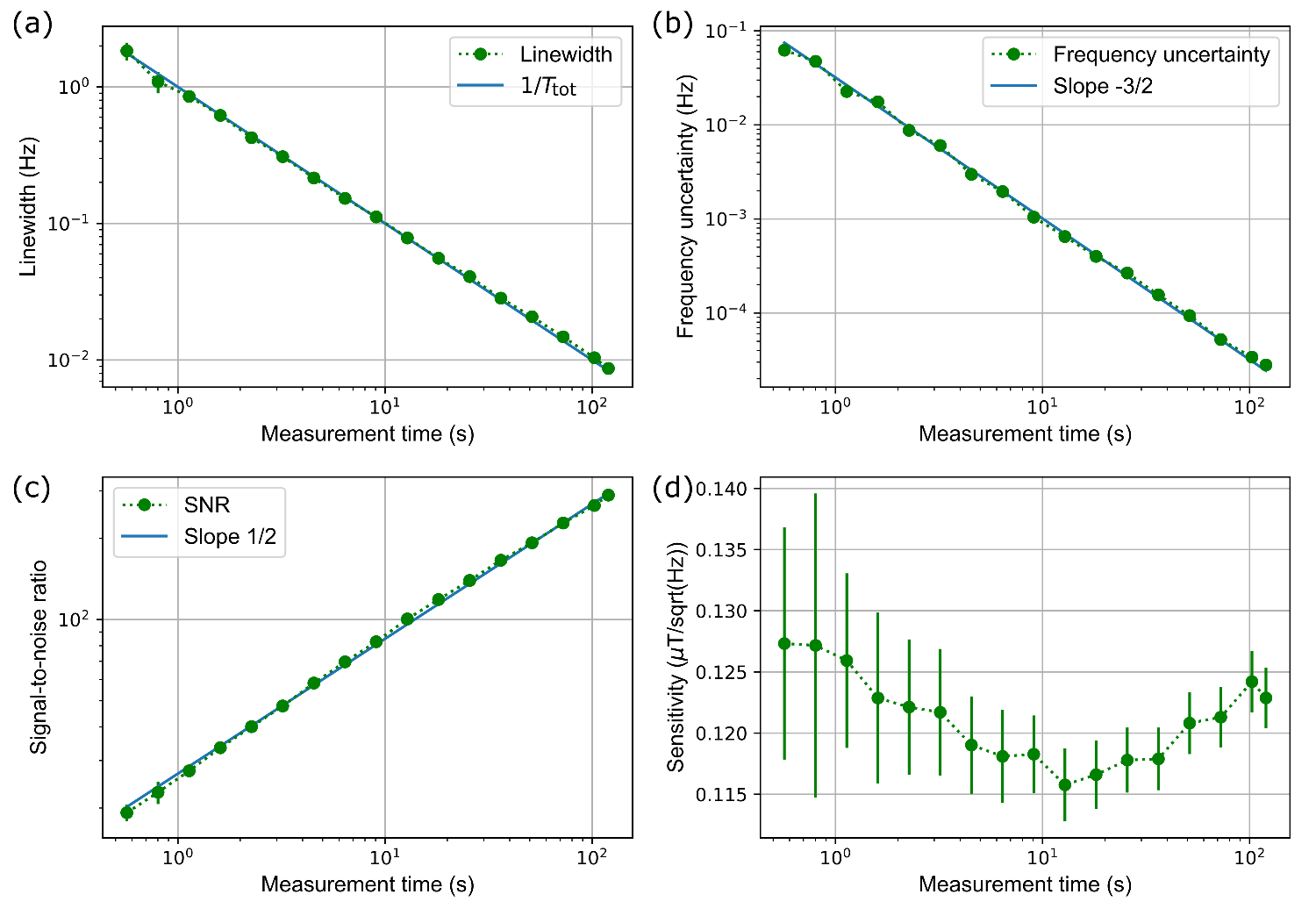}
\caption{
Results of Qdyne measurement with a $(3.33 \pm 0.04) \mu$T signal at 8~MHz. All results in dependence on the total measurement time $T_\mathrm{tot}$. (a) The linewidth of the measured signal shows an expected 1/$T_\mathrm{tot}$ behavior. (b) Precision of frequency estimation scales with $T_\mathrm{tot}^{-3/2}$. We achieve 28~$\mu$Hz uncertainty in 120~s measurement time and 32~mHz/Hz$^{3/2}$ frequency sensitivity. (c) Scaling of the signal-to-noise ratio (SNR) with $T_\mathrm{tot}^{1/2}$. (d) Measured sensitivity of the NV probe. The weighted average is $\eta$ = (118 $\pm$ 8) nT/$\sqrt{Hz}$.
}
\label{Fig:Supp_Qdyne}
\end{figure*}

Figure 4 in the main text shows the measured spectrum of the Qdyne measurement. In Figure~\ref{Fig:Supp_Qdyne} we show further analysis of the performance of the Qdyne measurement. The analysis is done equivalently to Ref. \cite{Staudenmaier2021pra}. The total measurement time is $T_\mathrm{tot}=120$ s. For the analysis we slice the data into smaller parts. The shortest time is 0.57 s (2$^{16}$ samples). For every measurement time the FFT of the corresponding time trace is calculated. A Lorentzian fit of the form
\begin{equation}
    L(\nu) = A \frac{\sigma^2}{(\nu_0 - \nu)^2 + \sigma^2} + \mathrm{offset}
\end{equation}
with amplitude $A$, center frequency $\nu_0$ and half-width at half-maximum $\sigma$ is performed on the data within a fixed frequency window around the frequency of the signal peak. We extract the center frequency $\nu_0$ together with its uncertainty $\delta\nu_0$, linewidth $2\sigma$ and amplitude $A$ directly from the fit parameters. We calculate the signal-to-noise ratio (SNR) by determining the noise $\delta A$ as the standard deviation of the noise floor of the FFT with $\mathrm{SNR} = \frac{A}{\delta A}$. While the SNR scales according to the standard quantum limit $\mathrm{SNR} \propto \sqrt{N_\mathrm{tot}}\propto T_\mathrm{tot}^{1/2}$, where $N_\mathrm{tot}$ is the total number of measurements, the frequency uncertainty shows a different scaling. As a result of reduced noise over measurement time ($\propto 1/\sqrt{N_\mathrm{tot}}\propto\,T_\mathrm{tot}^{-1/2}$) and the scaling of the linewidth as $2\sigma = 1/T_\mathrm{tot}$, %due to Fourier-sampling
the estimation precision of frequency scales as $\delta\nu_0 \propto T_\mathrm{tot}^{-3/2}$.

Figure \ref{Fig:Supp_Qdyne} demonstrates the typical time scaling for a Qdyne measurement \cite{SchmittScience2017,Staudenmaier2021pra}. The linewidth $2\sigma$ of the signal decreases exactly with 1/$T_\mathrm{tot}$ (see Fig. \ref{Fig:Supp_Qdyne}(a)), as expected from theory. Figure \ref{Fig:Supp_Qdyne}(b) shows that the precision $\delta\nu_0$ in frequency estimation scales with $T_\mathrm{tot}^{-3/2}$. Finally, SNR increases according to the standard quantum limit as $T_\mathrm{tot}^{1/2}$, as shown in Fig. \ref{Fig:Supp_Qdyne}(c).

The estimated frequency in the main text of $\nu_0 = 8000001.8883$\,Hz is determined from the beat frequency $f_b = |\nu_\mathrm{LO} - \nu_0|$ where $\nu_\mathrm{LO} = \frac{\mathrm{round}(\nu_0 T_L)}{T_L}$ ($T_L = 8.646\,\mu$s duration of a single measurement) is a reference frequency given by the undersampling of the signal. Deviation of about 1.9\,Hz from the expected 8\,MHz signal is due to different internal clock of the microwave source generating the test signal.

\subsection{Sensitivity and dynamic range of CPDD}\label{Appendix:Sensitivity}

We determine the frequency sensitivity of CPDD in combination with Qdyne by multiplying the estimated uncertainty $\delta\nu_0$ with the adequate time scaling, $\eta_{\nu_0} = \delta\nu_0\, T_\mathrm{tot}^{3/2}$, where $T_\mathrm{tot}$ is the total measurement time. This is characteristic for Qdyne, as shown in Fig. \ref{Fig:Supp_Qdyne}(b) and \cite{SchmittScience2017}, to obtain 32 mHz/Hz$^{3/2}$. Analogously, we calculate the amplitude sensitivity $\eta_B = \frac{B_0}{\mathrm{SNR}} T_\mathrm{tot}^{-1/2}$. We take the weighted average of the results in (d) and obtain $\eta$ = (118 $\pm$ 8) nT/$\sqrt{Hz}$.

Next, we note that formulas for the frequency and amplitude sensitivity for CPDD are the same as for pulsed dynamical decoupling, except for a difference in the signal attenuation factor $\kappa$ (see also Section \ref{Appendix:Slope_Variance_Detection} and Ref. \cite{DegenRMP2017}). One can obtain intuition by considering the photon shot noise limited sensitivity for detecting the amplitude of an oscillating magnetic field with an NV center, using standard pulsed DD sequences. The sensitivity is given by \cite{DegenRMP2017,Taylor2008high,Barry2020sensitivity}
\begin{equation}
    \eta(t)= \frac{2}{\gamma_{\text{NV}}\kappa C(t)}\frac{\sqrt{t+t_\text{oh}}}{\sqrt{N_{\text{ph}}}t},
    %\eta(t)= \frac{2}{\gamma_{\text{NV}}\kappa C(t)}\frac{1}{\sqrt{N_{\text{ph}}t}},
\label{eq:sensitivity_supp}
\end{equation}
where $\gamma_{\text{NV}}$ is the NV center gyromagnetic ratio, $t$ is the sensing time, $t_\text{oh}$ is the overhead time for readout and initialization, $N_\text{ph}$ is the
the average number of photons collected per measurement run, $C(t)=(a-b)\exp{\left[-\Gamma(t)\right]}$ is the expected normalized contrast with $\Gamma(t)$ characterizing the respective decay due to the NV spin decoherence, $a$ and $b$ denote the normalized signal, which corresponds to the bright and dark states of the NV electron spin, respectively.
For the NV center of the Qdyne measurements we estimate exponential decay of the signal with $\exp{\left[-\Gamma(t)\right]},\,\Gamma(t) = t/T_2$ and coherence time $T_2 \approx 250\,\mathrm{\mu s}$. The average number of fluorescence photons per readout is $N_\text{ph} = 0.164$ and the normalized contrast $a-b = 0.32$. The sensing time is $t = 5\,\mathrm{\mu s}$ corresponding to order 10 of a CXY8 at 8\,MHz resonance frequency, and the overhead time $t_\text{oh} = 3.646\,\mathrm{\mu s}$. With $\gamma_\text{NV} = 2\pi \times 28.03\,\mathrm{MHz/mT}$ and $\kappa = 1/2$ the theoretical photon shot noise limited sensitivity according to Eq.~\eqref{eq:sensitivity_supp} is $\eta \approx 105\,\mathrm{nT/\sqrt{Hz}}$. The experimental sensitivity is slightly worse due to additional noise, that is not accounted for in Eq.~\eqref{eq:sensitivity_supp}.

The photon shot noise limited amplitude sensitivity could be improved further by optimizing the sensing time $t$ to minimize the expression $\frac{\sqrt{t+t_\text{oh}}}{C(t)t}$. We obtain an optimal sensing time of
$t=\frac{1}{2} \left(\frac{T_2}{2}-t_{\text{oh}}+\sqrt{\frac{T_2^2}{4}+3 T_2
   t_{\text{oh}}+t_{\text{oh}}^2}\right)\approx \frac{T_2}{2}+t_{\text{oh}} -\frac{4
   t_{\text{oh}}^2}{T_2}\approx 128.45\,\mathrm{\mu s}$, where the first approximation is valid for $t_{\text{oh}}\ll T_2$, reaching the theoretical optimum of $\eta_\text{opt} = 26 \,\mathrm{nT/\sqrt{Hz}}$.
   Finally, we note that the reported sensitivities can be improved substantially by using NV centers with longer coherence times, which can reach up to several milliseconds \cite{Salhov2024prl}. Other possibilities include improving further the photon collection efficiency or using ensembles of NV centers where the average detected photon number per measurement is much higher and the effect of photon shot noise is reduced significantly \cite{DegenRMP2017,Barry2020sensitivity}.

The signal attenuation factor for CPDD is $\kappa=1/2$, similarly to the standard continuous dynamical decoupling protocol (see Eq. (3)) in the main text). In contrast, pulsed dynamical decoupling has a slightly better attenuation factor of $\kappa=2/\pi$ \cite{DegenRMP2017}. %The attenuation factor for CPDD is the same as with continuous dynamical decoupling and is slightly worse than with pulsed DD.
This effect can be compensated by a longer coherence time as the repetition rate of the spin flips can be higher with CPDD than with pulsed methods due to the negligible waiting time. CPDD also has higher fidelity in comparison to continuous decoupling, which results in improved sensitivity.

The dynamic range of CPDD is similar to the one of dressed states methods, e.g., continuous dynamical decoupling, i.e., 0.1-100 MHz for electronic spins or superconducting qubits \cite{DegenRMP2017} with the additional advantage of improved robustness. We note that the dynamic range can be expanded substantially by combining CPDD with other techniques, similar to the possibility of sensing GHz fields with double drive continuous decoupling \cite{CaiNJP2012,Salhov2024prl}.

%\input{./old_bibliography_SM}
%\bibliography{references}

\begin{thebibliography}{66}%
\makeatletter
\providecommand \@ifxundefined [1]{%
 \@ifx{#1\undefined}
}%
\providecommand \@ifnum [1]{%
 \ifnum #1\expandafter \@firstoftwo
 \else \expandafter \@secondoftwo
 \fi
}%
\providecommand \@ifx [1]{%
 \ifx #1\expandafter \@firstoftwo
 \else \expandafter \@secondoftwo
 \fi
}%
\providecommand \natexlab [1]{#1}%
\providecommand \enquote  [1]{``#1''}%
\providecommand \bibnamefont  [1]{#1}%
\providecommand \bibfnamefont [1]{#1}%
\providecommand \citenamefont [1]{#1}%
\providecommand \href@noop [0]{\@secondoftwo}%
\providecommand \href [0]{\begingroup \@sanitize@url \@href}%
\providecommand \@href[1]{\@@startlink{#1}\@@href}%
\providecommand \@@href[1]{\endgroup#1\@@endlink}%
\providecommand \@sanitize@url [0]{\catcode `\\12\catcode `\$12\catcode `\&12\catcode `\#12\catcode `\^12\catcode `\_12\catcode `\%12\relax}%
\providecommand \@@startlink[1]{}%
\providecommand \@@endlink[0]{}%
\providecommand \url  [0]{\begingroup\@sanitize@url \@url }%
\providecommand \@url [1]{\endgroup\@href {#1}{\urlprefix }}%
\providecommand \urlprefix  [0]{URL }%
\providecommand \Eprint [0]{\href }%
\providecommand \doibase [0]{https://doi.org/}%
\providecommand \selectlanguage [0]{\@gobble}%
\providecommand \bibinfo  [0]{\@secondoftwo}%
\providecommand \bibfield  [0]{\@secondoftwo}%
\providecommand \translation [1]{[#1]}%
\providecommand \BibitemOpen [0]{}%
\providecommand \bibitemStop [0]{}%
\providecommand \bibitemNoStop [0]{.\EOS\space}%
\providecommand \EOS [0]{\spacefactor3000\relax}%
\providecommand \BibitemShut  [1]{\csname bibitem#1\endcsname}%
\let\auto@bib@innerbib\@empty
%</preamble>
\bibitem [{\citenamefont {Viola}\ \emph {et~al.}(1999)\citenamefont {Viola}, \citenamefont {Knill},\ and\ \citenamefont {Lloyd}}]{Viola1999PRL}%
  \BibitemOpen
  \bibfield  {author} {\bibinfo {author} {\bibfnamefont {L.}~\bibnamefont {Viola}}, \bibinfo {author} {\bibfnamefont {E.}~\bibnamefont {Knill}},\ and\ \bibinfo {author} {\bibfnamefont {S.}~\bibnamefont {Lloyd}},\ }\bibfield  {title} {\bibinfo {title} {Dynamical decoupling of open quantum systems},\ }\href {https://doi.org/10.1103/PhysRevLett.82.2417} {\bibfield  {journal} {\bibinfo  {journal} {Phys. Rev. Lett.}\ }\textbf {\bibinfo {volume} {82}},\ \bibinfo {pages} {2417} (\bibinfo {year} {1999})}\BibitemShut {NoStop}%
\bibitem [{\citenamefont {Suter}\ and\ \citenamefont {\'Alvarez}(2016)}]{Suter2016RevModPhys}%
  \BibitemOpen
  \bibfield  {author} {\bibinfo {author} {\bibfnamefont {D.}~\bibnamefont {Suter}}\ and\ \bibinfo {author} {\bibfnamefont {G.~A.}\ \bibnamefont {\'Alvarez}},\ }\bibfield  {title} {\bibinfo {title} {Colloquium: Protecting quantum information against environmental noise},\ }\href {https://doi.org/10.1103/RevModPhys.88.041001} {\bibfield  {journal} {\bibinfo  {journal} {Rev. Mod. Phys.}\ }\textbf {\bibinfo {volume} {88}},\ \bibinfo {pages} {041001} (\bibinfo {year} {2016})}\BibitemShut {NoStop}%
\bibitem [{\citenamefont {Loretz}\ \emph {et~al.}(2015)\citenamefont {Loretz}, \citenamefont {Boss}, \citenamefont {Rosskopf}, \citenamefont {Mamin}, \citenamefont {Rugar},\ and\ \citenamefont {Degen}}]{LoretzPRX2015}%
  \BibitemOpen
  \bibfield  {author} {\bibinfo {author} {\bibfnamefont {M.}~\bibnamefont {Loretz}}, \bibinfo {author} {\bibfnamefont {J.~M.}\ \bibnamefont {Boss}}, \bibinfo {author} {\bibfnamefont {T.}~\bibnamefont {Rosskopf}}, \bibinfo {author} {\bibfnamefont {H.~J.}\ \bibnamefont {Mamin}}, \bibinfo {author} {\bibfnamefont {D.}~\bibnamefont {Rugar}},\ and\ \bibinfo {author} {\bibfnamefont {C.~L.}\ \bibnamefont {Degen}},\ }\bibfield  {title} {\bibinfo {title} {Spurious harmonic response of multipulse quantum sensing sequences},\ }\href {https://doi.org/10.1103/PhysRevX.5.021009} {\bibfield  {journal} {\bibinfo  {journal} {Phys. Rev. X}\ }\textbf {\bibinfo {volume} {5}},\ \bibinfo {pages} {021009} (\bibinfo {year} {2015})}\BibitemShut {NoStop}%
\bibitem [{\citenamefont {Neumann}\ \emph {et~al.}(2013)\citenamefont {Neumann}, \citenamefont {Jakobi}, \citenamefont {Dolde}, \citenamefont {Burk}, \citenamefont {Reuter}, \citenamefont {Waldherr}, \citenamefont {Honert}, \citenamefont {Wolf}, \citenamefont {Brunner}, \citenamefont {Shim}, \citenamefont {Suter}, \citenamefont {Sumiya}, \citenamefont {Isoya},\ and\ \citenamefont {Wrachtrup}}]{WrachrupNano2013}%
  \BibitemOpen
  \bibfield  {author} {\bibinfo {author} {\bibfnamefont {P.}~\bibnamefont {Neumann}}, \bibinfo {author} {\bibfnamefont {I.}~\bibnamefont {Jakobi}}, \bibinfo {author} {\bibfnamefont {F.}~\bibnamefont {Dolde}}, \bibinfo {author} {\bibfnamefont {C.}~\bibnamefont {Burk}}, \bibinfo {author} {\bibfnamefont {R.}~\bibnamefont {Reuter}}, \bibinfo {author} {\bibfnamefont {G.}~\bibnamefont {Waldherr}}, \bibinfo {author} {\bibfnamefont {J.}~\bibnamefont {Honert}}, \bibinfo {author} {\bibfnamefont {T.}~\bibnamefont {Wolf}}, \bibinfo {author} {\bibfnamefont {A.}~\bibnamefont {Brunner}}, \bibinfo {author} {\bibfnamefont {J.~H.}\ \bibnamefont {Shim}}, \bibinfo {author} {\bibfnamefont {D.}~\bibnamefont {Suter}}, \bibinfo {author} {\bibfnamefont {H.}~\bibnamefont {Sumiya}}, \bibinfo {author} {\bibfnamefont {J.}~\bibnamefont {Isoya}},\ and\ \bibinfo {author} {\bibfnamefont {J.}~\bibnamefont {Wrachtrup}},\ }\bibfield  {title} {\bibinfo {title} {High-precision nanoscale temperature sensing using single defects in diamond},\
  }\href {https://doi.org/10.1021/nl401216y} {\bibfield  {journal} {\bibinfo  {journal} {Nano Letters}\ }\textbf {\bibinfo {volume} {13}},\ \bibinfo {pages} {2738} (\bibinfo {year} {2013})},\ \bibinfo {note} {pMID: 23721106},\ \Eprint {https://arxiv.org/abs/https://doi.org/10.1021/nl401216y} {https://doi.org/10.1021/nl401216y} \BibitemShut {NoStop}%
\bibitem [{\citenamefont {Schirhagl}\ \emph {et~al.}(2014)\citenamefont {Schirhagl}, \citenamefont {Chang}, \citenamefont {Loretz},\ and\ \citenamefont {Degen}}]{DegenARPC2014}%
  \BibitemOpen
  \bibfield  {author} {\bibinfo {author} {\bibfnamefont {R.}~\bibnamefont {Schirhagl}}, \bibinfo {author} {\bibfnamefont {K.}~\bibnamefont {Chang}}, \bibinfo {author} {\bibfnamefont {M.}~\bibnamefont {Loretz}},\ and\ \bibinfo {author} {\bibfnamefont {C.~L.}\ \bibnamefont {Degen}},\ }\bibfield  {title} {\bibinfo {title} {Nitrogen-vacancy centers in diamond: Nanoscale sensors for physics and biology},\ }\href {https://doi.org/10.1146/annurev-physchem-040513-103659} {\bibfield  {journal} {\bibinfo  {journal} {Annual Review of Physical Chemistry}\ }\textbf {\bibinfo {volume} {65}},\ \bibinfo {pages} {83} (\bibinfo {year} {2014})},\ \bibinfo {note} {pMID: 24274702},\ \Eprint {https://arxiv.org/abs/https://doi.org/10.1146/annurev-physchem-040513-103659} {https://doi.org/10.1146/annurev-physchem-040513-103659} \BibitemShut {NoStop}%
\bibitem [{\citenamefont {Degen}\ \emph {et~al.}(2017)\citenamefont {Degen}, \citenamefont {Reinhard},\ and\ \citenamefont {Cappellaro}}]{DegenRMP2017}%
  \BibitemOpen
  \bibfield  {author} {\bibinfo {author} {\bibfnamefont {C.~L.}\ \bibnamefont {Degen}}, \bibinfo {author} {\bibfnamefont {F.}~\bibnamefont {Reinhard}},\ and\ \bibinfo {author} {\bibfnamefont {P.}~\bibnamefont {Cappellaro}},\ }\bibfield  {title} {\bibinfo {title} {Quantum sensing},\ }\href {https://doi.org/10.1103/RevModPhys.89.035002} {\bibfield  {journal} {\bibinfo  {journal} {Rev. Mod. Phys.}\ }\textbf {\bibinfo {volume} {89}},\ \bibinfo {pages} {035002} (\bibinfo {year} {2017})}\BibitemShut {NoStop}%
\bibitem [{\citenamefont {Balasubramanian}\ \emph {et~al.}(2009)\citenamefont {Balasubramanian}, \citenamefont {Neumann}, \citenamefont {Twitchen}, \citenamefont {Markham}, \citenamefont {Kolesov}, \citenamefont {Mizuochi}, \citenamefont {Isoya}, \citenamefont {Achard}, \citenamefont {Beck}, \citenamefont {Tissler}, \citenamefont {Jacques}, \citenamefont {Hemmer}, \citenamefont {Jelezko},\ and\ \citenamefont {Wrachtrup}}]{BalasubramanianNatMat2009}%
  \BibitemOpen
  \bibfield  {author} {\bibinfo {author} {\bibfnamefont {G.}~\bibnamefont {Balasubramanian}}, \bibinfo {author} {\bibfnamefont {P.}~\bibnamefont {Neumann}}, \bibinfo {author} {\bibfnamefont {D.}~\bibnamefont {Twitchen}}, \bibinfo {author} {\bibfnamefont {M.}~\bibnamefont {Markham}}, \bibinfo {author} {\bibfnamefont {R.}~\bibnamefont {Kolesov}}, \bibinfo {author} {\bibfnamefont {N.}~\bibnamefont {Mizuochi}}, \bibinfo {author} {\bibfnamefont {J.}~\bibnamefont {Isoya}}, \bibinfo {author} {\bibfnamefont {J.}~\bibnamefont {Achard}}, \bibinfo {author} {\bibfnamefont {J.}~\bibnamefont {Beck}}, \bibinfo {author} {\bibfnamefont {J.}~\bibnamefont {Tissler}}, \bibinfo {author} {\bibfnamefont {V.}~\bibnamefont {Jacques}}, \bibinfo {author} {\bibfnamefont {P.~R.}\ \bibnamefont {Hemmer}}, \bibinfo {author} {\bibfnamefont {F.}~\bibnamefont {Jelezko}},\ and\ \bibinfo {author} {\bibfnamefont {J.}~\bibnamefont {Wrachtrup}},\ }\bibfield  {title} {\bibinfo {title} {Ultralong spin coherence time in isotopically engineered
  diamond},\ }\href {https://doi.org/10.1038/nmat2420} {\bibfield  {journal} {\bibinfo  {journal} {Nature Materials 2009 8:5}\ }\textbf {\bibinfo {volume} {8}},\ \bibinfo {pages} {383} (\bibinfo {year} {2009})}\BibitemShut {NoStop}%
\bibitem [{\citenamefont {{De Lange}}\ \emph {et~al.}(2010)\citenamefont {{De Lange}}, \citenamefont {Wang}, \citenamefont {Rist{\`{e}}}, \citenamefont {Dobrovitski},\ and\ \citenamefont {Hanson}}]{deLangeScience2010}%
  \BibitemOpen
  \bibfield  {author} {\bibinfo {author} {\bibfnamefont {G.}~\bibnamefont {{De Lange}}}, \bibinfo {author} {\bibfnamefont {Z.~H.}\ \bibnamefont {Wang}}, \bibinfo {author} {\bibfnamefont {D.}~\bibnamefont {Rist{\`{e}}}}, \bibinfo {author} {\bibfnamefont {V.~V.}\ \bibnamefont {Dobrovitski}},\ and\ \bibinfo {author} {\bibfnamefont {R.}~\bibnamefont {Hanson}},\ }\bibfield  {title} {\bibinfo {title} {{Universal dynamical decoupling of a single solid-state spin from a spin bath}},\ }\href {https://doi.org/10.1126/science.1192739} {\bibfield  {journal} {\bibinfo  {journal} {Science}\ }\textbf {\bibinfo {volume} {330}},\ \bibinfo {pages} {60} (\bibinfo {year} {2010})}\BibitemShut {NoStop}%
\bibitem [{\citenamefont {Naydenov}\ \emph {et~al.}(2011)\citenamefont {Naydenov}, \citenamefont {Dolde}, \citenamefont {Hall}, \citenamefont {Shin}, \citenamefont {Fedder}, \citenamefont {Hollenberg}, \citenamefont {Jelezko},\ and\ \citenamefont {Wrachtrup}}]{NaydenovPRB2011}%
  \BibitemOpen
  \bibfield  {author} {\bibinfo {author} {\bibfnamefont {B.}~\bibnamefont {Naydenov}}, \bibinfo {author} {\bibfnamefont {F.}~\bibnamefont {Dolde}}, \bibinfo {author} {\bibfnamefont {L.~T.}\ \bibnamefont {Hall}}, \bibinfo {author} {\bibfnamefont {C.}~\bibnamefont {Shin}}, \bibinfo {author} {\bibfnamefont {H.}~\bibnamefont {Fedder}}, \bibinfo {author} {\bibfnamefont {L.~C.~L.}\ \bibnamefont {Hollenberg}}, \bibinfo {author} {\bibfnamefont {F.}~\bibnamefont {Jelezko}},\ and\ \bibinfo {author} {\bibfnamefont {J.}~\bibnamefont {Wrachtrup}},\ }\bibfield  {title} {\bibinfo {title} {Dynamical decoupling of a single-electron spin at room temperature},\ }\href {https://doi.org/10.1103/PhysRevB.83.081201} {\bibfield  {journal} {\bibinfo  {journal} {Phys. Rev. B}\ }\textbf {\bibinfo {volume} {83}},\ \bibinfo {pages} {081201(R)} (\bibinfo {year} {2011})}\BibitemShut {NoStop}%
\bibitem [{\citenamefont {Sriarunothai}\ \emph {et~al.}(2018)\citenamefont {Sriarunothai}, \citenamefont {W{\"o}lk}, \citenamefont {Giri}, \citenamefont {Friis}, \citenamefont {Dunjko}, \citenamefont {Briegel},\ and\ \citenamefont {Wunderlich}}]{SriarunothaiQST2018}%
  \BibitemOpen
  \bibfield  {author} {\bibinfo {author} {\bibfnamefont {T.}~\bibnamefont {Sriarunothai}}, \bibinfo {author} {\bibfnamefont {S.}~\bibnamefont {W{\"o}lk}}, \bibinfo {author} {\bibfnamefont {G.~S.}\ \bibnamefont {Giri}}, \bibinfo {author} {\bibfnamefont {N.}~\bibnamefont {Friis}}, \bibinfo {author} {\bibfnamefont {V.}~\bibnamefont {Dunjko}}, \bibinfo {author} {\bibfnamefont {H.~J.}\ \bibnamefont {Briegel}},\ and\ \bibinfo {author} {\bibfnamefont {C.}~\bibnamefont {Wunderlich}},\ }\bibfield  {title} {\bibinfo {title} {Speeding-up the decision making of a learning agent using an ion trap quantum processor},\ }\href {https://doi.org/10.1088/2058-9565/aaef5e} {\bibfield  {journal} {\bibinfo  {journal} {Quantum Science and Technology}\ }\textbf {\bibinfo {volume} {4}},\ \bibinfo {pages} {015014} (\bibinfo {year} {2018})}\BibitemShut {NoStop}%
\bibitem [{\citenamefont {Genov}\ \emph {et~al.}(2020)\citenamefont {Genov}, \citenamefont {Ben-Shalom}, \citenamefont {Jelezko}, \citenamefont {Retzker},\ and\ \citenamefont {Bar-Gill}}]{GenovPRR2020}%
  \BibitemOpen
  \bibfield  {author} {\bibinfo {author} {\bibfnamefont {G.~T.}\ \bibnamefont {Genov}}, \bibinfo {author} {\bibfnamefont {Y.}~\bibnamefont {Ben-Shalom}}, \bibinfo {author} {\bibfnamefont {F.}~\bibnamefont {Jelezko}}, \bibinfo {author} {\bibfnamefont {A.}~\bibnamefont {Retzker}},\ and\ \bibinfo {author} {\bibfnamefont {N.}~\bibnamefont {Bar-Gill}},\ }\bibfield  {title} {\bibinfo {title} {Efficient and robust signal sensing by sequences of adiabatic chirped pulses},\ }\href {https://doi.org/10.1103/PhysRevResearch.2.033216} {\bibfield  {journal} {\bibinfo  {journal} {Phys. Rev. Res.}\ }\textbf {\bibinfo {volume} {2}},\ \bibinfo {pages} {033216} (\bibinfo {year} {2020})}\BibitemShut {NoStop}%
\bibitem [{\citenamefont {Stark}\ \emph {et~al.}(2018)\citenamefont {Stark}, \citenamefont {Aharon}, \citenamefont {Huck}, \citenamefont {El-Ella}, \citenamefont {Retzker}, \citenamefont {Jelezko},\ and\ \citenamefont {Andersen}}]{StarkSciRep2018}%
  \BibitemOpen
  \bibfield  {author} {\bibinfo {author} {\bibfnamefont {A.}~\bibnamefont {Stark}}, \bibinfo {author} {\bibfnamefont {N.}~\bibnamefont {Aharon}}, \bibinfo {author} {\bibfnamefont {A.}~\bibnamefont {Huck}}, \bibinfo {author} {\bibfnamefont {H.~A.~R.}\ \bibnamefont {El-Ella}}, \bibinfo {author} {\bibfnamefont {A.}~\bibnamefont {Retzker}}, \bibinfo {author} {\bibfnamefont {F.}~\bibnamefont {Jelezko}},\ and\ \bibinfo {author} {\bibfnamefont {U.~L.}\ \bibnamefont {Andersen}},\ }\bibfield  {title} {\bibinfo {title} {Clock transition by continuous dynamical decoupling of a three-level system open},\ }\href {https://doi.org/10.1038/s41598-018-31984-4} {\bibfield  {journal} {\bibinfo  {journal} {Scientific Reports}\ }\textbf {\bibinfo {volume} {8}},\ \bibinfo {pages} {14807} (\bibinfo {year} {2018})}\BibitemShut {NoStop}%
\bibitem [{\citenamefont {Stark}\ \emph {et~al.}(2017)\citenamefont {Stark}, \citenamefont {Aharon}, \citenamefont {Unden}, \citenamefont {Louzon}, \citenamefont {Huck}, \citenamefont {Retzker}, \citenamefont {Andersen},\ and\ \citenamefont {Jelezko}}]{StarkNatComm2017}%
  \BibitemOpen
  \bibfield  {author} {\bibinfo {author} {\bibfnamefont {A.}~\bibnamefont {Stark}}, \bibinfo {author} {\bibfnamefont {N.}~\bibnamefont {Aharon}}, \bibinfo {author} {\bibfnamefont {T.}~\bibnamefont {Unden}}, \bibinfo {author} {\bibfnamefont {D.}~\bibnamefont {Louzon}}, \bibinfo {author} {\bibfnamefont {A.}~\bibnamefont {Huck}}, \bibinfo {author} {\bibfnamefont {A.}~\bibnamefont {Retzker}}, \bibinfo {author} {\bibfnamefont {U.~L.}\ \bibnamefont {Andersen}},\ and\ \bibinfo {author} {\bibfnamefont {F.}~\bibnamefont {Jelezko}},\ }\bibfield  {title} {\bibinfo {title} {{Narrow-bandwidth sensing of high-frequency fields with continuous dynamical decoupling}},\ }\href {https://doi.org/10.1038/s41467-017-01159-2} {\bibfield  {journal} {\bibinfo  {journal} {Nature Communications}\ }\textbf {\bibinfo {volume} {8}},\ \bibinfo {pages} {1} (\bibinfo {year} {2017})}\BibitemShut {NoStop}%
\bibitem [{\citenamefont {Knowles}\ \emph {et~al.}(2014)\citenamefont {Knowles}, \citenamefont {Kara},\ and\ \citenamefont {Atat{\"u}re}}]{KnowlesNatMat2014}%
  \BibitemOpen
  \bibfield  {author} {\bibinfo {author} {\bibfnamefont {H.~S.}\ \bibnamefont {Knowles}}, \bibinfo {author} {\bibfnamefont {D.~M.}\ \bibnamefont {Kara}},\ and\ \bibinfo {author} {\bibfnamefont {M.}~\bibnamefont {Atat{\"u}re}},\ }\bibfield  {title} {\bibinfo {title} {Observing bulk diamond spin coherence in high-purity nanodiamonds},\ }\bibfield  {journal} {\bibinfo  {journal} {Nature Materials}\ }\href {https://doi.org/10.1038/NMAT3805} {10.1038/NMAT3805} (\bibinfo {year} {2014})\BibitemShut {NoStop}%
\bibitem [{\citenamefont {Mcguinness}\ \emph {et~al.}(2011)\citenamefont {Mcguinness}, \citenamefont {Yan}, \citenamefont {Stacey}, \citenamefont {Simpson}, \citenamefont {Hall}, \citenamefont {Maclaurin}, \citenamefont {Prawer}, \citenamefont {Mulvaney}, \citenamefont {Wrachtrup}, \citenamefont {Caruso}, \citenamefont {Scholten},\ and\ \citenamefont {Hollenberg}}]{McguinnessNatNano2011}%
  \BibitemOpen
  \bibfield  {author} {\bibinfo {author} {\bibfnamefont {L.~P.}\ \bibnamefont {Mcguinness}}, \bibinfo {author} {\bibfnamefont {Y.}~\bibnamefont {Yan}}, \bibinfo {author} {\bibfnamefont {A.}~\bibnamefont {Stacey}}, \bibinfo {author} {\bibfnamefont {D.~A.}\ \bibnamefont {Simpson}}, \bibinfo {author} {\bibfnamefont {L.~T.}\ \bibnamefont {Hall}}, \bibinfo {author} {\bibfnamefont {D.}~\bibnamefont {Maclaurin}}, \bibinfo {author} {\bibfnamefont {S.}~\bibnamefont {Prawer}}, \bibinfo {author} {\bibfnamefont {P.}~\bibnamefont {Mulvaney}}, \bibinfo {author} {\bibfnamefont {J.}~\bibnamefont {Wrachtrup}}, \bibinfo {author} {\bibfnamefont {F.}~\bibnamefont {Caruso}}, \bibinfo {author} {\bibfnamefont {R.~E.}\ \bibnamefont {Scholten}},\ and\ \bibinfo {author} {\bibfnamefont {L.~C.~L.}\ \bibnamefont {Hollenberg}},\ }\bibfield  {title} {\bibinfo {title} {Quantum measurement and orientation tracking of fluorescent nanodiamonds inside living cells},\ }\bibfield  {journal} {\bibinfo  {journal} {Nature Nanotechnology}\ }\textbf
  {\bibinfo {volume} {6}},\ \href {https://doi.org/10.1038/NNANO.2011.64} {10.1038/NNANO.2011.64} (\bibinfo {year} {2011})\BibitemShut {NoStop}%
\bibitem [{\citenamefont {Solomon}(1959)}]{SolomonPRL1959}%
  \BibitemOpen
  \bibfield  {author} {\bibinfo {author} {\bibfnamefont {I.}~\bibnamefont {Solomon}},\ }\bibfield  {title} {\bibinfo {title} {Rotary spin echoes},\ }\href {https://doi.org/10.1103/PhysRevLett.2.301} {\bibfield  {journal} {\bibinfo  {journal} {Phys. Rev. Lett.}\ }\textbf {\bibinfo {volume} {2}},\ \bibinfo {pages} {301} (\bibinfo {year} {1959})}\BibitemShut {NoStop}%
\bibitem [{\citenamefont {Sage}\ \emph {et~al.}(2013{\natexlab{a}})\citenamefont {Sage}, \citenamefont {Arai}, \citenamefont {Glenn}, \citenamefont {DeVience}, \citenamefont {Pham}, \citenamefont {Rahn-Lee}, \citenamefont {Lukin}, \citenamefont {Yacoby}, \citenamefont {Komeili},\ and\ \citenamefont {Walsworth}}]{WalsworthNature2013}%
  \BibitemOpen
  \bibfield  {author} {\bibinfo {author} {\bibfnamefont {D.~L.}\ \bibnamefont {Sage}}, \bibinfo {author} {\bibfnamefont {K.}~\bibnamefont {Arai}}, \bibinfo {author} {\bibfnamefont {D.~R.}\ \bibnamefont {Glenn}}, \bibinfo {author} {\bibfnamefont {S.~J.}\ \bibnamefont {DeVience}}, \bibinfo {author} {\bibfnamefont {L.~M.}\ \bibnamefont {Pham}}, \bibinfo {author} {\bibfnamefont {L.}~\bibnamefont {Rahn-Lee}}, \bibinfo {author} {\bibfnamefont {M.~D.}\ \bibnamefont {Lukin}}, \bibinfo {author} {\bibfnamefont {A.}~\bibnamefont {Yacoby}}, \bibinfo {author} {\bibfnamefont {A.}~\bibnamefont {Komeili}},\ and\ \bibinfo {author} {\bibfnamefont {R.~L.}\ \bibnamefont {Walsworth}},\ }\bibfield  {title} {\bibinfo {title} {Optical magnetic imaging of living cells},\ }\bibfield  {journal} {\bibinfo  {journal} {Nature}\ }\href {https://doi.org/10.1038/nature12072} {10.1038/nature12072} (\bibinfo {year} {2013}{\natexlab{a}})\BibitemShut {NoStop}%
\bibitem [{\citenamefont {Kucsko}\ \emph {et~al.}(2013)\citenamefont {Kucsko}, \citenamefont {Maurer}, \citenamefont {Yao}, \citenamefont {Kubo}, \citenamefont {Noh}, \citenamefont {Lo}, \citenamefont {Park},\ and\ \citenamefont {Lukin}}]{KucskoNature2013}%
  \BibitemOpen
  \bibfield  {author} {\bibinfo {author} {\bibfnamefont {G.}~\bibnamefont {Kucsko}}, \bibinfo {author} {\bibfnamefont {P.~C.}\ \bibnamefont {Maurer}}, \bibinfo {author} {\bibfnamefont {N.~Y.}\ \bibnamefont {Yao}}, \bibinfo {author} {\bibfnamefont {M.}~\bibnamefont {Kubo}}, \bibinfo {author} {\bibfnamefont {H.~J.}\ \bibnamefont {Noh}}, \bibinfo {author} {\bibfnamefont {P.~K.}\ \bibnamefont {Lo}}, \bibinfo {author} {\bibfnamefont {H.}~\bibnamefont {Park}},\ and\ \bibinfo {author} {\bibfnamefont {M.~D.}\ \bibnamefont {Lukin}},\ }\bibfield  {title} {\bibinfo {title} {Nanometre-scale thermometry in a living cell},\ }\bibfield  {journal} {\bibinfo  {journal} {Nature}\ }\href {https://doi.org/10.1038/nature12373} {10.1038/nature12373} (\bibinfo {year} {2013})\BibitemShut {NoStop}%
\bibitem [{\citenamefont {Balasubramanian}\ \emph {et~al.}(2014)\citenamefont {Balasubramanian}, \citenamefont {Lazariev}, \citenamefont {Arumugam},\ and\ \citenamefont {Duan}}]{BalasubramanianOpinBio2014}%
  \BibitemOpen
  \bibfield  {author} {\bibinfo {author} {\bibfnamefont {G.}~\bibnamefont {Balasubramanian}}, \bibinfo {author} {\bibfnamefont {A.}~\bibnamefont {Lazariev}}, \bibinfo {author} {\bibfnamefont {R.}~\bibnamefont {Arumugam}},\ and\ \bibinfo {author} {\bibfnamefont {D.-W.}\ \bibnamefont {Duan}},\ }\bibfield  {title} {\bibinfo {title} {Nitrogen-vacancy color center in diamond-emerging nanoscale applications in bioimaging and biosensing},\ }\href {https://doi.org/10.1016/j.cbpa.2014.04.014} {\bibfield  {journal} {\bibinfo  {journal} {Current Opinion in Chemical Biology}\ }\textbf {\bibinfo {volume} {20}},\ \bibinfo {pages} {69} (\bibinfo {year} {2014})}\BibitemShut {NoStop}%
\bibitem [{\citenamefont {Hirose}\ \emph {et~al.}(2012)\citenamefont {Hirose}, \citenamefont {Aiello},\ and\ \citenamefont {Cappellaro}}]{HirosePRA2012}%
  \BibitemOpen
  \bibfield  {author} {\bibinfo {author} {\bibfnamefont {M.}~\bibnamefont {Hirose}}, \bibinfo {author} {\bibfnamefont {C.~D.}\ \bibnamefont {Aiello}},\ and\ \bibinfo {author} {\bibfnamefont {P.}~\bibnamefont {Cappellaro}},\ }\bibfield  {title} {\bibinfo {title} {Continuous dynamical decoupling magnetometry},\ }\href {https://doi.org/10.1103/PhysRevA.86.062320} {\bibfield  {journal} {\bibinfo  {journal} {Phys. Rev. A}\ }\textbf {\bibinfo {volume} {86}},\ \bibinfo {pages} {062320} (\bibinfo {year} {2012})}\BibitemShut {NoStop}%
\bibitem [{\citenamefont {Aiello}\ \emph {et~al.}(2013)\citenamefont {Aiello}, \citenamefont {Hirose},\ and\ \citenamefont {Cappellaro}}]{AielloNatComm2013}%
  \BibitemOpen
  \bibfield  {author} {\bibinfo {author} {\bibfnamefont {C.~D.}\ \bibnamefont {Aiello}}, \bibinfo {author} {\bibfnamefont {M.}~\bibnamefont {Hirose}},\ and\ \bibinfo {author} {\bibfnamefont {P.}~\bibnamefont {Cappellaro}},\ }\bibfield  {title} {\bibinfo {title} {Article composite-pulse magnetometry with a solid-state quantum sensor},\ }\bibfield  {journal} {\bibinfo  {journal} {Nature Communications}\ }\href {https://doi.org/10.1038/ncomms2375} {10.1038/ncomms2375} (\bibinfo {year} {2013})\BibitemShut {NoStop}%
\bibitem [{\citenamefont {Baumgart}\ \emph {et~al.}(2016)\citenamefont {Baumgart}, \citenamefont {Cai}, \citenamefont {Retzker}, \citenamefont {Plenio},\ and\ \citenamefont {Wunderlich}}]{Baumgart2016PRL}%
  \BibitemOpen
  \bibfield  {author} {\bibinfo {author} {\bibfnamefont {I.}~\bibnamefont {Baumgart}}, \bibinfo {author} {\bibfnamefont {J.-M.}\ \bibnamefont {Cai}}, \bibinfo {author} {\bibfnamefont {A.}~\bibnamefont {Retzker}}, \bibinfo {author} {\bibfnamefont {M.~B.}\ \bibnamefont {Plenio}},\ and\ \bibinfo {author} {\bibfnamefont {C.}~\bibnamefont {Wunderlich}},\ }\bibfield  {title} {\bibinfo {title} {Ultrasensitive magnetometer using a single atom},\ }\href {https://doi.org/10.1103/PhysRevLett.116.240801} {\bibfield  {journal} {\bibinfo  {journal} {Phys. Rev. Lett.}\ }\textbf {\bibinfo {volume} {116}},\ \bibinfo {pages} {240801} (\bibinfo {year} {2016})}\BibitemShut {NoStop}%
\bibitem [{\citenamefont {Cai}\ \emph {et~al.}(2012)\citenamefont {Cai}, \citenamefont {Naydenov}, \citenamefont {Pfeiffer}, \citenamefont {McGuinness}, \citenamefont {Jahnke}, \citenamefont {Jelezko}, \citenamefont {Plenio},\ and\ \citenamefont {Retzker}}]{CaiNJP2012}%
  \BibitemOpen
  \bibfield  {author} {\bibinfo {author} {\bibfnamefont {J.-M.}\ \bibnamefont {Cai}}, \bibinfo {author} {\bibfnamefont {B.}~\bibnamefont {Naydenov}}, \bibinfo {author} {\bibfnamefont {R.}~\bibnamefont {Pfeiffer}}, \bibinfo {author} {\bibfnamefont {L.~P.}\ \bibnamefont {McGuinness}}, \bibinfo {author} {\bibfnamefont {K.~D.}\ \bibnamefont {Jahnke}}, \bibinfo {author} {\bibfnamefont {F.}~\bibnamefont {Jelezko}}, \bibinfo {author} {\bibfnamefont {M.~B.}\ \bibnamefont {Plenio}},\ and\ \bibinfo {author} {\bibfnamefont {A.}~\bibnamefont {Retzker}},\ }\bibfield  {title} {\bibinfo {title} {Robust dynamical decoupling with concatenated continuous driving},\ }\href {https://doi.org/10.1088/1367-2630/14/11/113023} {\bibfield  {journal} {\bibinfo  {journal} {New Journal of Physics}\ }\textbf {\bibinfo {volume} {14}},\ \bibinfo {pages} {113023} (\bibinfo {year} {2012})}\BibitemShut {NoStop}%
\bibitem [{\citenamefont {Cohen}\ \emph {et~al.}(2017)\citenamefont {Cohen}, \citenamefont {Aharon},\ and\ \citenamefont {Retzker}}]{CohenFP2017}%
  \BibitemOpen
  \bibfield  {author} {\bibinfo {author} {\bibfnamefont {I.}~\bibnamefont {Cohen}}, \bibinfo {author} {\bibfnamefont {N.}~\bibnamefont {Aharon}},\ and\ \bibinfo {author} {\bibfnamefont {A.}~\bibnamefont {Retzker}},\ }\bibfield  {title} {\bibinfo {title} {Continuous dynamical decoupling utilizing time-dependent detuning},\ }\href {https://doi.org/https://doi.org/10.1002/prop.201600071} {\bibfield  {journal} {\bibinfo  {journal} {Fortschritte der Physik}\ }\textbf {\bibinfo {volume} {65}},\ \bibinfo {pages} {1600071} (\bibinfo {year} {2017})}\BibitemShut {NoStop}%
\bibitem [{\citenamefont {Genov}\ \emph {et~al.}(2019)\citenamefont {Genov}, \citenamefont {Aharon}, \citenamefont {Jelezko},\ and\ \citenamefont {Retzker}}]{Genov2019MDD}%
  \BibitemOpen
  \bibfield  {author} {\bibinfo {author} {\bibfnamefont {G.~T.}\ \bibnamefont {Genov}}, \bibinfo {author} {\bibfnamefont {N.}~\bibnamefont {Aharon}}, \bibinfo {author} {\bibfnamefont {F.}~\bibnamefont {Jelezko}},\ and\ \bibinfo {author} {\bibfnamefont {A.}~\bibnamefont {Retzker}},\ }\bibfield  {title} {\bibinfo {title} {Mixed dynamical decoupling},\ }\href {https://doi.org/10.1088/2058-9565/ab2afd} {\bibfield  {journal} {\bibinfo  {journal} {Quantum Science and Technology}\ }\textbf {\bibinfo {volume} {4}},\ \bibinfo {pages} {035010} (\bibinfo {year} {2019})}\BibitemShut {NoStop}%
\bibitem [{\citenamefont {Salhov}\ \emph {et~al.}(2024)\citenamefont {Salhov}, \citenamefont {Cao}, \citenamefont {Cai}, \citenamefont {Retzker}, \citenamefont {Jelezko},\ and\ \citenamefont {Genov}}]{Salhov2024prl}%
  \BibitemOpen
  \bibfield  {author} {\bibinfo {author} {\bibfnamefont {A.}~\bibnamefont {Salhov}}, \bibinfo {author} {\bibfnamefont {Q.}~\bibnamefont {Cao}}, \bibinfo {author} {\bibfnamefont {J.}~\bibnamefont {Cai}}, \bibinfo {author} {\bibfnamefont {A.}~\bibnamefont {Retzker}}, \bibinfo {author} {\bibfnamefont {F.}~\bibnamefont {Jelezko}},\ and\ \bibinfo {author} {\bibfnamefont {G.}~\bibnamefont {Genov}},\ }\bibfield  {title} {\bibinfo {title} {Protecting quantum information via destructive interference of correlated noise},\ }\href {https://doi.org/10.1103/PhysRevLett.132.223601} {\bibfield  {journal} {\bibinfo  {journal} {Phys. Rev. Lett.}\ }\textbf {\bibinfo {volume} {132}},\ \bibinfo {pages} {223601} (\bibinfo {year} {2024})}\BibitemShut {NoStop}%
\bibitem [{\citenamefont {Timoney}\ \emph {et~al.}(2011)\citenamefont {Timoney}, \citenamefont {Baumgart}, \citenamefont {Johanning}, \citenamefont {Var{\`o}n}, \citenamefont {Plenio}, \citenamefont {Retzker},\ and\ \citenamefont {Wunderlich}}]{TimoneyNature2011}%
  \BibitemOpen
  \bibfield  {author} {\bibinfo {author} {\bibfnamefont {N.}~\bibnamefont {Timoney}}, \bibinfo {author} {\bibfnamefont {I.}~\bibnamefont {Baumgart}}, \bibinfo {author} {\bibfnamefont {M.}~\bibnamefont {Johanning}}, \bibinfo {author} {\bibfnamefont {A.~F.}\ \bibnamefont {Var{\`o}n}}, \bibinfo {author} {\bibfnamefont {M.~B.}\ \bibnamefont {Plenio}}, \bibinfo {author} {\bibfnamefont {A.}~\bibnamefont {Retzker}},\ and\ \bibinfo {author} {\bibfnamefont {C.}~\bibnamefont {Wunderlich}},\ }\bibfield  {title} {\bibinfo {title} {Quantum gates and memory using microwave-dressed states},\ }\bibfield  {journal} {\bibinfo  {journal} {Nature}\ }\href {https://doi.org/10.1038/nature10319} {10.1038/nature10319} (\bibinfo {year} {2011})\BibitemShut {NoStop}%
\bibitem [{\citenamefont {Aharon}\ \emph {et~al.}(2013)\citenamefont {Aharon}, \citenamefont {Drewsen},\ and\ \citenamefont {Retzker}}]{AharonPRL2013}%
  \BibitemOpen
  \bibfield  {author} {\bibinfo {author} {\bibfnamefont {N.}~\bibnamefont {Aharon}}, \bibinfo {author} {\bibfnamefont {M.}~\bibnamefont {Drewsen}},\ and\ \bibinfo {author} {\bibfnamefont {A.}~\bibnamefont {Retzker}},\ }\bibfield  {title} {\bibinfo {title} {General scheme for the construction of a protected qubit subspace},\ }\href {https://doi.org/10.1103/PhysRevLett.111.230507} {\bibfield  {journal} {\bibinfo  {journal} {Phys. Rev. Lett.}\ }\textbf {\bibinfo {volume} {111}},\ \bibinfo {pages} {230507} (\bibinfo {year} {2013})}\BibitemShut {NoStop}%
\bibitem [{\citenamefont {Aharon}\ \emph {et~al.}(2016)\citenamefont {Aharon}, \citenamefont {Cohen}, \citenamefont {Jelezko},\ and\ \citenamefont {Retzker}}]{AharonNJP2016}%
  \BibitemOpen
  \bibfield  {author} {\bibinfo {author} {\bibfnamefont {N.}~\bibnamefont {Aharon}}, \bibinfo {author} {\bibfnamefont {I.}~\bibnamefont {Cohen}}, \bibinfo {author} {\bibfnamefont {F.}~\bibnamefont {Jelezko}},\ and\ \bibinfo {author} {\bibfnamefont {A.}~\bibnamefont {Retzker}},\ }\bibfield  {title} {\bibinfo {title} {Fully robust qubit in atomic and molecular three-level systems},\ }\href {https://doi.org/10.1088/1367-2630/aa4fd3} {\bibfield  {journal} {\bibinfo  {journal} {New Journal of Physics}\ }\textbf {\bibinfo {volume} {18}},\ \bibinfo {pages} {123012} (\bibinfo {year} {2016})}\BibitemShut {NoStop}%
\bibitem [{\citenamefont {Barfuss}\ \emph {et~al.}(2018)\citenamefont {Barfuss}, \citenamefont {K{\"o}lbl}, \citenamefont {Thiel}, \citenamefont {Teissier}, \citenamefont {Kasperczyk},\ and\ \citenamefont {Maletinsky}}]{BarfussNPhys2018}%
  \BibitemOpen
  \bibfield  {author} {\bibinfo {author} {\bibfnamefont {A.}~\bibnamefont {Barfuss}}, \bibinfo {author} {\bibfnamefont {J.}~\bibnamefont {K{\"o}lbl}}, \bibinfo {author} {\bibfnamefont {L.}~\bibnamefont {Thiel}}, \bibinfo {author} {\bibfnamefont {J.}~\bibnamefont {Teissier}}, \bibinfo {author} {\bibfnamefont {M.}~\bibnamefont {Kasperczyk}},\ and\ \bibinfo {author} {\bibfnamefont {P.}~\bibnamefont {Maletinsky}},\ }\bibfield  {title} {\bibinfo {title} {Phase-controlled coherent dynamics of a single spin under closed-contour interaction},\ }\bibfield  {journal} {\bibinfo  {journal} {Nature Physics}\ }\href {https://doi.org/10.1038/s41567-018-0231-8} {10.1038/s41567-018-0231-8} (\bibinfo {year} {2018})\BibitemShut {NoStop}%
\bibitem [{\citenamefont {Burum}\ \emph {et~al.}(1981)\citenamefont {Burum}, \citenamefont {Linder},\ and\ \citenamefont {Ernst}}]{BurumJMR1981}%
  \BibitemOpen
  \bibfield  {author} {\bibinfo {author} {\bibfnamefont {D.}~\bibnamefont {Burum}}, \bibinfo {author} {\bibfnamefont {M.}~\bibnamefont {Linder}},\ and\ \bibinfo {author} {\bibfnamefont {R.}~\bibnamefont {Ernst}},\ }\bibfield  {title} {\bibinfo {title} {Low-power multipulse line narrowing in solid-state nmr},\ }\href {https://doi.org/https://doi.org/10.1016/0022-2364(81)90200-6} {\bibfield  {journal} {\bibinfo  {journal} {Journal of Magnetic Resonance (1969)}\ }\textbf {\bibinfo {volume} {44}},\ \bibinfo {pages} {173} (\bibinfo {year} {1981})}\BibitemShut {NoStop}%
\bibitem [{\citenamefont {Nanz}\ \emph {et~al.}(1995)\citenamefont {Nanz}, \citenamefont {Ernst}, \citenamefont {Hong}, \citenamefont {Ziegeweid}, \citenamefont {Schmidtrohr},\ and\ \citenamefont {Pines}}]{NanzJMR1995}%
  \BibitemOpen
  \bibfield  {author} {\bibinfo {author} {\bibfnamefont {D.}~\bibnamefont {Nanz}}, \bibinfo {author} {\bibfnamefont {M.}~\bibnamefont {Ernst}}, \bibinfo {author} {\bibfnamefont {M.}~\bibnamefont {Hong}}, \bibinfo {author} {\bibfnamefont {M.}~\bibnamefont {Ziegeweid}}, \bibinfo {author} {\bibfnamefont {K.}~\bibnamefont {Schmidtrohr}},\ and\ \bibinfo {author} {\bibfnamefont {A.}~\bibnamefont {Pines}},\ }\bibfield  {title} {\bibinfo {title} {Low-power decoupling sequences for high-resolution chemical-shift and local-field nmr spectra of liquid crystals},\ }\href {https://doi.org/https://doi.org/10.1006/jmra.1995.1077} {\bibfield  {journal} {\bibinfo  {journal} {Journal of Magnetic Resonance, Series A}\ }\textbf {\bibinfo {volume} {113}},\ \bibinfo {pages} {169} (\bibinfo {year} {1995})}\BibitemShut {NoStop}%
\bibitem [{\citenamefont {Shaka}\ \emph {et~al.}(1983)\citenamefont {Shaka}, \citenamefont {Keeler},\ and\ \citenamefont {Freeman}}]{ShakaJMR1983}%
  \BibitemOpen
  \bibfield  {author} {\bibinfo {author} {\bibfnamefont {A.}~\bibnamefont {Shaka}}, \bibinfo {author} {\bibfnamefont {J.}~\bibnamefont {Keeler}},\ and\ \bibinfo {author} {\bibfnamefont {R.}~\bibnamefont {Freeman}},\ }\bibfield  {title} {\bibinfo {title} {Evaluation of a new broadband decoupling sequence: Waltz-16},\ }\href {https://doi.org/https://doi.org/10.1016/0022-2364(83)90035-5} {\bibfield  {journal} {\bibinfo  {journal} {Journal of Magnetic Resonance (1969)}\ }\textbf {\bibinfo {volume} {53}},\ \bibinfo {pages} {313} (\bibinfo {year} {1983})}\BibitemShut {NoStop}%
\bibitem [{\citenamefont {Levitt}(1986)}]{Levitt1986}%
  \BibitemOpen
  \bibfield  {author} {\bibinfo {author} {\bibfnamefont {M.~H.}\ \bibnamefont {Levitt}},\ }\bibfield  {title} {\bibinfo {title} {{Composite pulses}},\ }\href {https://doi.org/10.1016/0079-6565(86)80005-X} {\bibfield  {journal} {\bibinfo  {journal} {Progress in Nuclear Magnetic Resonance Spectroscopy}\ }\textbf {\bibinfo {volume} {18}},\ \bibinfo {pages} {61} (\bibinfo {year} {1986})}\BibitemShut {NoStop}%
\bibitem [{\citenamefont {Genov}\ \emph {et~al.}(2014)\citenamefont {Genov}, \citenamefont {Schraft}, \citenamefont {Halfmann},\ and\ \citenamefont {Vitanov}}]{GenovPRL2014}%
  \BibitemOpen
  \bibfield  {author} {\bibinfo {author} {\bibfnamefont {G.~T.}\ \bibnamefont {Genov}}, \bibinfo {author} {\bibfnamefont {D.}~\bibnamefont {Schraft}}, \bibinfo {author} {\bibfnamefont {T.}~\bibnamefont {Halfmann}},\ and\ \bibinfo {author} {\bibfnamefont {N.~V.}\ \bibnamefont {Vitanov}},\ }\bibfield  {title} {\bibinfo {title} {{Correction of Arbitrary Field Errors in Population Inversion of Quantum Systems by Universal Composite Pulses}},\ }\href {https://doi.org/10.1103/PhysRevLett.113.043001} {\bibfield  {journal} {\bibinfo  {journal} {Physical Review Letters}\ }\textbf {\bibinfo {volume} {113}},\ \bibinfo {pages} {043001} (\bibinfo {year} {2014})}\BibitemShut {NoStop}%
\bibitem [{\citenamefont {Souza}\ \emph {et~al.}(2012)\citenamefont {Souza}, \citenamefont {{\'A}lvarez},\ and\ \citenamefont {Suter}}]{RDD_review12Suter}%
  \BibitemOpen
  \bibfield  {author} {\bibinfo {author} {\bibfnamefont {A.~M.}\ \bibnamefont {Souza}}, \bibinfo {author} {\bibfnamefont {G.~A.}\ \bibnamefont {{\'A}lvarez}},\ and\ \bibinfo {author} {\bibfnamefont {D.}~\bibnamefont {Suter}},\ }\bibfield  {title} {\bibinfo {title} {Robust dynamical decoupling},\ }\href {https://doi.org/10.1098/RSTA.2011.0355} {\bibfield  {journal} {\bibinfo  {journal} {Philosophical Transactions of the Royal Society A: Mathematical, Physical and Engineering Sciences}\ }\textbf {\bibinfo {volume} {370}},\ \bibinfo {pages} {4748} (\bibinfo {year} {2012})}\BibitemShut {NoStop}%
\bibitem [{\citenamefont {Torosov}\ \emph {et~al.}(2011)\citenamefont {Torosov}, \citenamefont {Gu{\'{e}}rin},\ and\ \citenamefont {Vitanov}}]{TorosovPRL2011}%
  \BibitemOpen
  \bibfield  {author} {\bibinfo {author} {\bibfnamefont {B.~T.}\ \bibnamefont {Torosov}}, \bibinfo {author} {\bibfnamefont {S.}~\bibnamefont {Gu{\'{e}}rin}},\ and\ \bibinfo {author} {\bibfnamefont {N.~V.}\ \bibnamefont {Vitanov}},\ }\bibfield  {title} {\bibinfo {title} {{High-Fidelity Adiabatic Passage by Composite Sequences of Chirped Pulses}},\ }\href {https://doi.org/10.1103/PhysRevLett.106.233001} {\bibfield  {journal} {\bibinfo  {journal} {Physical Review Letters}\ }\textbf {\bibinfo {volume} {106}},\ \bibinfo {pages} {233001} (\bibinfo {year} {2011})}\BibitemShut {NoStop}%
\bibitem [{\citenamefont {Genov}\ \emph {et~al.}(2017)\citenamefont {Genov}, \citenamefont {Schraft}, \citenamefont {Vitanov},\ and\ \citenamefont {Halfmann}}]{GenovPRL2017}%
  \BibitemOpen
  \bibfield  {author} {\bibinfo {author} {\bibfnamefont {G.~T.}\ \bibnamefont {Genov}}, \bibinfo {author} {\bibfnamefont {D.}~\bibnamefont {Schraft}}, \bibinfo {author} {\bibfnamefont {N.~V.}\ \bibnamefont {Vitanov}},\ and\ \bibinfo {author} {\bibfnamefont {T.}~\bibnamefont {Halfmann}},\ }\bibfield  {title} {\bibinfo {title} {{Arbitrarily Accurate Pulse Sequences for Robust Dynamical Decoupling}},\ }\href {https://doi.org/10.1103/PhysRevLett.118.133202} {\bibfield  {journal} {\bibinfo  {journal} {Physical Review Letters}\ }\textbf {\bibinfo {volume} {118}},\ \bibinfo {pages} {133202} (\bibinfo {year} {2017})}\BibitemShut {NoStop}%
\bibitem [{\citenamefont {Gullion}\ \emph {et~al.}(1990)\citenamefont {Gullion}, \citenamefont {Baker},\ and\ \citenamefont {Conradi}}]{Gullion1990jmr}%
  \BibitemOpen
  \bibfield  {author} {\bibinfo {author} {\bibfnamefont {T.}~\bibnamefont {Gullion}}, \bibinfo {author} {\bibfnamefont {D.~B.}\ \bibnamefont {Baker}},\ and\ \bibinfo {author} {\bibfnamefont {M.~S.}\ \bibnamefont {Conradi}},\ }\bibfield  {title} {\bibinfo {title} {New, compensated carr-purcell sequences},\ }\href {https://doi.org/https://doi.org/10.1016/0022-2364(90)90331-3} {\bibfield  {journal} {\bibinfo  {journal} {Journal of Magnetic Resonance (1969)}\ }\textbf {\bibinfo {volume} {89}},\ \bibinfo {pages} {479} (\bibinfo {year} {1990})}\BibitemShut {NoStop}%
\bibitem [{\citenamefont {Maudsley}(1986)}]{Maudsley1986jmr}%
  \BibitemOpen
  \bibfield  {author} {\bibinfo {author} {\bibfnamefont {A.}~\bibnamefont {Maudsley}},\ }\bibfield  {title} {\bibinfo {title} {Modified carr-purcell-meiboom-gill sequence for nmr fourier imaging applications},\ }\href {https://doi.org/https://doi.org/10.1016/0022-2364(86)90160-5} {\bibfield  {journal} {\bibinfo  {journal} {Journal of Magnetic Resonance (1969)}\ }\textbf {\bibinfo {volume} {69}},\ \bibinfo {pages} {488} (\bibinfo {year} {1986})}\BibitemShut {NoStop}%
\bibitem [{\citenamefont {Schmitt}\ \emph {et~al.}(2017)\citenamefont {Schmitt}, \citenamefont {Gefen}, \citenamefont {St{\"u}rner}, \citenamefont {Unden}, \citenamefont {Wolff}, \citenamefont {M{\"u}ller}, \citenamefont {Scheuer}, \citenamefont {Naydenov}, \citenamefont {Markham}, \citenamefont {Pezzagna}, \citenamefont {Meijer}, \citenamefont {Schwarz}, \citenamefont {Plenio}, \citenamefont {Retzker}, \citenamefont {McGuinness},\ and\ \citenamefont {Jelezko}}]{SchmittScience2017}%
  \BibitemOpen
  \bibfield  {author} {\bibinfo {author} {\bibfnamefont {S.}~\bibnamefont {Schmitt}}, \bibinfo {author} {\bibfnamefont {T.}~\bibnamefont {Gefen}}, \bibinfo {author} {\bibfnamefont {F.~M.}\ \bibnamefont {St{\"u}rner}}, \bibinfo {author} {\bibfnamefont {T.}~\bibnamefont {Unden}}, \bibinfo {author} {\bibfnamefont {G.}~\bibnamefont {Wolff}}, \bibinfo {author} {\bibfnamefont {C.}~\bibnamefont {M{\"u}ller}}, \bibinfo {author} {\bibfnamefont {J.}~\bibnamefont {Scheuer}}, \bibinfo {author} {\bibfnamefont {B.}~\bibnamefont {Naydenov}}, \bibinfo {author} {\bibfnamefont {M.}~\bibnamefont {Markham}}, \bibinfo {author} {\bibfnamefont {S.}~\bibnamefont {Pezzagna}}, \bibinfo {author} {\bibfnamefont {J.}~\bibnamefont {Meijer}}, \bibinfo {author} {\bibfnamefont {I.}~\bibnamefont {Schwarz}}, \bibinfo {author} {\bibfnamefont {M.}~\bibnamefont {Plenio}}, \bibinfo {author} {\bibfnamefont {A.}~\bibnamefont {Retzker}}, \bibinfo {author} {\bibfnamefont {L.~P.}\ \bibnamefont {McGuinness}},\ and\ \bibinfo {author} {\bibfnamefont
  {F.}~\bibnamefont {Jelezko}},\ }\bibfield  {title} {\bibinfo {title} {Submillihertz magnetic spectroscopy performed with a nanoscale quantum sensor},\ }\href {https://doi.org/10.1126/science.aam5532} {\bibfield  {journal} {\bibinfo  {journal} {Science}\ }\textbf {\bibinfo {volume} {356}},\ \bibinfo {pages} {832} (\bibinfo {year} {2017})}\BibitemShut {NoStop}%
\bibitem [{\citenamefont {Hartmann}\ and\ \citenamefont {Hahn}(1962)}]{Hartmann1962pr}%
  \BibitemOpen
  \bibfield  {author} {\bibinfo {author} {\bibfnamefont {S.~R.}\ \bibnamefont {Hartmann}}\ and\ \bibinfo {author} {\bibfnamefont {E.~L.}\ \bibnamefont {Hahn}},\ }\bibfield  {title} {\bibinfo {title} {Nuclear double resonance in the rotating frame},\ }\href {https://doi.org/10.1103/PhysRev.128.2042} {\bibfield  {journal} {\bibinfo  {journal} {Phys. Rev.}\ }\textbf {\bibinfo {volume} {128}},\ \bibinfo {pages} {2042} (\bibinfo {year} {1962})}\BibitemShut {NoStop}%
\bibitem [{\citenamefont {Sage}\ \emph {et~al.}(2013{\natexlab{b}})\citenamefont {Sage}, \citenamefont {Arai}, \citenamefont {Glenn}, \citenamefont {DeVience}, \citenamefont {Pham}, \citenamefont {Rahn-Lee}, \citenamefont {Lukin}, \citenamefont {Yacoby}, \citenamefont {Komeili},\ and\ \citenamefont {Walsworth}}]{LeSageNature2013}%
  \BibitemOpen
  \bibfield  {author} {\bibinfo {author} {\bibfnamefont {D.~L.}\ \bibnamefont {Sage}}, \bibinfo {author} {\bibfnamefont {K.}~\bibnamefont {Arai}}, \bibinfo {author} {\bibfnamefont {D.~R.}\ \bibnamefont {Glenn}}, \bibinfo {author} {\bibfnamefont {S.~J.}\ \bibnamefont {DeVience}}, \bibinfo {author} {\bibfnamefont {L.~M.}\ \bibnamefont {Pham}}, \bibinfo {author} {\bibfnamefont {L.}~\bibnamefont {Rahn-Lee}}, \bibinfo {author} {\bibfnamefont {M.~D.}\ \bibnamefont {Lukin}}, \bibinfo {author} {\bibfnamefont {A.}~\bibnamefont {Yacoby}}, \bibinfo {author} {\bibfnamefont {A.}~\bibnamefont {Komeili}},\ and\ \bibinfo {author} {\bibfnamefont {R.~L.}\ \bibnamefont {Walsworth}},\ }\bibfield  {title} {\bibinfo {title} {Optical magnetic imaging of living cells},\ }\bibfield  {journal} {\bibinfo  {journal} {Nature}\ }\href {https://doi.org/10.1038/nature12072} {10.1038/nature12072} (\bibinfo {year} {2013}{\natexlab{b}})\BibitemShut {NoStop}%
\bibitem [{Sup()}]{Supplemental}%
  \BibitemOpen
  \href@noop {} {\bibinfo {title} {Supplemental material at link, which includes refs. \cite{Bruns2018PRA,Findler2020SciRep,Binder2017qudi,Staudenmaier2023prl,Myers2016prl,Gefen2019,haberle2017nuclear,Taylor2008high} and more details on the theory of continuous phased dynamical decoupling; an analysis of the effect of pulse errors; slope and variance detection for quantum sensing; details on spurious harmonics and how to suppress them using randomized phases; a detailed description of the sample preparation and experimental setup; additional experimental results, including an analysis of the sensitivity and dynamic range.}}\BibitemShut {Stop}%
\bibitem [{\citenamefont {Bruns}\ \emph {et~al.}(2018)\citenamefont {Bruns}, \citenamefont {Genov}, \citenamefont {Hain}, \citenamefont {Vitanov},\ and\ \citenamefont {Halfmann}}]{Bruns2018PRA}%
  \BibitemOpen
  \bibfield  {author} {\bibinfo {author} {\bibfnamefont {A.}~\bibnamefont {Bruns}}, \bibinfo {author} {\bibfnamefont {G.~T.}\ \bibnamefont {Genov}}, \bibinfo {author} {\bibfnamefont {M.}~\bibnamefont {Hain}}, \bibinfo {author} {\bibfnamefont {N.~V.}\ \bibnamefont {Vitanov}},\ and\ \bibinfo {author} {\bibfnamefont {T.}~\bibnamefont {Halfmann}},\ }\bibfield  {title} {\bibinfo {title} {Experimental demonstration of composite stimulated raman adiabatic passage},\ }\href {https://doi.org/10.1103/PhysRevA.98.053413} {\bibfield  {journal} {\bibinfo  {journal} {Phys. Rev. A}\ }\textbf {\bibinfo {volume} {98}},\ \bibinfo {pages} {053413} (\bibinfo {year} {2018})}\BibitemShut {NoStop}%
\bibitem [{\citenamefont {Findler}\ \emph {et~al.}(2020)\citenamefont {Findler}, \citenamefont {Lang}, \citenamefont {Osterkamp}, \citenamefont {Nesládek},\ and\ \citenamefont {Jelezko}}]{Findler2020SciRep}%
  \BibitemOpen
  \bibfield  {author} {\bibinfo {author} {\bibfnamefont {C.}~\bibnamefont {Findler}}, \bibinfo {author} {\bibfnamefont {J.}~\bibnamefont {Lang}}, \bibinfo {author} {\bibfnamefont {C.}~\bibnamefont {Osterkamp}}, \bibinfo {author} {\bibfnamefont {M.}~\bibnamefont {Nesládek}},\ and\ \bibinfo {author} {\bibfnamefont {F.}~\bibnamefont {Jelezko}},\ }\bibfield  {title} {\bibinfo {title} {Indirect overgrowth as a synthesis route for superior diamond nano sensors},\ }\bibfield  {journal} {\bibinfo  {journal} {Scientific Reports}\ }\href {https://doi.org/10.1038/s41598-020-79943-2} {10.1038/s41598-020-79943-2} (\bibinfo {year} {2020})\BibitemShut {NoStop}%
\bibitem [{\citenamefont {Binder}\ \emph {et~al.}(2017)\citenamefont {Binder}, \citenamefont {Stark}, \citenamefont {Tomek}, \citenamefont {Scheuer}, \citenamefont {Frank}, \citenamefont {Jahnke}, \citenamefont {M{\"u}ller}, \citenamefont {Schmitt}, \citenamefont {Metsch}, \citenamefont {Unden}, \citenamefont {Gehring}, \citenamefont {Huck}, \citenamefont {Andersen}, \citenamefont {Rogers},\ and\ \citenamefont {Jelezko}}]{Binder2017qudi}%
  \BibitemOpen
  \bibfield  {author} {\bibinfo {author} {\bibfnamefont {J.~M.}\ \bibnamefont {Binder}}, \bibinfo {author} {\bibfnamefont {A.}~\bibnamefont {Stark}}, \bibinfo {author} {\bibfnamefont {N.}~\bibnamefont {Tomek}}, \bibinfo {author} {\bibfnamefont {J.}~\bibnamefont {Scheuer}}, \bibinfo {author} {\bibfnamefont {F.}~\bibnamefont {Frank}}, \bibinfo {author} {\bibfnamefont {K.~D.}\ \bibnamefont {Jahnke}}, \bibinfo {author} {\bibfnamefont {C.}~\bibnamefont {M{\"u}ller}}, \bibinfo {author} {\bibfnamefont {S.}~\bibnamefont {Schmitt}}, \bibinfo {author} {\bibfnamefont {M.~H.}\ \bibnamefont {Metsch}}, \bibinfo {author} {\bibfnamefont {T.}~\bibnamefont {Unden}}, \bibinfo {author} {\bibfnamefont {T.}~\bibnamefont {Gehring}}, \bibinfo {author} {\bibfnamefont {A.}~\bibnamefont {Huck}}, \bibinfo {author} {\bibfnamefont {U.~L.}\ \bibnamefont {Andersen}}, \bibinfo {author} {\bibfnamefont {L.~J.}\ \bibnamefont {Rogers}},\ and\ \bibinfo {author} {\bibfnamefont {F.}~\bibnamefont {Jelezko}},\ }\bibfield  {title} {\bibinfo {title}
  {Qudi: A modular python suite for experiment control and data processing},\ }\href {https://doi.org/10.1016/J.SOFTX.2017.02.001} {\bibfield  {journal} {\bibinfo  {journal} {SoftwareX}\ }\textbf {\bibinfo {volume} {6}},\ \bibinfo {pages} {85} (\bibinfo {year} {2017})}\BibitemShut {NoStop}%
\bibitem [{\citenamefont {Staudenmaier}\ \emph {et~al.}(2023)\citenamefont {Staudenmaier}, \citenamefont {Vijayakumar-Sreeja}, \citenamefont {Genov}, \citenamefont {Cohen}, \citenamefont {Findler}, \citenamefont {Lang}, \citenamefont {Retzker}, \citenamefont {Jelezko},\ and\ \citenamefont {Oviedo-Casado}}]{Staudenmaier2023prl}%
  \BibitemOpen
  \bibfield  {author} {\bibinfo {author} {\bibfnamefont {N.}~\bibnamefont {Staudenmaier}}, \bibinfo {author} {\bibfnamefont {A.}~\bibnamefont {Vijayakumar-Sreeja}}, \bibinfo {author} {\bibfnamefont {G.}~\bibnamefont {Genov}}, \bibinfo {author} {\bibfnamefont {D.}~\bibnamefont {Cohen}}, \bibinfo {author} {\bibfnamefont {C.}~\bibnamefont {Findler}}, \bibinfo {author} {\bibfnamefont {J.}~\bibnamefont {Lang}}, \bibinfo {author} {\bibfnamefont {A.}~\bibnamefont {Retzker}}, \bibinfo {author} {\bibfnamefont {F.}~\bibnamefont {Jelezko}},\ and\ \bibinfo {author} {\bibfnamefont {S.}~\bibnamefont {Oviedo-Casado}},\ }\bibfield  {title} {\bibinfo {title} {Optimal sensing protocol for statistically polarized nano-nmr with nv centers},\ }\href {https://doi.org/10.1103/PhysRevLett.131.150801} {\bibfield  {journal} {\bibinfo  {journal} {Phys. Rev. Lett.}\ }\textbf {\bibinfo {volume} {131}},\ \bibinfo {pages} {150801} (\bibinfo {year} {2023})}\BibitemShut {NoStop}%
\bibitem [{\citenamefont {Myers}\ \emph {et~al.}(2017)\citenamefont {Myers}, \citenamefont {Ariyaratne},\ and\ \citenamefont {Jayich}}]{Myers2016prl}%
  \BibitemOpen
  \bibfield  {author} {\bibinfo {author} {\bibfnamefont {B.~A.}\ \bibnamefont {Myers}}, \bibinfo {author} {\bibfnamefont {A.}~\bibnamefont {Ariyaratne}},\ and\ \bibinfo {author} {\bibfnamefont {A.~C.~B.}\ \bibnamefont {Jayich}},\ }\bibfield  {title} {\bibinfo {title} {Double-quantum spin-relaxation limits to coherence of near-surface nitrogen-vacancy centers},\ }\href {https://doi.org/10.1103/PhysRevLett.118.197201} {\bibfield  {journal} {\bibinfo  {journal} {Phys. Rev. Lett.}\ }\textbf {\bibinfo {volume} {118}},\ \bibinfo {pages} {197201} (\bibinfo {year} {2017})}\BibitemShut {NoStop}%
\bibitem [{\citenamefont {Gefen}\ \emph {et~al.}(2019)\citenamefont {Gefen}, \citenamefont {Rotem},\ and\ \citenamefont {Retzker}}]{Gefen2019}%
  \BibitemOpen
  \bibfield  {author} {\bibinfo {author} {\bibfnamefont {T.}~\bibnamefont {Gefen}}, \bibinfo {author} {\bibfnamefont {A.}~\bibnamefont {Rotem}},\ and\ \bibinfo {author} {\bibfnamefont {A.}~\bibnamefont {Retzker}},\ }\bibfield  {title} {\bibinfo {title} {Overcoming resolution limits with quantum sensing},\ }\href {https://doi.org/10.1038/s41467-019-12817-y} {\bibfield  {journal} {\bibinfo  {journal} {Nature Communications}\ }\textbf {\bibinfo {volume} {10}},\ \bibinfo {pages} {4992} (\bibinfo {year} {2019})}\BibitemShut {NoStop}%
\bibitem [{\citenamefont {H{\"a}berle}\ \emph {et~al.}(2017)\citenamefont {H{\"a}berle}, \citenamefont {Oeckinghaus}, \citenamefont {Schmid-Lorch}, \citenamefont {Pfender}, \citenamefont {de~Oliveira}, \citenamefont {Momenzadeh}, \citenamefont {Finkler},\ and\ \citenamefont {Wrachtrup}}]{haberle2017nuclear}%
  \BibitemOpen
  \bibfield  {author} {\bibinfo {author} {\bibfnamefont {T.}~\bibnamefont {H{\"a}berle}}, \bibinfo {author} {\bibfnamefont {T.}~\bibnamefont {Oeckinghaus}}, \bibinfo {author} {\bibfnamefont {D.}~\bibnamefont {Schmid-Lorch}}, \bibinfo {author} {\bibfnamefont {M.}~\bibnamefont {Pfender}}, \bibinfo {author} {\bibfnamefont {F.~F.}\ \bibnamefont {de~Oliveira}}, \bibinfo {author} {\bibfnamefont {S.~A.}\ \bibnamefont {Momenzadeh}}, \bibinfo {author} {\bibfnamefont {A.}~\bibnamefont {Finkler}},\ and\ \bibinfo {author} {\bibfnamefont {J.}~\bibnamefont {Wrachtrup}},\ }\bibfield  {title} {\bibinfo {title} {Nuclear quantum-assisted magnetometer},\ }\href@noop {} {\bibfield  {journal} {\bibinfo  {journal} {Review of Scientific Instruments}\ }\textbf {\bibinfo {volume} {88}},\ \bibinfo {pages} {013702} (\bibinfo {year} {2017})}\BibitemShut {NoStop}%
\bibitem [{\citenamefont {Taylor}\ \emph {et~al.}(2008)\citenamefont {Taylor}, \citenamefont {Cappellaro}, \citenamefont {Childress}, \citenamefont {Jiang}, \citenamefont {Budker}, \citenamefont {Hemmer}, \citenamefont {Yacoby}, \citenamefont {Walsworth},\ and\ \citenamefont {Lukin}}]{Taylor2008high}%
  \BibitemOpen
  \bibfield  {author} {\bibinfo {author} {\bibfnamefont {J.~M.}\ \bibnamefont {Taylor}}, \bibinfo {author} {\bibfnamefont {P.}~\bibnamefont {Cappellaro}}, \bibinfo {author} {\bibfnamefont {L.}~\bibnamefont {Childress}}, \bibinfo {author} {\bibfnamefont {L.}~\bibnamefont {Jiang}}, \bibinfo {author} {\bibfnamefont {D.}~\bibnamefont {Budker}}, \bibinfo {author} {\bibfnamefont {P.}~\bibnamefont {Hemmer}}, \bibinfo {author} {\bibfnamefont {A.}~\bibnamefont {Yacoby}}, \bibinfo {author} {\bibfnamefont {R.}~\bibnamefont {Walsworth}},\ and\ \bibinfo {author} {\bibfnamefont {M.}~\bibnamefont {Lukin}},\ }\bibfield  {title} {\bibinfo {title} {High-sensitivity diamond magnetometer with nanoscale resolution},\ }\href@noop {} {\bibfield  {journal} {\bibinfo  {journal} {Nature Physics}\ }\textbf {\bibinfo {volume} {4}},\ \bibinfo {pages} {810} (\bibinfo {year} {2008})}\BibitemShut {NoStop}%
\bibitem [{\citenamefont {Wang}\ \emph {et~al.}(2019)\citenamefont {Wang}, \citenamefont {Lang}, \citenamefont {Schmitt}, \citenamefont {Lang}, \citenamefont {Casanova}, \citenamefont {McGuinness}, \citenamefont {Monteiro}, \citenamefont {Jelezko},\ and\ \citenamefont {Plenio}}]{WangPRL2019}%
  \BibitemOpen
  \bibfield  {author} {\bibinfo {author} {\bibfnamefont {Z.-Y.}\ \bibnamefont {Wang}}, \bibinfo {author} {\bibfnamefont {J.~E.}\ \bibnamefont {Lang}}, \bibinfo {author} {\bibfnamefont {S.}~\bibnamefont {Schmitt}}, \bibinfo {author} {\bibfnamefont {J.}~\bibnamefont {Lang}}, \bibinfo {author} {\bibfnamefont {J.}~\bibnamefont {Casanova}}, \bibinfo {author} {\bibfnamefont {L.}~\bibnamefont {McGuinness}}, \bibinfo {author} {\bibfnamefont {T.~S.}\ \bibnamefont {Monteiro}}, \bibinfo {author} {\bibfnamefont {F.}~\bibnamefont {Jelezko}},\ and\ \bibinfo {author} {\bibfnamefont {M.~B.}\ \bibnamefont {Plenio}},\ }\bibfield  {title} {\bibinfo {title} {Randomization of pulse phases for unambiguous and robust quantum sensing},\ }\href {https://doi.org/10.1103/PhysRevLett.122.200403} {\bibfield  {journal} {\bibinfo  {journal} {Phys. Rev. Lett.}\ }\textbf {\bibinfo {volume} {122}},\ \bibinfo {pages} {200403} (\bibinfo {year} {2019})}\BibitemShut {NoStop}%
\bibitem [{\citenamefont {Doherty}\ \emph {et~al.}(2012)\citenamefont {Doherty}, \citenamefont {Dolde}, \citenamefont {Fedder}, \citenamefont {Jelezko}, \citenamefont {Wrachtrup}, \citenamefont {Manson},\ and\ \citenamefont {Hollenberg}}]{DohertyPRB2012}%
  \BibitemOpen
  \bibfield  {author} {\bibinfo {author} {\bibfnamefont {M.~W.}\ \bibnamefont {Doherty}}, \bibinfo {author} {\bibfnamefont {F.}~\bibnamefont {Dolde}}, \bibinfo {author} {\bibfnamefont {H.}~\bibnamefont {Fedder}}, \bibinfo {author} {\bibfnamefont {F.}~\bibnamefont {Jelezko}}, \bibinfo {author} {\bibfnamefont {J.}~\bibnamefont {Wrachtrup}}, \bibinfo {author} {\bibfnamefont {N.~B.}\ \bibnamefont {Manson}},\ and\ \bibinfo {author} {\bibfnamefont {L.~C.~L.}\ \bibnamefont {Hollenberg}},\ }\bibfield  {title} {\bibinfo {title} {Theory of the ground-state spin of the nv${}^{\ensuremath{-}}$ center in diamond},\ }\href {https://doi.org/10.1103/PhysRevB.85.205203} {\bibfield  {journal} {\bibinfo  {journal} {Phys. Rev. B}\ }\textbf {\bibinfo {volume} {85}},\ \bibinfo {pages} {205203} (\bibinfo {year} {2012})}\BibitemShut {NoStop}%
\bibitem [{\citenamefont {L{\"u}hmann}\ \emph {et~al.}(2011)\citenamefont {L{\"u}hmann}, \citenamefont {Raatz}, \citenamefont {John}, \citenamefont {al}, \citenamefont {Shao}, \citenamefont {Zhang}, \citenamefont {Robledo}, \citenamefont {Bernien}, \citenamefont {van~der Sar},\ and\ \citenamefont {Hanson}}]{RobledoNJP2011}%
  \BibitemOpen
  \bibfield  {author} {\bibinfo {author} {\bibfnamefont {T.}~\bibnamefont {L{\"u}hmann}}, \bibinfo {author} {\bibfnamefont {N.}~\bibnamefont {Raatz}}, \bibinfo {author} {\bibfnamefont {R.}~\bibnamefont {John}}, \bibinfo {author} {\bibnamefont {al}}, \bibinfo {author} {\bibfnamefont {T.-J.}\ \bibnamefont {Shao}}, \bibinfo {author} {\bibfnamefont {Q.-L.}\ \bibnamefont {Zhang}}, \bibinfo {author} {\bibfnamefont {L.}~\bibnamefont {Robledo}}, \bibinfo {author} {\bibfnamefont {H.}~\bibnamefont {Bernien}}, \bibinfo {author} {\bibfnamefont {T.}~\bibnamefont {van~der Sar}},\ and\ \bibinfo {author} {\bibfnamefont {R.}~\bibnamefont {Hanson}},\ }\bibfield  {title} {\bibinfo {title} {Spin dynamics in the optical cycle of single nitrogen-vacancy centres in diamond control of high-harmonic generation from periodic asymmetric lattices spin dynamics in the optical cycle of single nitrogen-vacancy centres in diamond},\ }\href {https://doi.org/10.1088/1367-2630/13/2/025013} {\bibfield  {journal} {\bibinfo  {journal} {New
  Journal of Physics}\ }\textbf {\bibinfo {volume} {13}},\ \bibinfo {pages} {25013} (\bibinfo {year} {2011})}\BibitemShut {NoStop}%
\bibitem [{\citenamefont {Harrison}\ \emph {et~al.}(2006)\citenamefont {Harrison}, \citenamefont {Sellars},\ and\ \citenamefont {Manson}}]{HarrisonDRM2006}%
  \BibitemOpen
  \bibfield  {author} {\bibinfo {author} {\bibfnamefont {J.}~\bibnamefont {Harrison}}, \bibinfo {author} {\bibfnamefont {M.~J.}\ \bibnamefont {Sellars}},\ and\ \bibinfo {author} {\bibfnamefont {N.~B.}\ \bibnamefont {Manson}},\ }\bibfield  {title} {\bibinfo {title} {Measurement of the optically induced spin polarisation of n-v centres in diamond},\ }\href {https://doi.org/10.1016/J.DIAMOND.2005.12.027} {\bibfield  {journal} {\bibinfo  {journal} {Diamond and Related Materials}\ }\textbf {\bibinfo {volume} {15}},\ \bibinfo {pages} {586} (\bibinfo {year} {2006})}\BibitemShut {NoStop}%
\bibitem [{\citenamefont {M{\"u}ller}\ \emph {et~al.}(2014)\citenamefont {M{\"u}ller}, \citenamefont {Kong}, \citenamefont {Cai}, \citenamefont {Melentijevi{\'c}}, \citenamefont {Stacey}, \citenamefont {Markham}, \citenamefont {Twitchen}, \citenamefont {Isoya}, \citenamefont {Pezzagna}, \citenamefont {Meijer}, \citenamefont {Du}, \citenamefont {Plenio}, \citenamefont {Naydenov}, \citenamefont {Mcguinness},\ and\ \citenamefont {Jelezko}}]{MuellerNatComm2014}%
  \BibitemOpen
  \bibfield  {author} {\bibinfo {author} {\bibfnamefont {C.}~\bibnamefont {M{\"u}ller}}, \bibinfo {author} {\bibfnamefont {X.}~\bibnamefont {Kong}}, \bibinfo {author} {\bibfnamefont {J.-M.}\ \bibnamefont {Cai}}, \bibinfo {author} {\bibfnamefont {K.}~\bibnamefont {Melentijevi{\'c}}}, \bibinfo {author} {\bibfnamefont {A.}~\bibnamefont {Stacey}}, \bibinfo {author} {\bibfnamefont {M.}~\bibnamefont {Markham}}, \bibinfo {author} {\bibfnamefont {D.}~\bibnamefont {Twitchen}}, \bibinfo {author} {\bibfnamefont {J.}~\bibnamefont {Isoya}}, \bibinfo {author} {\bibfnamefont {S.}~\bibnamefont {Pezzagna}}, \bibinfo {author} {\bibfnamefont {J.}~\bibnamefont {Meijer}}, \bibinfo {author} {\bibfnamefont {J.~F.}\ \bibnamefont {Du}}, \bibinfo {author} {\bibfnamefont {M.~B.}\ \bibnamefont {Plenio}}, \bibinfo {author} {\bibfnamefont {B.}~\bibnamefont {Naydenov}}, \bibinfo {author} {\bibfnamefont {L.~P.}\ \bibnamefont {Mcguinness}},\ and\ \bibinfo {author} {\bibfnamefont {F.}~\bibnamefont {Jelezko}},\ }\bibfield  {title} {\bibinfo
  {title} {Article nuclear magnetic resonance spectroscopy with single spin sensitivity},\ }\bibfield  {journal} {\bibinfo  {journal} {Nature Communications}\ }\href {https://doi.org/10.1038/ncomms5703} {10.1038/ncomms5703} (\bibinfo {year} {2014})\BibitemShut {NoStop}%
\bibitem [{\citenamefont {Vetter}\ \emph {et~al.}(2022)\citenamefont {Vetter}, \citenamefont {Marshall}, \citenamefont {Genov}, \citenamefont {Weiss}, \citenamefont {Striegler}, \citenamefont {Gro\ss{}mann}, \citenamefont {Oviedo-Casado}, \citenamefont {Cerrillo}, \citenamefont {Prior}, \citenamefont {Neumann},\ and\ \citenamefont {Jelezko}}]{Vetter2022prapplied}%
  \BibitemOpen
  \bibfield  {author} {\bibinfo {author} {\bibfnamefont {P.~J.}\ \bibnamefont {Vetter}}, \bibinfo {author} {\bibfnamefont {A.}~\bibnamefont {Marshall}}, \bibinfo {author} {\bibfnamefont {G.~T.}\ \bibnamefont {Genov}}, \bibinfo {author} {\bibfnamefont {T.~F.}\ \bibnamefont {Weiss}}, \bibinfo {author} {\bibfnamefont {N.}~\bibnamefont {Striegler}}, \bibinfo {author} {\bibfnamefont {E.~F.}\ \bibnamefont {Gro\ss{}mann}}, \bibinfo {author} {\bibfnamefont {S.}~\bibnamefont {Oviedo-Casado}}, \bibinfo {author} {\bibfnamefont {J.}~\bibnamefont {Cerrillo}}, \bibinfo {author} {\bibfnamefont {J.}~\bibnamefont {Prior}}, \bibinfo {author} {\bibfnamefont {P.}~\bibnamefont {Neumann}},\ and\ \bibinfo {author} {\bibfnamefont {F.}~\bibnamefont {Jelezko}},\ }\bibfield  {title} {\bibinfo {title} {Zero- and low-field sensing with nitrogen-vacancy centers},\ }\href {https://doi.org/10.1103/PhysRevApplied.17.044028} {\bibfield  {journal} {\bibinfo  {journal} {Phys. Rev. Appl.}\ }\textbf {\bibinfo {volume} {17}},\ \bibinfo {pages}
  {044028} (\bibinfo {year} {2022})}\BibitemShut {NoStop}%
\bibitem [{\citenamefont {Fedder}\ \emph {et~al.}(2011)\citenamefont {Fedder}, \citenamefont {Dolde}, \citenamefont {Rempp}, \citenamefont {Wolf}, \citenamefont {Hemmer}, \citenamefont {Jelezko},\ and\ \citenamefont {Wrachtrup}}]{FedderAPB2011}%
  \BibitemOpen
  \bibfield  {author} {\bibinfo {author} {\bibfnamefont {H.}~\bibnamefont {Fedder}}, \bibinfo {author} {\bibfnamefont {F.}~\bibnamefont {Dolde}}, \bibinfo {author} {\bibfnamefont {F.}~\bibnamefont {Rempp}}, \bibinfo {author} {\bibfnamefont {T.}~\bibnamefont {Wolf}}, \bibinfo {author} {\bibfnamefont {P.}~\bibnamefont {Hemmer}}, \bibinfo {author} {\bibfnamefont {F.}~\bibnamefont {Jelezko}},\ and\ \bibinfo {author} {\bibfnamefont {J.}~\bibnamefont {Wrachtrup}},\ }\bibfield  {title} {\bibinfo {title} {Towards t 1-limited magnetic resonance imaging using rabi beats},\ }\href {https://doi.org/10.1007/s00340-011-4408-4} {\bibfield  {journal} {\bibinfo  {journal} {Appl Phys B}\ }\textbf {\bibinfo {volume} {102}},\ \bibinfo {pages} {497} (\bibinfo {year} {2011})}\BibitemShut {NoStop}%
\bibitem [{\citenamefont {Staudenmaier}\ \emph {et~al.}(2022)\citenamefont {Staudenmaier}, \citenamefont {Vijayakumar-Sreeja}, \citenamefont {Oviedo-Casado}, \citenamefont {Genov}, \citenamefont {Cohen}, \citenamefont {Dulog}, \citenamefont {Unden}, \citenamefont {Striegler}, \citenamefont {Marshall}, \citenamefont {Scheuer}, \citenamefont {Findler}, \citenamefont {Lang}, \citenamefont {Schwartz}, \citenamefont {Neumann}, \citenamefont {Retzker},\ and\ \citenamefont {Jelezko}}]{Staudenmaier2022}%
  \BibitemOpen
  \bibfield  {author} {\bibinfo {author} {\bibfnamefont {N.}~\bibnamefont {Staudenmaier}}, \bibinfo {author} {\bibfnamefont {A.}~\bibnamefont {Vijayakumar-Sreeja}}, \bibinfo {author} {\bibfnamefont {S.}~\bibnamefont {Oviedo-Casado}}, \bibinfo {author} {\bibfnamefont {G.}~\bibnamefont {Genov}}, \bibinfo {author} {\bibfnamefont {D.}~\bibnamefont {Cohen}}, \bibinfo {author} {\bibfnamefont {D.}~\bibnamefont {Dulog}}, \bibinfo {author} {\bibfnamefont {T.}~\bibnamefont {Unden}}, \bibinfo {author} {\bibfnamefont {N.}~\bibnamefont {Striegler}}, \bibinfo {author} {\bibfnamefont {A.}~\bibnamefont {Marshall}}, \bibinfo {author} {\bibfnamefont {J.}~\bibnamefont {Scheuer}}, \bibinfo {author} {\bibfnamefont {C.}~\bibnamefont {Findler}}, \bibinfo {author} {\bibfnamefont {J.}~\bibnamefont {Lang}}, \bibinfo {author} {\bibfnamefont {I.}~\bibnamefont {Schwartz}}, \bibinfo {author} {\bibfnamefont {P.}~\bibnamefont {Neumann}}, \bibinfo {author} {\bibfnamefont {A.}~\bibnamefont {Retzker}},\ and\ \bibinfo {author} {\bibfnamefont
  {F.}~\bibnamefont {Jelezko}},\ }\bibfield  {title} {\bibinfo {title} {Power-law scaling of correlations in statistically polarised nano-nmr},\ }\href {https://doi.org/10.1038/s41534-022-00632-1} {\bibfield  {journal} {\bibinfo  {journal} {npj Quantum Information}\ }\textbf {\bibinfo {volume} {8}},\ \bibinfo {pages} {120} (\bibinfo {year} {2022})}\BibitemShut {NoStop}%
\bibitem [{\citenamefont {Staudenmaier}\ \emph {et~al.}(2021)\citenamefont {Staudenmaier}, \citenamefont {Schmitt}, \citenamefont {McGuinness},\ and\ \citenamefont {Jelezko}}]{Staudenmaier2021pra}%
  \BibitemOpen
  \bibfield  {author} {\bibinfo {author} {\bibfnamefont {N.}~\bibnamefont {Staudenmaier}}, \bibinfo {author} {\bibfnamefont {S.}~\bibnamefont {Schmitt}}, \bibinfo {author} {\bibfnamefont {L.~P.}\ \bibnamefont {McGuinness}},\ and\ \bibinfo {author} {\bibfnamefont {F.}~\bibnamefont {Jelezko}},\ }\bibfield  {title} {\bibinfo {title} {Phase-sensitive quantum spectroscopy with high-frequency resolution},\ }\href {https://doi.org/10.1103/PhysRevA.104.L020602} {\bibfield  {journal} {\bibinfo  {journal} {Phys. Rev. A}\ }\textbf {\bibinfo {volume} {104}},\ \bibinfo {pages} {L020602} (\bibinfo {year} {2021})}\BibitemShut {NoStop}%
\bibitem [{\citenamefont {Orwa}\ \emph {et~al.}(2011)\citenamefont {Orwa}, \citenamefont {Santori}, \citenamefont {Fu}, \citenamefont {Gibson}, \citenamefont {Simpson}, \citenamefont {Aharonovich}, \citenamefont {Stacey}, \citenamefont {Cimmino}, \citenamefont {Balog}, \citenamefont {Markham}, \citenamefont {Twitchen}, \citenamefont {Greentree}, \citenamefont {Beausoleil},\ and\ \citenamefont {Prawer}}]{orwa2011annealing}%
  \BibitemOpen
  \bibfield  {author} {\bibinfo {author} {\bibfnamefont {J.~O.}\ \bibnamefont {Orwa}}, \bibinfo {author} {\bibfnamefont {C.}~\bibnamefont {Santori}}, \bibinfo {author} {\bibfnamefont {K.~M.~C.}\ \bibnamefont {Fu}}, \bibinfo {author} {\bibfnamefont {B.}~\bibnamefont {Gibson}}, \bibinfo {author} {\bibfnamefont {D.}~\bibnamefont {Simpson}}, \bibinfo {author} {\bibfnamefont {I.}~\bibnamefont {Aharonovich}}, \bibinfo {author} {\bibfnamefont {A.}~\bibnamefont {Stacey}}, \bibinfo {author} {\bibfnamefont {A.}~\bibnamefont {Cimmino}}, \bibinfo {author} {\bibfnamefont {P.}~\bibnamefont {Balog}}, \bibinfo {author} {\bibfnamefont {M.}~\bibnamefont {Markham}}, \bibinfo {author} {\bibfnamefont {D.}~\bibnamefont {Twitchen}}, \bibinfo {author} {\bibfnamefont {A.~D.}\ \bibnamefont {Greentree}}, \bibinfo {author} {\bibfnamefont {R.~G.}\ \bibnamefont {Beausoleil}},\ and\ \bibinfo {author} {\bibfnamefont {S.}~\bibnamefont {Prawer}},\ }\bibfield  {title} {\bibinfo {title} {Engineering of nitrogen-vacancy color centers in high
  purity diamond by ion implantation and annealing},\ }\href {https://doi.org/10.1063/1.3573768} {\bibfield  {journal} {\bibinfo  {journal} {Journal of Applied Physics}\ }\textbf {\bibinfo {volume} {109}},\ \bibinfo {pages} {083530} (\bibinfo {year} {2011})}\BibitemShut {NoStop}%
\bibitem [{\citenamefont {Osterkamp}\ \emph {et~al.}(2015)\citenamefont {Osterkamp}, \citenamefont {Lang}, \citenamefont {Scharpf}, \citenamefont {M{\"u}ller}, \citenamefont {McGuinness}, \citenamefont {Diemant}, \citenamefont {Behm}, \citenamefont {Naydenov},\ and\ \citenamefont {Jelezko}}]{osterkamp2015stabilizing}%
  \BibitemOpen
  \bibfield  {author} {\bibinfo {author} {\bibfnamefont {C.}~\bibnamefont {Osterkamp}}, \bibinfo {author} {\bibfnamefont {J.}~\bibnamefont {Lang}}, \bibinfo {author} {\bibfnamefont {J.}~\bibnamefont {Scharpf}}, \bibinfo {author} {\bibfnamefont {C.}~\bibnamefont {M{\"u}ller}}, \bibinfo {author} {\bibfnamefont {L.~P.}\ \bibnamefont {McGuinness}}, \bibinfo {author} {\bibfnamefont {T.}~\bibnamefont {Diemant}}, \bibinfo {author} {\bibfnamefont {R.~J.}\ \bibnamefont {Behm}}, \bibinfo {author} {\bibfnamefont {B.}~\bibnamefont {Naydenov}},\ and\ \bibinfo {author} {\bibfnamefont {F.}~\bibnamefont {Jelezko}},\ }\bibfield  {title} {\bibinfo {title} {Stabilizing shallow color centers in diamond created by nitrogen delta-doping using sf6 plasma treatment},\ }\href@noop {} {\bibfield  {journal} {\bibinfo  {journal} {Applied Physics Letters}\ }\textbf {\bibinfo {volume} {106}},\ \bibinfo {pages} {113109} (\bibinfo {year} {2015})}\BibitemShut {NoStop}%
\bibitem [{\citenamefont {Hadden}\ \emph {et~al.}(2010)\citenamefont {Hadden}, \citenamefont {Harrison}, \citenamefont {Stanley-Clarke}, \citenamefont {Marseglia}, \citenamefont {Ho}, \citenamefont {Patton}, \citenamefont {O\'Brien},\ and\ \citenamefont {Rarity}}]{hadden2010strongly}%
  \BibitemOpen
  \bibfield  {author} {\bibinfo {author} {\bibfnamefont {J.}~\bibnamefont {Hadden}}, \bibinfo {author} {\bibfnamefont {J.}~\bibnamefont {Harrison}}, \bibinfo {author} {\bibfnamefont {A.~C.}\ \bibnamefont {Stanley-Clarke}}, \bibinfo {author} {\bibfnamefont {L.}~\bibnamefont {Marseglia}}, \bibinfo {author} {\bibfnamefont {Y.-L.}\ \bibnamefont {Ho}}, \bibinfo {author} {\bibfnamefont {B.}~\bibnamefont {Patton}}, \bibinfo {author} {\bibfnamefont {J.~L.}\ \bibnamefont {O\'Brien}},\ and\ \bibinfo {author} {\bibfnamefont {J.}~\bibnamefont {Rarity}},\ }\bibfield  {title} {\bibinfo {title} {Strongly enhanced photon collection from diamond defect centers under microfabricated integrated solid immersion lenses},\ }\href@noop {} {\bibfield  {journal} {\bibinfo  {journal} {Applied Physics Letters}\ }\textbf {\bibinfo {volume} {97}},\ \bibinfo {pages} {241901} (\bibinfo {year} {2010})}\BibitemShut {NoStop}%
\bibitem [{\citenamefont {Siyushev}\ \emph {et~al.}(2010)\citenamefont {Siyushev}, \citenamefont {Kaiser}, \citenamefont {Jacques}, \citenamefont {Gerhardt}, \citenamefont {Bischof}, \citenamefont {Fedder}, \citenamefont {Dodson}, \citenamefont {Markham}, \citenamefont {Twitchen}, \citenamefont {Jelezko} \emph {et~al.}}]{siyushev2010monolithic}%
  \BibitemOpen
  \bibfield  {author} {\bibinfo {author} {\bibfnamefont {P.}~\bibnamefont {Siyushev}}, \bibinfo {author} {\bibfnamefont {F.}~\bibnamefont {Kaiser}}, \bibinfo {author} {\bibfnamefont {V.}~\bibnamefont {Jacques}}, \bibinfo {author} {\bibfnamefont {I.}~\bibnamefont {Gerhardt}}, \bibinfo {author} {\bibfnamefont {S.}~\bibnamefont {Bischof}}, \bibinfo {author} {\bibfnamefont {H.}~\bibnamefont {Fedder}}, \bibinfo {author} {\bibfnamefont {J.}~\bibnamefont {Dodson}}, \bibinfo {author} {\bibfnamefont {M.}~\bibnamefont {Markham}}, \bibinfo {author} {\bibfnamefont {D.}~\bibnamefont {Twitchen}}, \bibinfo {author} {\bibfnamefont {F.}~\bibnamefont {Jelezko}}, \emph {et~al.},\ }\bibfield  {title} {\bibinfo {title} {Monolithic diamond optics for single photon detection},\ }\href@noop {} {\bibfield  {journal} {\bibinfo  {journal} {Applied Physics Letters}\ }\textbf {\bibinfo {volume} {97}},\ \bibinfo {pages} {241902} (\bibinfo {year} {2010})}\BibitemShut {NoStop}%
\bibitem [{\citenamefont {Barry}\ \emph {et~al.}(2020)\citenamefont {Barry}, \citenamefont {Schloss}, \citenamefont {Bauch}, \citenamefont {Turner}, \citenamefont {Hart}, \citenamefont {Pham},\ and\ \citenamefont {Walsworth}}]{Barry2020sensitivity}%
  \BibitemOpen
  \bibfield  {author} {\bibinfo {author} {\bibfnamefont {J.~F.}\ \bibnamefont {Barry}}, \bibinfo {author} {\bibfnamefont {J.~M.}\ \bibnamefont {Schloss}}, \bibinfo {author} {\bibfnamefont {E.}~\bibnamefont {Bauch}}, \bibinfo {author} {\bibfnamefont {M.~J.}\ \bibnamefont {Turner}}, \bibinfo {author} {\bibfnamefont {C.~A.}\ \bibnamefont {Hart}}, \bibinfo {author} {\bibfnamefont {L.~M.}\ \bibnamefont {Pham}},\ and\ \bibinfo {author} {\bibfnamefont {R.~L.}\ \bibnamefont {Walsworth}},\ }\bibfield  {title} {\bibinfo {title} {Sensitivity optimization for nv-diamond magnetometry},\ }\href@noop {} {\bibfield  {journal} {\bibinfo  {journal} {Reviews of Modern Physics}\ }\textbf {\bibinfo {volume} {92}},\ \bibinfo {pages} {015004} (\bibinfo {year} {2020})}\BibitemShut {NoStop}%
\end{thebibliography}

\end{document}